\documentclass[aps,twocolumn,showpacs,floatfix]{revtex4}
\usepackage{amsmath}
\usepackage{amssymb}
\usepackage{graphicx}
\usepackage{subfigure}
\usepackage{dcolumn}
\usepackage{bm}
\begin{document}

\title{Stretching Semiflexible Polymer Chains: Evidence for the Importance 
of Excluded Volume Effects from Monte Carlo Simulation}
\author{Hsiao-Ping Hsu}
\email{hsu@uni-mainz.de}
\affiliation{Institut f\"ur Physik, Johannes Gutenberg-Universit\"at Mainz, 
Staudinger Weg 7, D-55099 Mainz, Germany}
\author{Kurt Binder}
\affiliation{Institut f\"ur Physik, Johannes Gutenberg-Universit\"at Mainz, 
Staudinger Weg 7, D-55099 Mainz, Germany}
\date{\today}
\begin{abstract}

Semiflexible macromolecules in dilute solution under very good solvent 
conditions are modeled by self-avoiding walks on the simple cubic lattice 
($d=3$ dimensions) and square lattice ($d=2$ dimensions), varying chain 
stiffness by an energy penalty $\epsilon_b$ for chain bending. 
In the absence of excluded volume interactions, the persistence length 
$\ell_p$ of the polymers would then simply be 
$\ell_p=\ell_b(2d-2)^{-1}q_b^{-1}$ with 
$q_b= \exp(-\epsilon_b/k_BT)$, the bond length $\ell_b$ being the lattice 
spacing, and $k_BT$ is the thermal energy. 
Using Monte Carlo simulations applying 
the pruned-enriched Rosenbluth method (PERM), both $q_b$ and the chain 
length $N$ are varied over a wide range 
$(0.005 \leq q_b \leq 1, \; N \leq 50000$), and also a stretching force $f$ 
is applied to one chain end (fixing the other end at the origin). In the 
absence of this force, in $d=2$ a single crossover from rod-like behavior 
(for contour lengths less than $\ell_p$) to swollen coils occurs, invalidating 
the Kratky-Porod model, while in $d=3$ a double crossover occurs, from rods to 
Gaussian coils (as implied by the Kratky-Porod model) and then to coils that 
are swollen due to the excluded volume interaction. If the stretching force is 
applied, excluded volume interactions matter for the force versus extension 
relation irrespective of chain stiffness in $d=2$, while theories based on the 
Kratky-Porod model are found to work in $d=3$ for stiff chains in an 
intermediate regime of chain extensions. While for $q_b \ll 1$ in this model 
a persistence length can be estimated from the initial decay of 
bond-orientational correlations, it is argued that this is not possible for 
more complex wormlike chains (e.g. bottle-brush polymers). Consequences for 
the proper interpretation of experiments are briefly discussed.
\end{abstract}
\maketitle

\section{Introduction}
The response of macromolecules with linear chemical architecture to mechanical 
forces pulling at their ends has been a longstanding problem in the statistical
mechanics of 
polymers~\cite{1,1a,2,3,4,5,6,7,8,9,10,11,12,13,14,15,15A,16,17,18,19,20A,20B,20,21,22,22A,23,24}.
Particular interest in this problem is due to advances in experimental 
techniques of single molecule measurements, probing the tension-induced 
stretching of biological macromolecules such as DNA~\cite{25}, RNA~\cite{26}, 
proteins~\cite{27} and polysaccharides~\cite{28}. But also insight into the 
structure-property relationships of synthetic polymers, e.g. 
bottle brushes~\cite{29}, has been gained by such experiments. However, 
despite extensive work on these problems, important aspects are still not well 
understood, even for the relatively simple case of macromolecules in dilute 
solutions of good solvent quality, disregarding the interesting interplay of 
chain stretching and collapse that occurs in poor 
solvents~\cite{22,30,31,32,33}, and also the interplay of chain stretching and 
adsorption on substrates~\cite{34,35A,36A,37A}.

An important aspect of these problems is local chain stiffness. Traditionally, 
chain stiffness is characterized by ``the'' persistence length~\cite{1,8} 
but evidence has been presented~\cite{35,36,37} that the traditional 
definitions are not useful under good solvent conditions, where excluded volume
interactions create long range correlations with respect to the conformational 
properties of a macromolecule~\cite{5,38}. For stretched flexible polymers 
under a force $f$ the standard theory uses the concept of 
``Pincus blobs''~\cite{4}, of size $\xi_P=k_BT/f$, $k_BT$ being the thermal 
energy, predicting a crossover from a Hookean regime, where the extension 
$\langle X \rangle$ for a force (applied in $x$-direction) scales as 
$\langle X \rangle = \langle R^2\rangle (f/dk_BT)$ in $d$ dimensions, to a 
nonlinear power law $\langle X \rangle \propto f ^{1/\nu-1}$ ($\nu$ being the 
Flory exponent, $\nu \approx 3/5$ in $d=3$ dimensions). In contrast, for 
stretched semiflexible polymers excluded volume is widely neglected in the 
literature~\cite{11,12,16,17,18} and using the 
Kratky-Porod model (K-P model)~\cite{39,40} 
simple analytic relations between force $f$ and relative extension 
$\langle X\rangle/L$, $L$ being the contour length of the polymer, are 
derived~\cite{11,20}. We recall that $L=N_b\ell_b$ where $N_b$ is the number 
of effective bonds of length $\ell _b$ connecting the effective monomeric 
units of the macromolecule, and if excluded volume interactions were absent, 
we would have, 
for $N_b \rightarrow \infty$ in the absence of the force $f$, the
end-to-end distance of the polymer chains as (the index ``0" refers to $f=0$)
\begin{equation}\label{eq1}
\langle R^2\rangle_{0} = \ell_kL=2 \ell _p L = \ell^2_k n \,,
\end{equation}
$\ell_k = 2 \ell _p$ being the length of a Kuhn segment, $n=N_b\ell_b/\ell_k$ 
being the number of such Kuhn segments forming the equivalent freely jointed 
chain~\cite{1,5}.
However, neither $\ell_k$ nor $\ell_p$ can be defined 
straightforwardly in the presence of excluded volume 
forces~\cite{35,36,37,38,41}.

With recent large scale computer simulations, we have studied the combined 
effects of local ``intrinsic'' chain stiffness and excluded volume interactions
on the conformational properties of polymers in the absence of stretching 
forces, both for $d=3$~\cite{35,36} and $d=2$~\cite{37} dimensions. 
The present study extends this work, giving a detailed study of force-extension
relations for both $d=2$ and $d=3$, complementing our results also by 
investigating fluctuations 
$\langle X^2\rangle - \langle X\rangle ^2, \; \langle R_\bot^2\rangle$ of the 
chain linear dimensions in the direction of the force and perpendicular to it. 
Whenever possible, a comparison with theoretical predictions will be given.
{
Since our study is based on modelling polymers as self-avoiding walks on
square and simple cubic lattices, the main focus of our work is on the
regime of low and intermediate forces (for very high forces, a more realistic
description of the local structure and energetics of a polymer chains, such as
bond length, bond angle and torsional potentials, becomes 
important~\cite{20B}).}

The outline of our paper is as follows: in Sec.~II, we summarize 
the state of the art, discussing in particular the theoretical results we 
want to compare to. Sec.~III briefly describes our model and the simulation 
technique, while Sec.~IV reviews the properties of chains in the absence
of stretching forces. Sec.~V describes the effects of stretching forces on 
conformational properties in $d=2$ and Sec.~VI in $d=3$ dimensions. 
Finally Sec.~VII
gives a summary and an outlook on experimental work, as well as on the related 
but more complicated problem of the ``electrostatic persistence length'' in 
polyelectrolytes (see~\cite{42} for a review and further references).

\section{Theoretical Background}
\subsection{Force-Extension Curves for Flexible Chains}
Suppose we fix one end of an isolated polymer chain at the origin and apply a 
force $\vec{f}=(f,0,0)$ acting along the x-axis to the other chain end. 
This means, we add to the Hamiltonian of the chain a potential
\begin{equation}\label{eq2}
U = - fX
\end{equation}
where X is the x-component of the end-to-end vector $\vec{R}$ of the chain. 
Noting that $\vec{R} = \sum \limits _{i=1} ^{N_b}  \vec{a}_i$, where 
$\vec{a}_i = \vec{r}_{i+1} - \vec{r}_i$ is the bond vector connecting monomers,
at sites $\vec{r}_i$ and $\vec{r}_{i+1}$, with $|\vec{a}_i|= \ell_b$ the bond 
length which we assume as a constant, X can be rewritten as
\begin{equation}\label{3}
X= \ell_b \sum \limits ^{N_b}_{i=1} \cos {\vartheta}_i \,,
\end{equation}
where ${\vartheta}_i$ is the angle between $\vec{a}_i$ and the x-axis. 
For a model 
of freely jointed chains (FJC), a straightforward calculation of the partition 
function of the chain yields the force versus extension curve in terms of the 
Langevin function~\cite{2,3,8,43} (in $d=3$ dimensions)
\begin{equation}\label{eq4}
\langle X \rangle = \ell_b N_b \mathcal{L} (f \ell_b/k_BT) \enspace {\rm (FJC)}\,,
\end{equation}
\begin{equation}\label{eq5}
\mathcal{L}(\zeta) = \coth (\zeta) - 1/\zeta \,, \enspace \zeta \equiv f 
\ell_b/k_BT\,.
\end{equation}
Note that $\mathcal{L} (\zeta \ll 1)\approx \zeta/3$ and hence one finds for 
small forces that 
\begin{eqnarray}
&& \langle X \rangle \approx \frac 1 3 \ell _b ^2 N_b f/k_BT = 
\langle R^2 \rangle_0 f/(3k_BT) \,, \nonumber \\
&&\langle X\rangle /L \approx f \ell_b/(3k_BT) \enspace {\rm (FJC)} \,, \label{eq6}
\end{eqnarray}
where $\langle R^2 \rangle_0 = \ell_b^2 N_b$ for a freely joined chain.
Eq.~(\ref{eq6}) would apply for Gaussian chains in the continuum for arbitrary 
large extensions, while for the freely jointed chain 
Eqs.~(\ref{eq4}), (\ref{eq5}) 
describe a saturation behavior for large $f$, when 
$\langle X \rangle $ approaches the contour length $L = N_b\ell_b$, as
\begin{equation}\label{eq7}
\langle X \rangle / L \approx 1 - k_BT / f \ell_b \enspace {\rm (FJC)} \,.
\end{equation}
Of course, excluded volume interactions significantly modify the behavior 
described by Eqs.~(\ref{eq4})-(\ref{eq6}). Already in the absence of the force, 
Eq.~(\ref{eq1}) is replaced by 
\begin{equation}\label{eq8}
\langle R ^2 \rangle _0 = C \ell_b ^2 N _b ^{2 \nu} \,, \enspace 
N_b \rightarrow \infty \enspace {\rm (SAW)} \,,
\end{equation}
where $C$ is a (non-universal, i.e. model - or system-dependent) constant of 
order unity, and the exponent $\nu \approx 0.588$ in $d=3$ 
dimensions~\cite{44,45} while $\nu = 3/4$ in $d=2$ dimensions~\cite{5,8,45}
\{Polymer chains behave like self-avoiding walks (SAWs)\}. 
Treating the potential, Eq.~(\ref{eq2}), as a small perturbation in linear 
response one can generally show that
\begin{equation}\label{eq9}
\langle X \rangle = \langle R^2\rangle _0 f / (dk_BT)\qquad
\textrm{for \; small} \,\, f 
\end{equation}
and hence the relative extension becomes in this regime
\begin{equation}\label{eq10}
\langle X \rangle / L = C(L/\ell_b)^{2\nu-1} (f \ell_b)/(dk_BT) 
\enspace {\rm (SAW)} \,.
\end{equation}
Comparing to Eq.~(\ref{eq6}) we note that the relative extension is enhanced by 
a factor $C(L/\ell_b)^{2\nu-1}$ in comparison with the result for the freely 
jointed chain.

While for the freely jointed chain the linear behavior, Eq.~(\ref{eq6}), 
smoothly crosses over to the saturation behavior 
$\langle X \rangle /L\rightarrow 1$, Eq.~(\ref{eq7}), for the 
swollen coil there
occurs an intermediate regime with a nonlinear relation between extension and 
force. This regime was first discussed by Pincus~\cite{4} in terms of the 
scaling ansatz
\begin{equation}\label{eq11}
\langle X \rangle = \langle R^2\rangle _0^{1/2} F (\langle
R^2\rangle _0^{1/2} / \xi _p) \enspace {\rm (SAW)} \,,
\end{equation}
where $\xi_P $ is the radius of a ``tensile blob'' 
(also called now ``Pincus blob''),
\begin{equation}\label{eq12}
\xi_P=k_BT/f \,,
\end{equation}
and $F$ is a scaling function. Of course, this description makes only sense if
\begin{equation}\label{eq13}
\ell_b \ll \xi_P \ll \langle R^2\rangle ^{1/2}_0\,,
\end{equation}
since the scaling law for a blob $(\xi_P \approx \ell_b g ^\nu$ with $g$ 
monomers per blob) breaks down when $g$ is no longer very large; then a 
gradual crossover to the behavior of a strongly stretched freely jointed 
chain must occur (excluded volume then becomes irrelevant). For $\xi_P$ 
of order $\langle R^2 \rangle_0 ^{1/2}$, $F(\eta)$ behaves as 
$(\eta \equiv \langle R ^2\rangle _0^{1/2}/\xi_P$)
\begin{equation}\label{eq14}
F(\eta) = \eta/d
\end{equation}
so that Eq.~(\ref{eq11}) reduces to the linear response results, 
Eq.~(\ref{eq9}). In the regime where 
Eq.~(\ref{eq13}) holds, the conformation of the chain is a stretched string 
of $N_b/g$ blobs, i.e. in order to obtain $\langle X \rangle \propto L$ 
one must require that $F(\eta) \propto \eta^{1/\nu-1}$ and hence
\begin{equation}\label{eq15}
\langle X \rangle \propto \xi_P N_b/g \propto (k_BT/f \ell_b)^{1-1/\nu} L
\enspace {\rm (SAW)} \,,
\end{equation}
and hence one finds that in this regime the relative extension varies as
\begin{equation}\label{eq16}
\langle X \rangle /L \propto (f\ell_b/k_BT) ^{1/\nu-1} \enspace {\rm (SAW)}\,.
\end{equation}
{
In a biopolymer context, this old result due to Pincus~\cite{4} was recently
``rediscovered" by Lam~\cite{Lam}.}
This scaling behavior can be made somewhat more explicit using the scaling 
description of the distribution $P_{N_b}(X)$ in which (in the absence of a 
stretching force $f$) a displacement $X$ between two end monomers 
occurs~\cite{9}
\begin{equation}\label{eq17}
\langle X \rangle = k_BT\partial \ln Z(f)/\partial f \enspace {\rm (SAW)} \,, 
\end{equation}
where the partition function $Z(f)$ is ($Z_0$ is a normalization constant 
out of interest here)
\begin{equation}\label{eq18}
Z(f)= Z_0\int d^d \vec{R}P_{N_b}(X) \exp(fX/k_BT)\,.
\end{equation}
Using the ansatz~\cite{45}
\begin{equation}\label{eq19}
P_{N_b} (X) \propto h (y) \propto y^\phi \exp[-Dy^{1/(1-\nu)}]\,,
\end{equation}
where $y \equiv X/\langle R^2 \rangle_0 ^{1/2}, \; \phi =
(1-\gamma +\nu d-d/2)/(1-\nu)$, $\gamma$ is a standard critical 
exponent~\cite{5} and $D$ is a constant, Eqs.~(\ref{eq17}), (\ref{eq18})
can be worked out numerically.

Wittkop et al.~\cite{9} derived Eq.~(\ref{eq16}) without explicit
recourse to a blob picture. Wittkop et al.~\cite{9} tried also to
provide Monte Carlo evidence for Eq.~(\ref{eq16}), both in $d=2$ and
$d=3$, but they had to restrict their study to very short chains
$(20 \leq N_b\leq 100$). Morrison et al.~\cite{21} argued that
chain lengths of at least $N_b = 10^3$ are necessary to provide
clear simulation evidence for the Pincus tensile blob regime
(described by Eqs.~(\ref{eq13}), (\ref{eq16})). In fact, using $N_b =
6000$ Pierleoni et al.~\cite{13} succeeded to obtain evidence in
favor of the Pincus theory~\cite{4} for the chain structure factor
under stretch.
However, we are not aware of systematic tests of Eq.~(\ref{eq16})
for very large $N_b$, as shall be presented here.

A very interesting issue are also the longitudinal and transverse
fluctuations, in the extensions of stretched chains. For freely
jointed chains Titantah et al.~\cite{15} derived
\begin{equation}\label{eq20}
\langle X^2\rangle - \langle X\rangle^2 = 
N_b \ell_b^2 [1-2 \mathcal{L} (\zeta)/\zeta - \mathcal{L}^2(\zeta)] 
\enspace {\rm (FJC)}
\end{equation}
which for large $\zeta = f\ell_b/k_BT $ reduces to
\begin{equation}\label{eq21}
\langle X^2\rangle - \langle X \rangle^2 \approx
N_b\ell_b^2\zeta^{-2}=(k_BT/f\ell_b)^2N_b\ell_b^2 \enspace {\rm (FJC)} \,.
\end{equation}
The transverse fluctuations becomes
\begin{eqnarray}\label{eq22}
\langle R^2_\bot \rangle &=&\langle Y^2\rangle + 
\langle Z^2\rangle \nonumber \\
&=& 2N_b\ell_b^2\mathcal{L}(\zeta)/\zeta \approx
2N_b\ell_b^2(k_BT/f\ell_b) \enspace {\rm (FJC)} \,,
\end{eqnarray}
where the last expression again refers to $f \rightarrow \infty$.
Of course, Eq.~(\ref{eq22}) differs substantially from a continuum
Gaussian model of a chain (there the transverse linear dimensions
are not affected by the pulling force at all).

In the case where excluded volume is taken into account, one
obtains using again an approach based on
Eqs.~(\ref{eq17})-(\ref{eq19}) the approximate expressions~\cite{15}
\begin{equation}\label{eq23}
(\langle X^2\rangle - \langle X \rangle ^2)/\langle R^2\rangle _0
=\frac {s_3}{s_1} - (\frac{c_2}{s_1})^2 + \zeta ^{-2} \enspace {\rm (SAW)} \,,
\end{equation}
\begin{equation}\label{eq24}
\langle R_\bot^2\rangle /\langle R^2\rangle _0 = \zeta^{-2}[\zeta
c_2/s_1-1] \enspace {\rm (SAW)} \,,
\end{equation}
where the functions $s_i(\zeta)$ and $c_i(\zeta)$ are defined as
\begin{equation}\label{eq25}
s_i(\zeta) = \int \limits _0^\infty dy \sinh (\zeta y)y^{i+\phi}
\exp[-D y^{1/(1-\nu)}]\,, \enspace i=1,3\,,
\end{equation}
and
\begin{equation}\label{eq26}
c_i(\zeta)= \int \limits _0^\infty d y \cosh (\zeta y) y^{i+\phi}
\exp [-Dy^{1/(1-\nu)}]\,, \enspace i=1,2\,.
\end{equation}
Alternative approximate expressions were derived by Morrison et
al.~\cite{21} using the self-consistent variational method due to
Edwards and Singh~\cite{46}.

\subsection{Semiflexible chains in the absence of stretching
forces: Chain linear dimensions and bond vector orientational
correlations}

Following Winkler~\cite{18,47} we first consider a chain with fixed
bond length $\ell _b$ but successive bonds are correlated with
respect to their relative orientations,
\begin{equation}\label{eq27}
\langle \vec{a}_i^2\rangle = \ell ^2_b, \enspace \langle \vec{a}_i
\cdot \vec{a}_{i+1}\rangle = \ell _b^2t, \enspace t \equiv \langle
\cos \theta \rangle \,,
\end{equation}
$\theta$ being the angle between the orientation of two successive
bond vectors. For this model, in the absence of excluded volume effects,
the end-to-end distance is well-known~\cite{1,47}
\begin{equation}\label{eq28}
\langle R^2\rangle _0 = N_b \ell _b^2 (\frac {1+t}{1-t} + \frac
{2t}{N_b} \frac {t^{N_b}-1}{(t-1)^2})
\end{equation}
In the limit $N_b \rightarrow \infty$ the correlation function of
bond vectors decays exponentially as a function of their chemical
distance $s$,
\begin{equation}\label{eq29}
\langle \vec{a}_i \cdot \vec{a}_{i+s}\rangle = \ell _b^2 \langle
\cos \theta (s)\rangle = \ell_b^2 \langle \cos \theta \rangle ^s =
\ell _b^2 \exp (-\ell_bs/\ell_p)\,,
\end{equation}
where we have introduced the notion of the persistence length
$\ell_p$~\cite{8,43} which becomes in this case
\begin{equation}\label{eq30}
\ell_b/\ell_p = - \ln (\langle \cos \theta \rangle) \,.
\end{equation}
In the case of semiflexible chains one has $\langle \cos \theta
\rangle \approx 1 - \langle \theta ^2\rangle /2$, since $\langle
\theta ^2\rangle$ then is small, and hence one finds (for $N_b
\rightarrow \infty$)~\cite{43}
\begin{equation}\label{eq31}
\ell_p \approx 2 \ell_b /\langle \theta ^2\rangle \,, \enspace \langle
R^2\rangle _0 \approx 4 N_b \ell _b^2/\langle \theta ^2\rangle = 2
\ell_p\ell_bN_b\,,
\end{equation}
so in this limit the Kuhn length $\ell_k$ \{Eq.~(\ref{eq1})\} becomes
$\ell_k=2\ell_p$, as was anticipated.

When one considers now the limit $\ell_b \rightarrow 0$, $N_b
\rightarrow \infty$, keeping $L = \ell_bN_b$ as well as $\ell_p$ finite, one obtains from Eq.~(\ref{eq28})~\cite{47}

\begin{equation}\label{eq32}
\langle R^2 \rangle _0 = 2 \ell _p L \{1- \frac {\ell_p}{L}  
[1-\exp(-L/\ell_p)]\} \,,
\end{equation}
which is nothing but the result that one could have derived directly from the Kratky-Porod model~\cite{39,40} for wormlike chains,
\begin{equation}\label{eq33}
\mathcal{H} =  \frac \kappa 2 \int \limits _0^L 
(\frac {\partial ^2 \vec{r}}{\partial s^2})^2 ds\,,
\end{equation}
where the polymer chain is described by the contour $\vec{r}(s)$ in continuous 
space. The bending stiffness $\kappa$ is related to the persistence length 
$\ell_p$ as
\begin{equation}\label{eq34}
\kappa = \ell_pk_BT \,, \enspace d=3 \,, \enspace \textrm{or} \enspace 
\kappa = \frac{\ell_p}{2} k_BT \,, \enspace d=2 \,.
\end{equation}
We note in this context the connection to the lattice models that will be 
studied in the present work, where we study self-avoiding walks on square 
$(d=2)$ and simple cubic $(d=3)$  lattices, using a ``penalty energy'' 
$\epsilon_b$ if the chain makes a bend (by a 90$^\circ$ angle). Relaxing the 
excluded volume constraint by considering a 
``non-reversal random walk''~\cite{48}, where only immediate reversals of a 
simple random walk model would be forbidden, we immediately conclude that
\begin{equation}\label{eq35}
\langle \cos \theta\rangle = 1/[1+(2d-2))q_b]\,, \enspace
q_b = \exp(-\epsilon_b/k_BT) \,,
\end{equation}
and hence one would obtain for $q_b\rightarrow 0$ from Eq.~(\ref{eq31}) that
\begin{equation}\label{eq36}
\ell_p / \ell_b=1/(2q_b) \, \enspace d = 2\,, \enspace {\rm or} \enspace  
\ell_p / \ell_b=1/(4q_b) \, \enspace d=3\,.
\end{equation}

{
At this point, it is interesting to recall that the present lattice model can
be described by the Hamiltonian ${\cal H}=\epsilon_b \sum_i 
(1-\vec{a}_i \cdot \vec{a}_j/\ell_b^2)=\epsilon \sum_i(1-\cos \theta_i)$,
plus excluded volume interaction, i.e. it is a discretized version of the 
Kratky-Porod model plus excluded volume, with angles $\theta_i$ restricted
to $\theta_i=0$ and $\theta_i=\pm 90^o$, respectively. For a 
corresponding continuum model for large $\epsilon$
small angles would dominate, however: putting 
$1-\cos \theta_i \approx \theta_i^2/2$ the corresponding Hamiltonian would be
${\cal H}=(\epsilon_b/2) \sum_i \theta_i^2$ so one would conclude that 
$\kappa$ ( and hence $\ell_p$) are simply proportional to $\epsilon_b$.
Eqs.~(\ref{eq35}), (\ref{eq36}) rather imply 
$\ell_p \propto \exp(\epsilon/k_BT)$; this effect is due to the fact
that only large nonzero angles $\pm 90^o$ are permitted.}

Eq.~(\ref{eq32}) describes the crossover from the behavior of a rigid rod for 
$L <\ell_p$, where $\langle R^2\rangle _0 = L^2$, to Gaussian chains for 
$L \gg \ell_p$, where Eq.~(\ref{eq1}) holds, 
$\langle R^2\rangle _0 = 2 \ell_p L$. However, neither the exponential decay 
of the correlation function of bond vectors \{Eq.~(\ref{eq29})\} nor the 
Gaussian behavior implied by Eq.~(\ref{eq32}) remain valid when excluded 
volume effects are considered.

On a qualitative level, insight into the effects of excluded volume on 
semiflexible chains can be gotten by Flory-type free energy minimization 
arguments~\cite{5,8,49,50}. Consider a model where rods of length 
$\ell_k$ and diameter $D$ are jointed together, such that the contour length 
$L=N_b\ell_b= n\ell_k$. Apart from prefactors of order unity, the second 
virial coefficient then can be estimated as (in $d=3$ dimensions)
\begin{equation}\label{eq37}
v_2=\ell_k^2D \,.
\end{equation}
The free energy of a chain now contains two terms, the elastic energy and 
the energy due to the excluded volume interactions embodied in 
Eq.~(\ref{eq37}). 
The elastic energy is taken as that of a free Gaussian chain, 
i.e. $F_{el}\approx R^2/(\ell_kL)$. The repulsive interactions are treated 
in mean field approximation, i.e. proportional to the square of the density 
$n/R^3$ and the volume $R^3$. Thus
\begin{equation}\label{eq38}
\Delta F/k_BT \approx R^2/(\ell_kL) + v_2R^3[(L/\ell_k)/R^3]^2
\end{equation}
Minimizing $\Delta F$ with respect to $R$, we find for 
$L \rightarrow \infty$ the standard Flory result
\begin{equation}\label{eq39}
R \approx (v_2/\ell_k)^{1/5}L^{3/5}  =  (\ell_kD)^{1/5}(N_b\ell_b)^{3/5}\,.
\end{equation}
However, for finite $L$ (or finite $N_b$, respectively), Eq.~(\ref{eq39}) 
applies only when the chain length $N_b$ exceeds the crossover length 
$N_b^*$ or when $R$ exceeds the associate radius $R^*$,
\begin{equation}\label{eq40}
N_b >N_b^*\,, \enspace N_b^*=\ell_k^3/(\ell_bD^2)\, , \enspace R^*=\ell_k^2/D
\, .
\end{equation}
If $N_b<N_b^*$ the contribution of the second term in Eq.~(\ref{eq38}) would 
still be negligible for $R^2\approx \ell_kL$, where $\Delta F/k_BT$ is of 
order unity, and hence for $N_b<N^*$ the first term in Eq.~(\ref{eq38}) 
dominates, and hence the chain behaves like a Gaussian coil. However, 
this Gaussian regime only exists if 
$N_b^*\gg N_b^{\textrm{rod}}=\ell_k/\ell_b$, the number of monomers per Kuhn 
length. For $N_b<N_b^{\textrm{rod}}$, the chain resembles a rigid rod. 
Thus we predict (in $d=3$) two subsequent crossovers:
\begin{equation}\label{eq41}
R \approx L \,,  \enspace  N_b <N_b^{\textrm{rod}} = \ell_k/\ell_b
\thinspace  {\rm (rod-like \; chain)}\,, \\
\end{equation}
\begin{equation}\label{eq42}
R \approx (\ell_kL)^{1/2}\,,  \enspace  N _b^{\textrm{rod}} <N_b <N_b^* 
\thinspace  {\rm (Gaussian \; coil)}\,, \\
\end{equation}
\begin{equation} \label{eq43}
R \approx (\ell_kD)^{1/5}L^{3/5}\, , \enspace  N_b >N_b^* \, (R>R^*)
\thinspace  {\rm (SAW) } \,.
\end{equation}
Of course, the intermediate Gaussian regime of Eq.~(\ref{eq42}) only exists 
if $N_b^* \gg N_b^{\textrm{rod}}$, or alternatively
\begin{equation}\label{eq44}
\ell_k \gg D\,.
\end{equation}
E.g., in the case of bottle brush polymers with flexible backbone chains 
and flexible side chains under good solvent conditions evidence has been 
presented for the fact that the stiffness of these wormlike chains is only 
due to their thickness~\cite{35,36}, and hence $\ell_k$ and $\ell_p$ are of 
the same order as $D$ (disregarding the difficulty to define either $\ell_k$ 
or $\ell_p$ in this case properly) and thus the intermediate Gaussian regime 
does not occur. On the other hand, for a large number of real semiflexible 
macromolecules under good solvent conditions the double crossover described 
by Eqs.~(\ref{eq41})-(\ref{eq43}) has been clearly observed~\cite{51}.

On the other hand, the situation is completely different in $d=2$ dimensions, 
where Eq.~(\ref{eq37}) is replaced by
\begin{equation}\label{eq45}
v_2=\ell_k^2
\end{equation}
since a rod of length $\ell_k$ blocks an area of order $\ell_k^2$ by 
occupation from a (differently oriented) second rod. Eq.~(\ref{eq38}) in $d=2$ 
becomes~\cite{37}

\begin{equation}\label{eq46}
\Delta F/k_BT \approx R^2/(\ell_kL) + v_2R^2[(L/\ell_k)/R^2]^2\,.
\end{equation}
Minimizing again $\Delta F $ with respect to $R$ yields now
\begin{equation}\label{eq47}
R \approx (v_2/\ell_k)^{1/4}L^{3/4}\approx \ell_k^{1/4} L ^{3/4}\,, \enspace
d=2\, ,
\end{equation}
where now the size $L^*=\ell_bN_b^*$ where Eq.~(\ref{eq47}) starts to hold is
\begin{equation}\label{eq48}
L^* = \ell_k, \enspace \textrm{i.e.} \enspace N_b^*= N_b^{\textrm{rod}}\,.
\end{equation}
Thus we note that a direct crossover occurs from rods to swollen coils 
(exhibiting statistical properties of two-dimensional self-avoiding walks), 
and no regime with intermediate Gaussian behavior occurs.

An important consequence of the excluded volume interaction is that they 
cause a much slower asymptotic decay of orientational correlations than 
described by Eq.~(\ref{eq29}) occurs. For fully flexible chains one has a power 
law~\cite{38}
\begin{equation}\label{eq49}
\langle \vec{a}_i \cdot \vec{a}_{i+s}\rangle \propto s^{-\beta}\,, \enspace
\beta =2(1-\nu)\,, \enspace 1\ll s \ll N_b\, .
\end{equation}
Eq.~(\ref{eq49}) has been verified by extensive Monte Carlo simulations of 
self-avoiding walks on simple cubic $(d=3)$~\cite{35,36} and square 
$(d=2)$~\cite{37} lattices; we shall recall these results and extend 
them in Sec.~IV below.

For semiflexible chains we expect a crossover from exponential decay 
\{Eq.~(\ref{eq29})\} to the power law \{Eq.~(\ref{eq51})\} to occur near 
$s=N^*_b$, i.e. we make the speculative assumption that
\begin{equation}\label{eq50}
\langle \vec{a}_i\cdot \vec{a}_{i+s}\rangle \approx \ell_b^2 
\exp(-s\ell_b/\ell_p)\,, \enspace 1 \ll s \ll N_b^*\,,
\end{equation}
\begin{equation}\label{eq51}
\langle \vec{a}_i\cdot \vec{a}_{i+s}\rangle \approx \exp(-N_b^*\ell_b/\ell_p) 
\ell_b^2 (\frac{s\ell_b}{L^*})^{-\beta}\,, \enspace N_b^* \ll s \ll N_b \,.
\end{equation}
Note that the prefactor of the power law in Eq.~(\ref{eq51}) was chosen such 
that for $s=N_b^*$ (where $s \ell_b=L^*$) a smooth crossover to 
Eq.~(\ref{eq50}) is possible. In $d=2$, where $N_b^*= N_b^{\textrm{rod}}$, 
and hence $N_b^*\ell_b=\ell_k=2\ell_p$, we note that the factor 
$\exp(-N_b^*\ell_b/\ell_p)\approx 0.14$ and using $\beta = 1/2$ in $d=2$ 
one finds that
\begin{equation}\label{eq52}
\langle \vec{a}_i\cdot \vec{a}_{i+s}\rangle /\ell_b^2 \approx 
0.14 (2 \ell_p/\ell_b)^{1/2} s^{-1/2}, \enspace N_b^* \ll s \ll N_b
\end{equation}
and hence there occurs an increase of the amplitude of the power law with 
$\ell_p$. However, in $d=3$, where $N_b^*/N_b^{\textrm{rod}}=(2\ell_p/D)^2$ 
the factor $\exp(-N_b^* \ell_b/\ell_p)=\exp (-2N_b^*/N_b^{\textrm{rod}})= 
\exp[-8(\ell_p/D)^2]$ for large $\ell_p $ will dominate and hence lead to a 
strong decrease of the amplitude of the power law in Eq.~(\ref{eq51}). 
Of course, the crossover at $N_b^*$ is not at all sharp but rather spread out 
over several decades in $s$, and hence the observability of 
Eqs.~(\ref{eq50}), (\ref{eq51}) is rather restricted. Note, however, 
that $\langle \vec{a}_i \cdot \vec{a}_{i+s}\rangle$ always exhibits a single 
crossover only (for $N_b^* \rightarrow \infty)$, near $s=N_b^*$: while the 
radius exhibits two crossovers in $d=3$ (at $N_b=N_b^{\textrm{rod}}$ and at 
$N_b=N_b^*$), Eq.~(\ref{eq50}) does not exhibit any change in behavior when 
$s\approx N_b^{\textrm{rod}}=2\ell_p/\ell_b$. Thus we predict that for very 
stiff and thin chains (for which $\ell_p/D$ and hence 
$N_b^*/N_b^{\textrm{rod}}$ are large numbers) one can follow the exponential 
decay $\exp(-s\ell_b/\ell_p)$ over several decades. In $d=2$, where 
$N_b^*=N_b^{\textrm{rod}}$, this is predicted to be impossible; rather one 
can follow the exponential decay only from unity to about $1/e$. We shall 
discuss in Sec.~IV a Monte Carlo test of these predictions.

\subsection{Stretching of semiflexible chains}
There exists a rich literature~\cite{11,12,16,17,18,20,21,22,23} where 
a force term, Eq.~(\ref{eq2}), is added to the Kratky-Porod Hamiltonian, 
Eq.~(\ref{eq33}), to obtain
\begin{equation}\label{eq53}
\mathcal{H} =  \frac \kappa 2 \int \limits _0^L 
(\frac {\partial ^2\vec{r}(s)}{\partial s^2} )^2 ds - f 
\int \limits _0 ^L \frac {\partial x (s)}{\partial s} ds \,.
\end{equation}
We shall not dwell here on the exact numerical methods by which force versus 
extension curves can be derived from Eq.~(\ref{eq53}), but simply quote 
approximate interpolation formulas~\cite{11,20} (which are known to deviate 
from the numerically exact solutions at most by a few percent),
\begin{equation}\label{eq54}
\frac {f\ell_p} {k_BT} = \frac{\langle X\rangle}{L} + 
\frac {1} {4(1-\langle X\rangle/L)^2} - 
\frac 1 4 \,, \enspace d=3 \enspace {\rm (K-P\, model)}\,,
\end{equation}
\begin{equation}\label{eq55}
  \frac {f\ell_p}{k_BT} = \frac 3 4 \frac {\langle X \rangle}{L} + 
\frac {1} {8(1-\langle X\rangle/L)^2} - 
\frac 1 8 \,, \enspace d =2 \enspace {\rm (K-P\, model)}\,.
\end{equation}
At this point, we remind the reader that $\kappa = \ell_pk_BT$ in $d=3$ 
while $\kappa = \ell_pk_BT/2$ in $d=2 $ \{Eq.~(\ref{eq34})\}. For small $f$, 
Eqs.~(\ref{eq54}), (\ref{eq55}) are compatible with the relations that 
one can derive by treating $f$ via linear response,
$\langle X \rangle = f \langle X^2 \rangle_0 /k_BT$ and hence 
$(\langle X^2 \rangle_0 = \langle R^2 \rangle_0/ d=2\ell_p L/d)$
\begin{equation}\label{eq57}
\frac {f\ell_p} {k_BT} = \frac d 2 
\frac {\langle X \rangle}{L}  \,,
\enspace d=2,\;3 \enspace {\rm (K-P\, model)}\,,
\end{equation}
while for large $f$ we find
\begin{equation}\label{eq58}
\langle X \rangle/L \approx 1-1/\sqrt{4f\ell_p/k_BT}\,, 
\enspace d=3 \enspace {\rm (K-P\, model)}\, ,
\end{equation}
or
\begin{equation}\label{eq59}
\langle X \rangle/L \approx 1-1/\sqrt{8f\ell_p/k_BT}\,, 
\enspace d=2 \enspace {\rm (K-P\, model)}\,.
\end{equation}

A further quantity of interest is the ``deflection length''~\cite{52,53}, i.e. the correlation length of fluctuations along a semiflexible polymer. In the presence of a strong force it is given by~\cite{17}
\begin{equation}\label{eq60}
\lambda = (f/k_BT\ell_ p)^{-1/2} \,,\enspace \textrm{or} \enspace 
\lambda / \ell_p = (f\ell_p/k_BT)^{-1/2} \,.
\end{equation}
When $f\ell_p$ exceeds $k_BT$, $\lambda$ hence becomes smaller than $\ell_p$, 
and in this limit one expects that excluded volume indeed becomes negligible. 
However, when $\lambda$ becomes of the order of the bond length $\ell_b$, 
it is clear that the continuum description in terms of Eq.~(\ref{eq53}) breaks 
down, the discreteness of the chain molecule becomes 
relevant~\cite{17,20A,21},
and a crossover to the behavior of a freely jointed polymer occurs, as was 
described by Eq.~(\ref{eq7}). Toan and Thirumalai~\cite{23} have emphasized 
that all polymers under sufficiently high stretching forces should show a 
crossover from a force law of the type of the Kratky-Porod model, 
$1-\langle X \rangle/L \propto f^{-1/2}$ \{Eqs.~(\ref{eq58}), (\ref{eq59})\}, 
to the law of the freely jointed model, 
$1-\langle X \rangle/L \propto f^{-1}$ \{Eq.~(\ref{eq7})\}, and they argued 
that the crossover force between both descriptions is obtained by putting 
$\lambda = \ell_b$, providing evidence for this concept both by an analysis 
of experimental data and by a study of various models. However, here we are 
mostly interested in the regime where $\langle X \rangle /L$ is significantly 
smaller than unity, and excluded volume effects are still relevant.

In particular, for weak forces we can combine Eq.~(\ref{eq9}) with the proper 
relations for the linear dimensions of the semiflexible chains, as discussed 
in Eq.~(\ref{eq39}) for $d=3$ and Eq.~(\ref{eq47}) for $d=2$, respectively. 
Thus, Eq.~(\ref{eq57}) gets replaced by
\begin{equation}\label{eq61}
\langle X \rangle /L \propto (f \ell_p/k_BT)(D/\ell_p)^{2/5} 
(L/\ell_p)^{1/5}\, , \enspace L >L^*=\ell_bN_b^*
\end{equation}
in $d=3$, while in $d=2$ we have
\begin{equation}\label{eq62}
\langle X \rangle/L \propto (f \ell_p/k_BT) (L/\ell_p)^{1/2}\,, 
\enspace L >\ell_p\,,
\end{equation}
where factors of order unity have been disregarded throughout. We
note that in this regime the relations $\langle X \rangle/L$
versus $f \ell_p/k_BT$ vary much more steeply than predicted by
Eq.~(\ref{eq57}) if $L/\ell_p$ is very large. Furthermore we note
from Eqs.~(\ref{eq9})-(\ref{eq16}) that the nonlinear Pincus
force-extension relation, Eq.~(\ref{eq16}), sets in when the size of
the Pincus blob is of the same order as the coil size 
$\sqrt{\langle R^2 \rangle_0}$ ( Eq.~(\ref{eq39}) for $d=3$, or
Eq.~(\ref{eq47}) for $d=2$, in the
absence of a force), defining a crossover length $\xi_{P,c}$ and
associated force $f_c$,
\begin{eqnarray}\label{eq63}
&& \xi_{P,c} = \frac {k_BT}{f_c} = (\ell_pD)^{1/5} L^{3/5}\,, \enspace d=3 \,,
\nonumber \\
&&\enspace {\rm or}  \nonumber  \\
&& \xi_{P,c}=\ell_p^{1/4} L ^{3/4}\,, \enspace d=2\,, 
\end{eqnarray}
i.e. for a force for which the extension $\langle X \rangle$ is of the same 
order as the coil size $\sqrt{\langle R^2\rangle_o}$.
As one expected for scaling theories, the crossover between the various regimes 
occur when all characteristic lengths $\sqrt{\langle R^2 \rangle }$, 
$\langle X \rangle$, $\xi_{P,c}$ are of the same order. In the non-linear 
regime, according to the scaling ansatz, Eq.~(\ref{eq11}), 
Eq.~(\ref{eq15}) gets replaced by (taking $\nu=\frac 3 5 $ in $d=3$)
\begin{equation}\label{eq64}
\langle X \rangle \propto (k_BT/f\ell_b)^{-2/3} (\ell_pD/\ell_b^2)^{1/3}L, \enspace d=3\;,
\end{equation}
or (recall $\nu = 3/4$ in $d=2$)
\begin{equation}\label{eq65}
\langle X \rangle \propto (k_BT/f\ell_b)^{-1/3} (\ell_p/\ell_b)^{1/3} L \,, 
\enspace d=2\,.
\end{equation}
Thus we see that in the nonlinear regime the persistence of the chains leads 
to an enhancement of the chain extension by a factor $\ell_p^{1/3}$. Of course,
the relation Eq.~(\ref{eq65}) can only hold if a Pincus blob contains many 
Kuhn segments, i.e. now $\xi_P=k_BT/f \gg \ell_p$ is required. This condition 
is nothing but the condition $\langle X \rangle /L \ll 1$, in the case $d=2$, 
as expected. In $d=3$, however, we expect that the Pincus regime, as described 
by Eq.~(\ref{eq64}), already ends when the size of a Pincus blob, 
$\xi_P=k_BT/f$, equals the crossover radius $R^*$, Eq.~(\ref{eq40}) 
(remember that only for radii exceeding $R^*$ the excluded volume effects 
dominate). For the crossover force
\begin{equation}\label{eq66}
f^*=k_BT/R^*\propto k_BTD/\ell _p^2
\end{equation}
we find from Eq.~(\ref{eq64}) that
\begin{equation}\label{eq67}
\langle X \rangle /L \propto D/\ell_p \propto f^* \ell_p/k_BT \, .
\end{equation}
Comparing Eq.~(\ref{eq67}) with Eqs.~(\ref{eq54}), (\ref{eq57}), we see
that indeed for $f \approx f^*$ a smooth crossover from the Pincus
behavior, as described by Eq.~(\ref{eq64}), to the Kratky-Porod law
for wormlike chains for which excluded volume is negligible, can
occur. We also note from Eq.~(\ref{eq66}) that in cases such as
occur for bottle brush polymers~\cite{35,36} where chain stiffness
is due to chain thickness, $\ell_p \propto D$, we would have $f^*=
k_BT/\ell_p$ as in the two-dimensional case, and then the Pincus
regime (which applies for $k_BT/\sqrt{\langle R^2\rangle _0}<
f<f^*$) becomes much broader and easier observable.

\begin{figure*}[t]
\begin{center}
(a)\includegraphics[scale=0.27,angle=270]{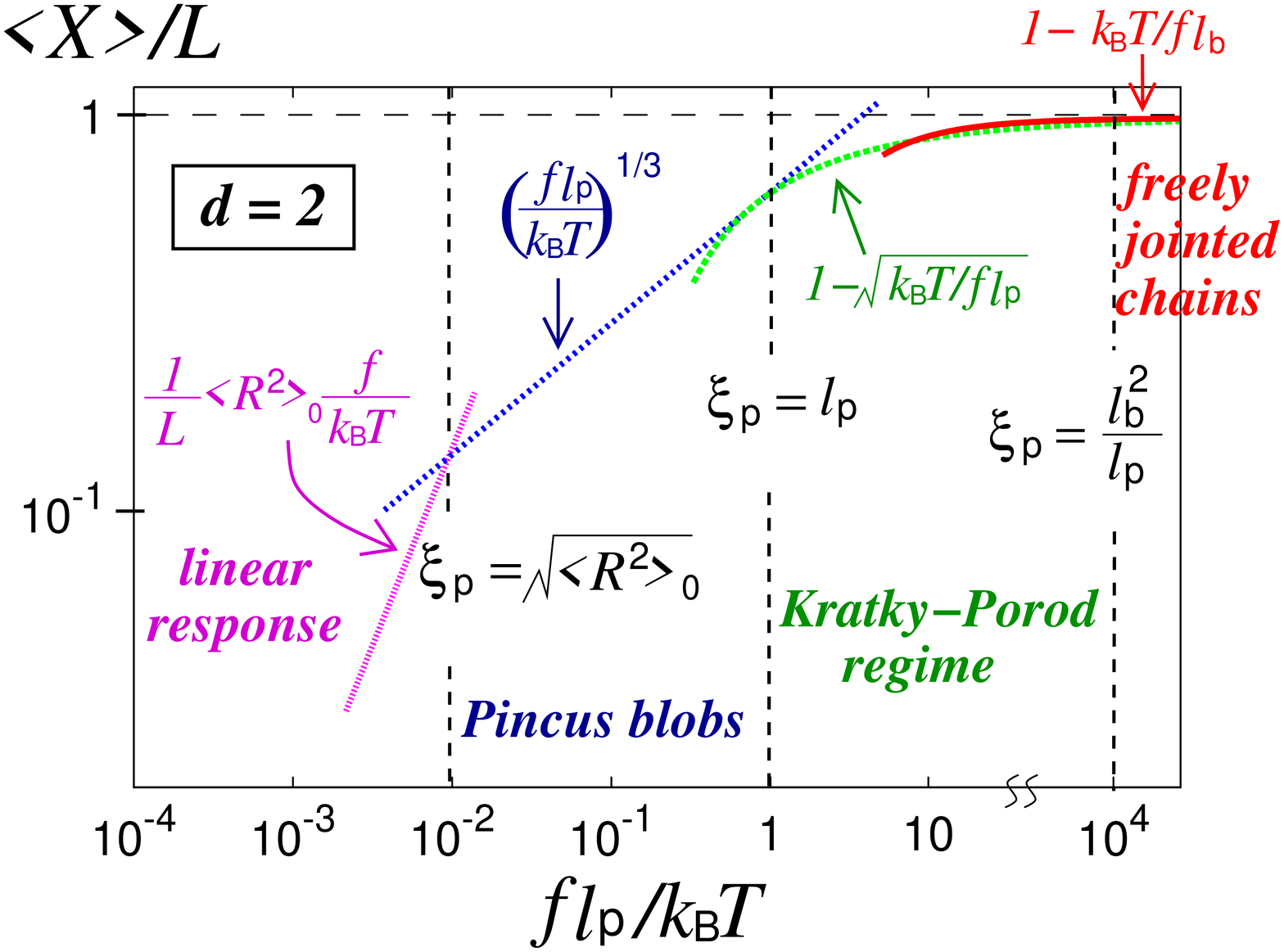}\hspace{0.4cm}
(b)\includegraphics[scale=0.27,angle=270]{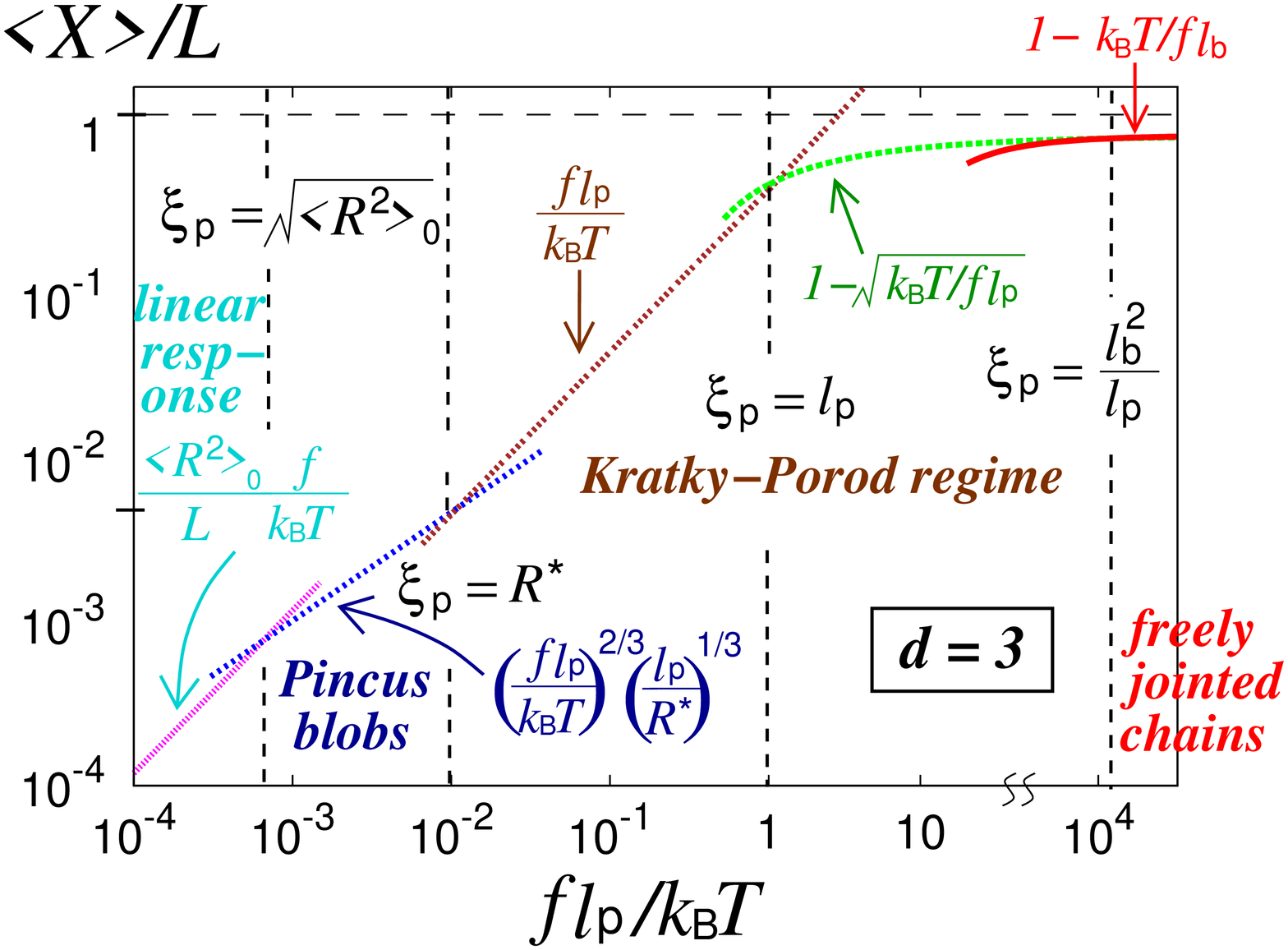}\\
\caption{Schematic plot of the relative chain extension,
$(\langle X \rangle/L)$, versus the scaled force,
$(f\ell_p/k_BT)$, in a log-log scale for $d=2$ dimensions (a)
and $d=3$ dimensions (b). Broken vertical straight lines indicate
various (smooth, not sharp!) crossovers in the response to the stretching
force $f$. The first crossover occurs at very small forces, when the tensile
length $\xi_p=k_BT/f$ becomes equal to the chain size
$\sqrt{\langle R^2 \rangle _0}$ in the absence of forces. In the first regime
(to the left of this crossover) the extension is proportional to the force,
$\langle X \rangle \propto \langle R^2\rangle _0f/k_BT$
(linear response regime). To the right of this crossover, the extension
versus force curve follows the Pincus power law,
$\langle X \rangle \propto f^{1/\nu-1}=f^{1/3} (d=2) $ or
$\approx f ^{2/3} (d=3)$, respectively. In $d=2$ this power law regime
extends up to the crossover where the tensile length equals the persistence
length $\ell_p$, while in $d=3$ the power law ends already at an earlier
crossover, $\xi_p=R^*$, $R^*$ being the crossover radius where excluded
volume statistics comes into play for semiflexible chains: then there
exists a regime described by the Kratky-Porod model,
$\langle X \rangle /L \propto f \ell_p/k_BT$. For $\xi_p$ smaller than
$\ell_p$, the extension approaches saturation according to the Kratky-Porod
relation, $1-\sqrt{k_BT/f\ell_p}$, while for still larger forces (when the
deflection length becomes comparable to the bond length, a further crossover
to the behavior expected for freely jointed chains
$(\langle X \rangle /L \approx 1-k_BT/f\ell_b)$) occurs.}
\label{fig1}
\end{center}
\end{figure*}

Finally, we consider again the fluctuations in the extensions of stretched 
chains (which were considered for flexible chains in 
Eqs.~(\ref{eq20})-(\ref{eq26})
already). However, the results known to us for semiflexible chains are 
somewhat scarce (although the end-to-end distribution function of the 
Kratky-Porod wormlike chain has been discussed~\cite{18,54,55,56,57,58}). 
Marko and Siggia~\cite{11} obtained for the side-to-side excursions of the 
chain over a contour length $s$ the result
\begin{equation}\label{eq68}
\langle [\vec{r}_\bot (s) - \vec{r}_\bot (0)]^2 \rangle = \frac
{2k_BT}{f} \{ s\ell_b- \frac
{1-\exp[-s(f/\kappa k_BT)^{1/2}]}{(f/\kappa k_BT)^{1/2}} \}
\end{equation}
where $\kappa$ is the coupling constant of the Kratky-Porod model 
\{Eq.~(\ref{eq34})\}, of course. For $s\ell_b=L=N_b\ell_b$ the result 
$\langle R_\bot ^2\rangle = N_b\ell_b(2 k_BT/f)$ is identical to the large 
force limit for the flexible chains, Eq.~(\ref{eq22}).
For small $s$ Eq.~(\ref{eq68}) yields 
$\langle [\vec{r}_\bot (s) - \vec{r}_\bot (0)]^2 \rangle
=s^2\ell_b^2 (k_BT/\kappa f)^{1/2}$.

Given the fact that in the Pincus regime the picture of the chain conformation
essentially is a stretched string of Pincus blobs (inside a blob excluded
volume statistics prevails), we know that there occur of the order of 
$n=N_b(\ell_b/\ell_p)/g$ such blobs per string where $g$ is the number of
Kuhn steps (containing $\ell_p/\ell_b$ monomers each) per blob.
Remember that the Pincus blob has the radius $\xi_p$, and built as 
a self-avoiding walk of $g$ steps of length $\ell_p$ so that
$\xi_P=\ell_p g^\nu$, i.e. $g=(\xi_P/\ell_p)^{1/\nu}=(k_BT/f\ell_p)^{1/\nu}$
(here we disregard the factor $2$ between the effective Kuhn step
length $\ell_k$ and the persistence length $\ell_p$. 
Thus $n=N_b(\ell_b/\ell_p)(f\ell_p/k_BT)^{1/\nu}$.)
If this
string of blobs would be completely stretched in a rod-like 
configuration, its lateral width would be simply 
$\langle R_{\perp}^2 \rangle \approx \xi_P^2 = \ell_p^2(k_BT/f\ell_p)^2$.
However, this estimate neglects the random statistical fluctuations
that the string of blobs may exhibit in the transverse directions.
We may consider the problem as a directed random walk where each step has a 
component $\xi_P$ in the $+x$-direction and a transverse component 
$\pm c \xi_P$, where $c$ is a constant ($c\ll 1$). 
If we have $n$ such steps ($n=N_b(g\ell_p/\ell_b)^{-1}$),
we hence predict 
$\langle R_\perp^2 \rangle=c^2 \xi_P^2 n= c^2 
\ell_p (k_BT/f\ell_p)^{2-1/\nu}N_b\ell_b$. 
Of course, this result can only apply if $n$ is large enough so that 
$cn^2 >1$, because $\langle R_\perp^2\rangle$ cannot be smaller then 
$\xi_P^2$, of course. Hence we would predict from these speculative
scaling arguments
\begin{eqnarray}\label{eq69}
  \langle R_\perp^2 \rangle \propto \xi_P^2n=\ell_p \ell_b N_b 
(k_BT/f\ell_p)^{2-1/\nu} \, , \enspace n \rightarrow \infty  
\end{eqnarray}
\begin{equation}
\langle R_\perp^2 \rangle  =\xi_p^2 = \ell_p^2 (k_BT/f\ell_p)^2 \,, 
\enspace {\rm small} \enspace  n \,.
\end{equation}
In $d=2$, this result should hold up to a force $f=k_BT/\ell_p$,
where $\xi_P=\ell_p$. Then Eq.~(\ref{eq69}) predicts
$\langle R_\perp^2 \rangle \propto \ell_p \ell_b N_b=\ell_p L$,
i.e. there a smooth crossover to the result 
$\langle R_\perp^2 \rangle \propto k_BTL/f$ derived from Eq.~(\ref{eq68})
occurs. In $d=3$, however, Eq.~(\ref{eq69}) is supposed to hold only 
for $\xi_P > R^* \propto \ell_p^2/D$.

At the end of this section, we summarize our findings for the force 
extension curves (Fig.~\ref{fig1}). The key to identify the various regimes 
is A COMPARISON OF LENGTHS, namely the ``tensile length'' $\xi_P= k_BT/f$ needs
to be compared to the various characteristic lengths of the unperturbed chain.

The simplest case actually occurs in $d=2$ (Fig.~\ref{fig1}a).

For $\xi_P >\sqrt{\langle R^2\rangle_0}$ we are in the regime of
linear response, the extension $\langle X \rangle$ scales linearly
with the force \{Eq.~(\ref{eq62})\}. For 
$\xi_P\approx \sqrt{\langle R ^2\rangle _0}$
the extension $\langle X \rangle $ and $\sqrt{\langle R^2 \rangle
_0}$ are of the same order, linear response breaks down, and a
(smooth!) crossover to the nonlinear Pincus regime occurs,
$\langle X \rangle/L \propto (f\ell_p/k_BT)^{1/3}$ 
\{Eq.~(\ref{eq65})\}. The chain can
be viewed as a stretched string of ``Pincus blobs'' of diameter
$\xi_P$ (inside the blobs excluded volume statistics prevails).
Near $\xi_P= \ell_p$ the extension $\langle X \rangle $ already is
no longer much smaller than the contour length $L$ itself. Only
the regime where the extension approaches its saturation value,
from $\xi_P= \ell_p $ down to $\xi_P =\ell_b^2/\ell_p$, when the deflection 
length becomes equal to the bond length, the Kratky-Porod
model holds \{Eq.~(\ref{eq59})\}, while for still larger forces 
(where the deflection length would
be less than a bond length) the discreteness of the polymer chain
causes a different relation between force and extension \{Eq.~(\ref{eq7})\}, as
indicated in the figure. Of course, for flexible chains
$\ell_p=\ell_b$ (actually $\ell_p$ is completely ill-defined then)
and the Kratky-Porod regime disappears altogether.

For $d=3$ dimensions the situation is more complicated, since for
semiflexible chains without force another regime appears, for
distances in between $\ell_p$ and $R^*=\ell_p^2/D$, where Gaussian
statistics prevails, and this regime finds its correspondence in
the force versus extension curve. Thus, for very long
semiflexible thin chains there are three regimes, where the
force-extension curve exhibits power laws: for very weak forces
$(\xi_P>\sqrt{\langle R^2\rangle _0})$ the linear response regime
occurs with $\langle X \rangle \propto \langle R^2 \rangle
_0f/k_BT$ \{Eq.~(\ref{eq61})\}, then a nonlinear regime with 
$\langle X \rangle /L
\propto (f \ell_p / k_BT)^{2/3}(\ell_p/R^*)^{1/3}$
\{Eq.~(\ref{eq64})\} follows for
$\sqrt{\langle R^2\rangle _0} >\xi _p > R^* \propto \ell^2_p/D$,
and then the linear regime as described by the Kratky-Porod model
follows, $\langle X \rangle/L\propto f \ell_p/k_BT$ 
\{Eq.~(\ref{eq57})\}, for
$R^*>\xi_P >\ell_p$. For $\xi_P <\ell_p$ the approach of $\langle
X \rangle /L$ to its saturation value unity proceeds in a similar
manner as in $d=2$ \{Eqs.~(\ref{eq58}), (\ref{eq7})\}.

Of course, the description in Fig.~\ref{fig1} contains the
force-extension curves of fully flexible polymers as a limiting
case: there both $R^*$ and $\ell_p$ tend to $\ell_b$, and the
regime where the Kratky-Porod model is applicable gradually
disappears. The same statement applies to semiflexible chains in
$d=3$ when stiffness is due to chain thickness, so that $R^*$
tends towards $\ell_p$, and the nonlinear Pincus blob regime
$(\langle X \rangle /L \propto (f \ell_p/k_BT)^{2/3}$ takes over
and holds down to $\xi _p \approx \ell_p$. Conversely, if one
considers not very long chains, such that $\langle R^2\rangle _0
\leq R^{*2}$, the regime dominated by excluded volume effects
disappears from the picture, and the Kratky-Porod model
description becomes valid down to arbitrarily small forces.

We end this section with several caveats: (i) All crossovers in
Fig.~\ref{fig1} are smooth and we do not expect any sharp kinks at
the crossover values of $\xi_P$ that are indicated by the vertical
broken lines; rather gradual changes will occur on the
log-log plot $\langle X \rangle /L$ vs. $f\ell_p/k_BT$, spread out
over (at least) a decade in $f \ell _p/k_BT$. Consequently, 
$\sqrt{\langle R ^2 \rangle _0}$ must exceed $\ell _p$ by
four (or more!) decades, in order to resolve the
multiple crossovers of Fig.~\ref{fig1} in $d=3$. (ii) We have
disregarded any special structures of the polymer such as
$\alpha$-helices known for proteins, double helix-portions of
copolymers formed from double-stranded DNA and other biopolymers,
etc.~\cite{59,60}. Any such special structures of biopolymers will
lead to non-universal special features of the force-extension
curve, in particular in the regime of rather low forces, but these are
outside of consideration here. Rather only a generic description
of the universal behavior of very long flexible or semiflexible
polymers is within our focus. (iii) Very long polymers exhibiting
a (swollen) random coil configuration are expected to contain
knots (to precisely define them, one can transform the polymer
configuration into a closed loop by adding the end-to-end vector
as an extra special bond)~\cite{61}. Pulling such a chain at both
ends will have the effect that the knots tighten, and the knots
can only be made to disappear by moving them to the chain ends.
Effects due to knots~\cite{62,63} clearly are beyond the realm of
our scaling description (while in the computer simulations
presented in Secs.~IV-VI knots are automatically included implicitly
in our models, though we do not make any explicit attempt to study
their effects). In view of all these caveats, the extent to which
the scaling theory sketched in Fig.~1 is practically useful is a
nontrivial matter.

\begin{figure*}
\begin{center}
(a)\includegraphics[scale=0.29,angle=270]{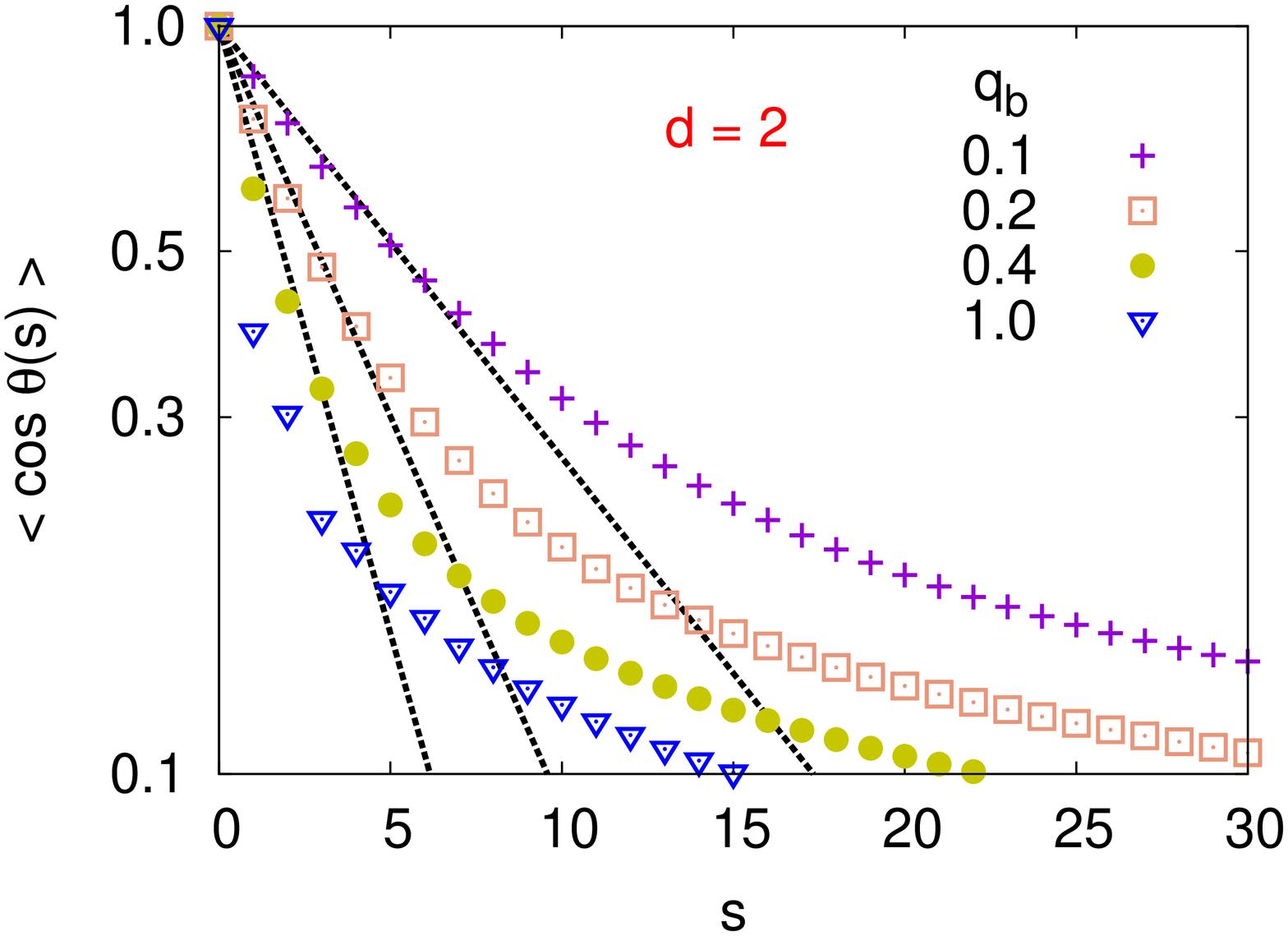}\hspace{0.4cm}
(b)\includegraphics[scale=0.29,angle=270]{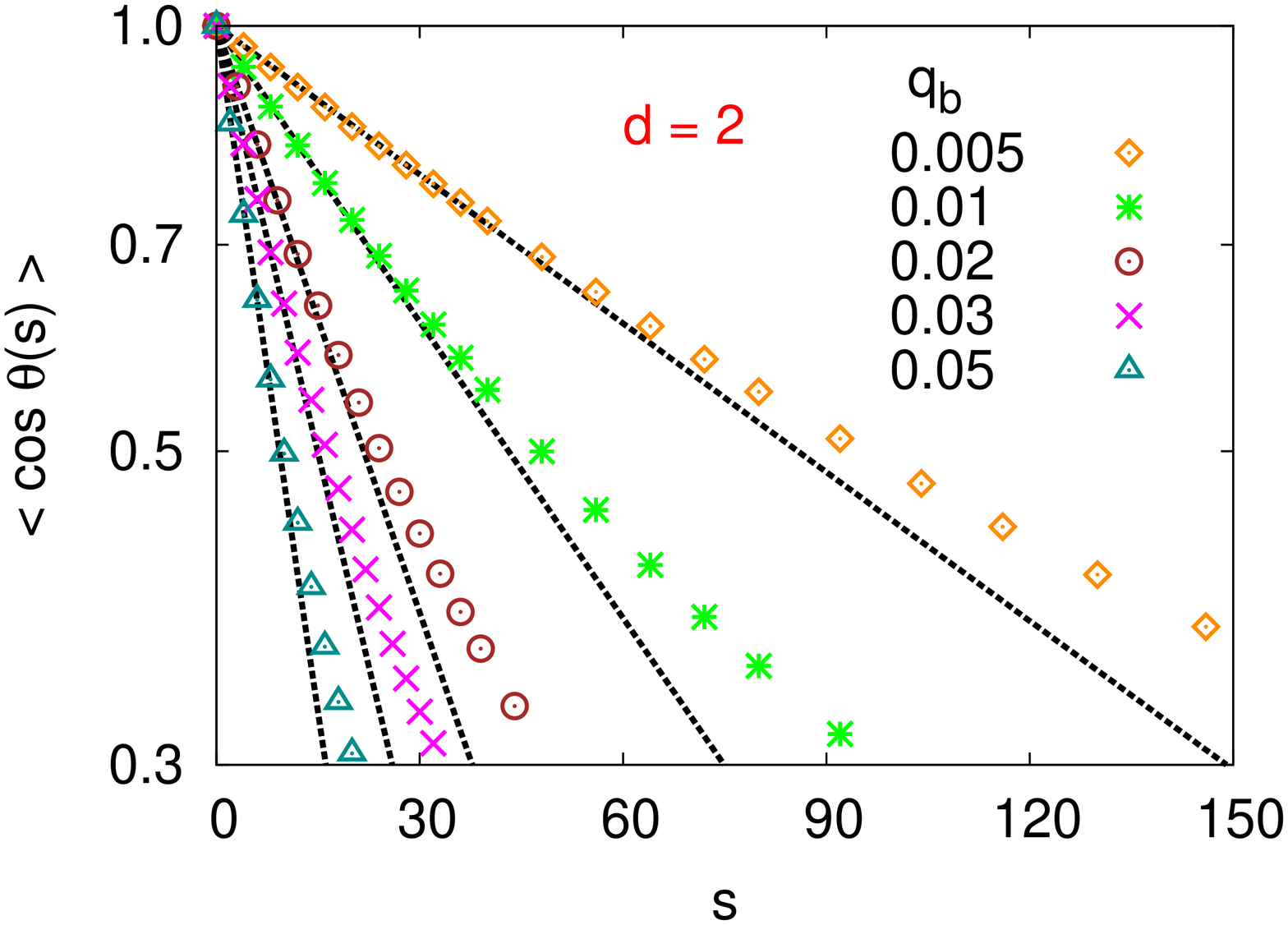}\\
(c)\includegraphics[scale=0.29,angle=270]{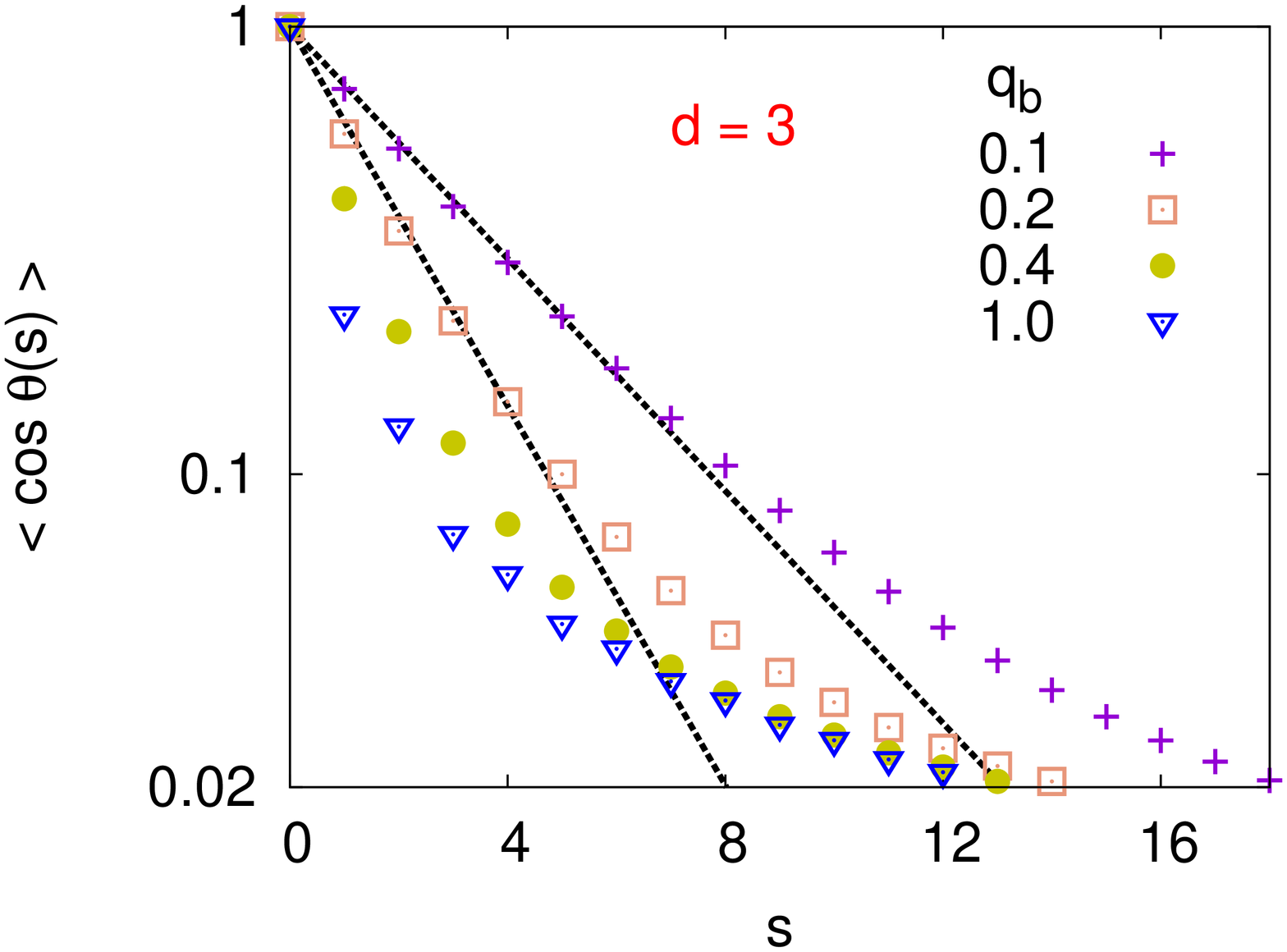}\hspace{0.4cm}
(d)\includegraphics[scale=0.29,angle=270]{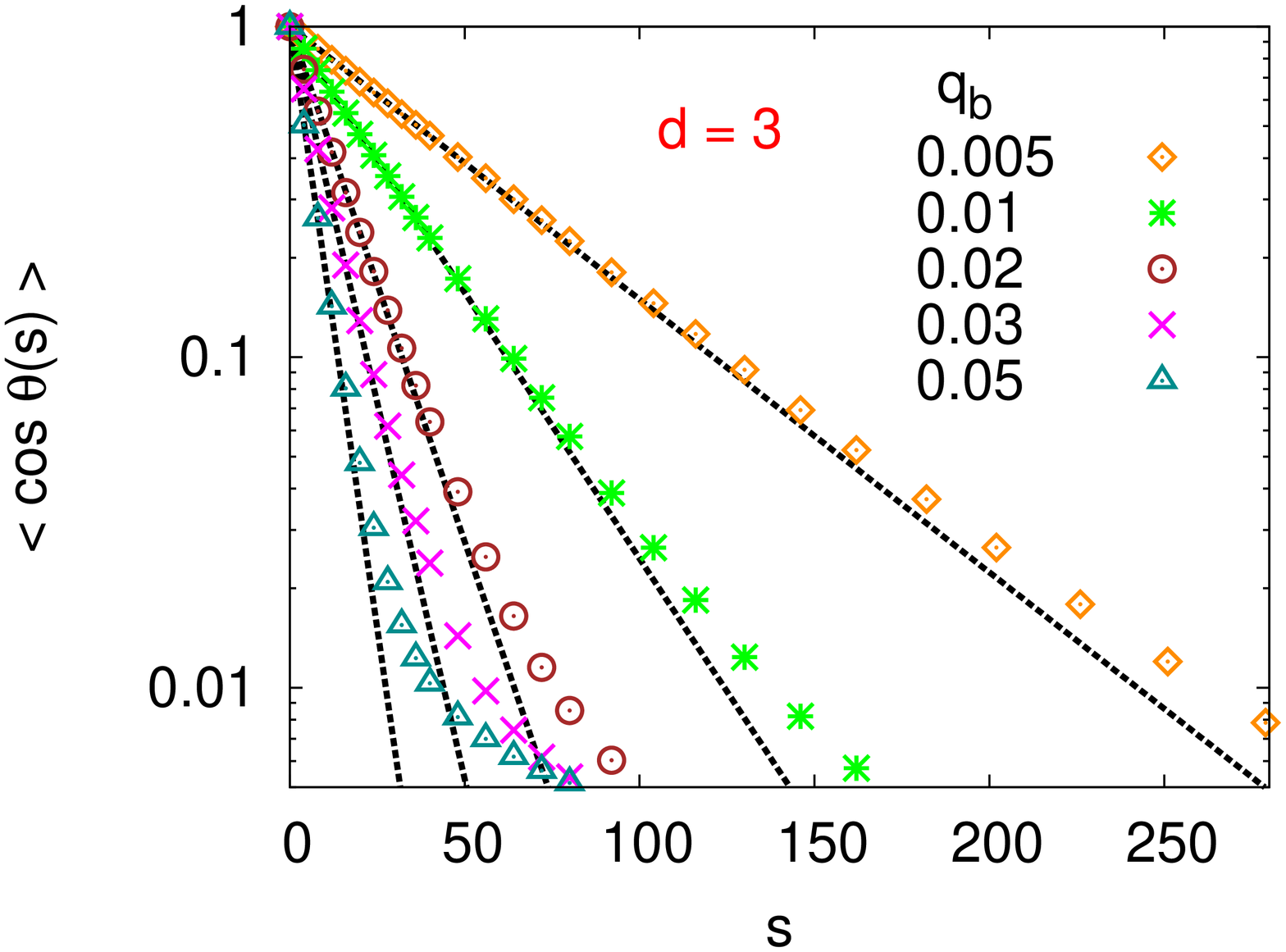}\\
\caption{Semi-log plot of $\langle \cos \theta (s) \rangle$ versus the
contour length $s$, for $q_b$ in the range from $q_b=0.1$ to
$q_b=1.0$ (a) (c), and for rather stiff chains, $0.005 \leq q_b\leq 0.05$,
(b) (d). Data are taken for $N_b=25600$ in $d=2$ (a) (b) and for $N_b=50000$
in $d=3$ (c) (d). The straight lines indicate fits of the initial decay to
Eq.~(\ref{eq29}), $\langle \cos \theta (s)\rangle \propto e^{-s\ell_b/\ell_p}$;
for flexible chains ($q_b=1.0$ and $q_b=0.4$) meaningful
fits are not possible. Estimates for $\ell_p/\ell_b$ are listed in
Table~\ref{table1} and \ref{table2}.
Note the difference in ordinate scales between $d=2$
and $d=3$: in $d=2$, deviations from Eq.~(\ref{eq29}) have set in when
$\langle \cos \theta (s) \rangle$  has decayed to $1/e$, in $d=3$, however,
for very stiff chains one can follow the exponential decay for almost
two decades.}
\label{fig2}
\end{center}
\end{figure*}

\section{Model and Simulation Technique}
The model that we study in this paper is the classical
self-avoiding walk (SAW)~\cite{5,8,48} on square and simple cubic
lattices, where bonds connect nearest neighbor sites on the
lattice, and the excluded volume interaction is realized by the
constraint that every lattice site can be taken only once by an
effective monomer occupying that site. 
{We take the lattice spacing 
as our unit of length, $\ell _b = 1$. Variable chain stiffness (or
flexibility) then is introduced into the model by an energy
$\epsilon_b$ that occurs for any kink (that is at a right angle 
and costs $\epsilon_b$).}

No energy arises for $\theta = 0^\circ$, of course, and hence in the
statistical weight of a SAW configuration on the lattice every
kink will contribute a factor $q_b= \exp(-\epsilon/k_BT)$. In the
presence of a force $f$, the potential $U$ written in
Eq.~(\ref{eq2}) yields another factor $b^X$ with $b = \exp (f/k_BT)$
to the statistical weight (as in Sec.~II, the force $f$ is assumed
to act in the positive x-direction, and X is the x-component of
the end-to-end vector $\vec{R}$ of the chain). So the partition
function of a SAW with $N_b$ bends $(N_b+1$ effective monomers)
and $N_{\textrm{bend}}$ local bends by $\pm 90^\circ$ is
\begin{equation}\label{eq70}
Z_{N,N_{\textrm{bend}}}(q_b,b) = \sum \limits _{\textrm{config}}
C(N_b,N_{\textrm{bend}}, X) q_b^{N_\textrm{bend}} b^X\,.
\end{equation}
We have carried out Monte Carlo simulations applying the
pruned-enriched Rosenbluth Method~\cite{64,65,66} (PERM algorithm)
using chain lengths up to $N_b=50000$ in $d=3$ and $N_b= 25600$ in
$d=2$, varying also the chain stiffness over a wide range $(0.005
\leq q_b \leq 1.0$). As mentioned already in Sec.~II
\{Eq.~(\ref{eq36})\}, this means that the persistence length
$\ell_p$ varies over about two orders of magnitude (note, however,
that in the presence of excluded volume one has to be very careful
with the notion of a persistence length, particularly in $d=2$
dimensions~\cite{35,36,37}).

\begin{figure*}
\begin{center}
(a)\includegraphics[scale=0.29,angle=270]{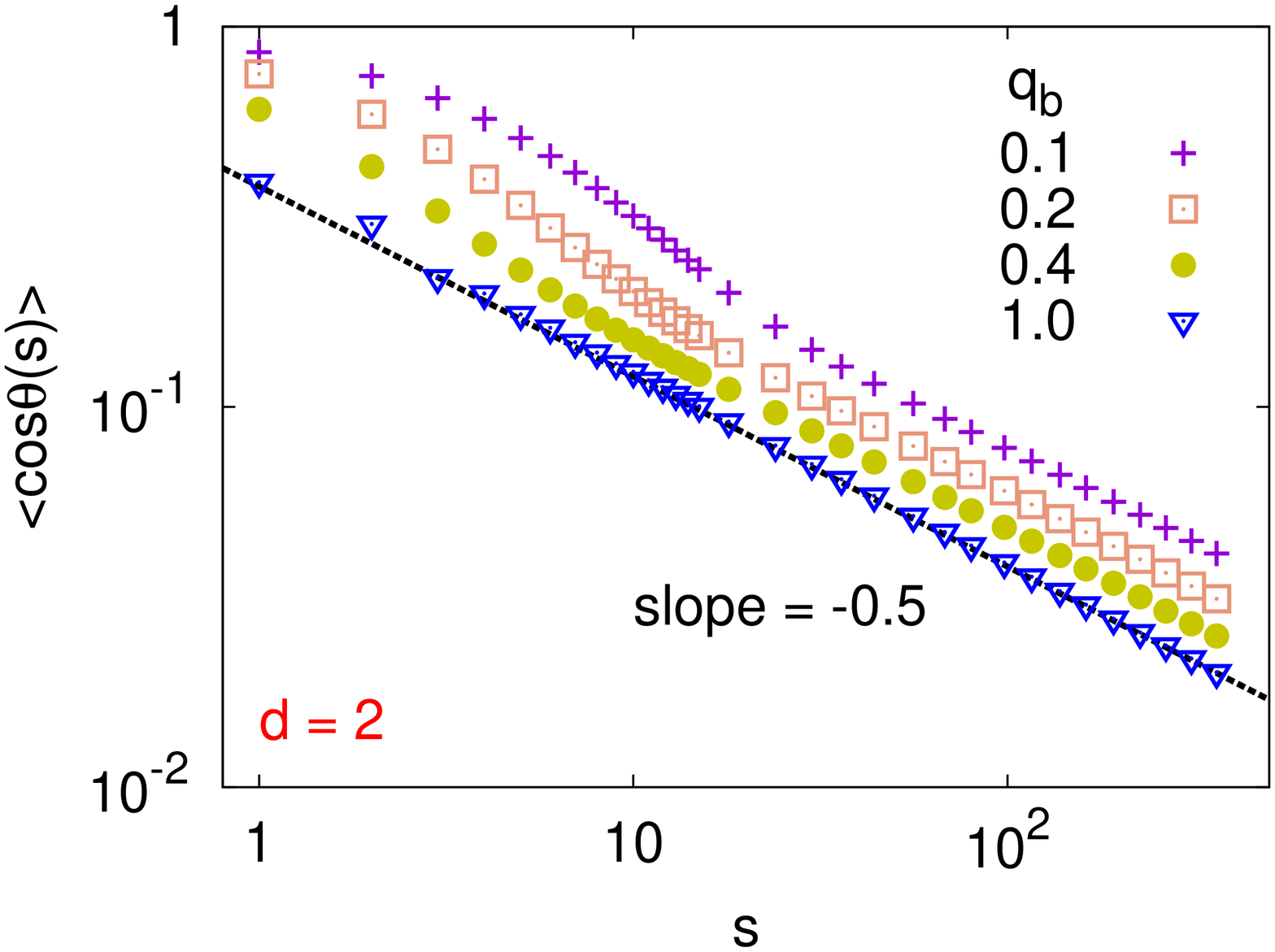}\hspace{0.4cm}
(b)\includegraphics[scale=0.29,angle=270]{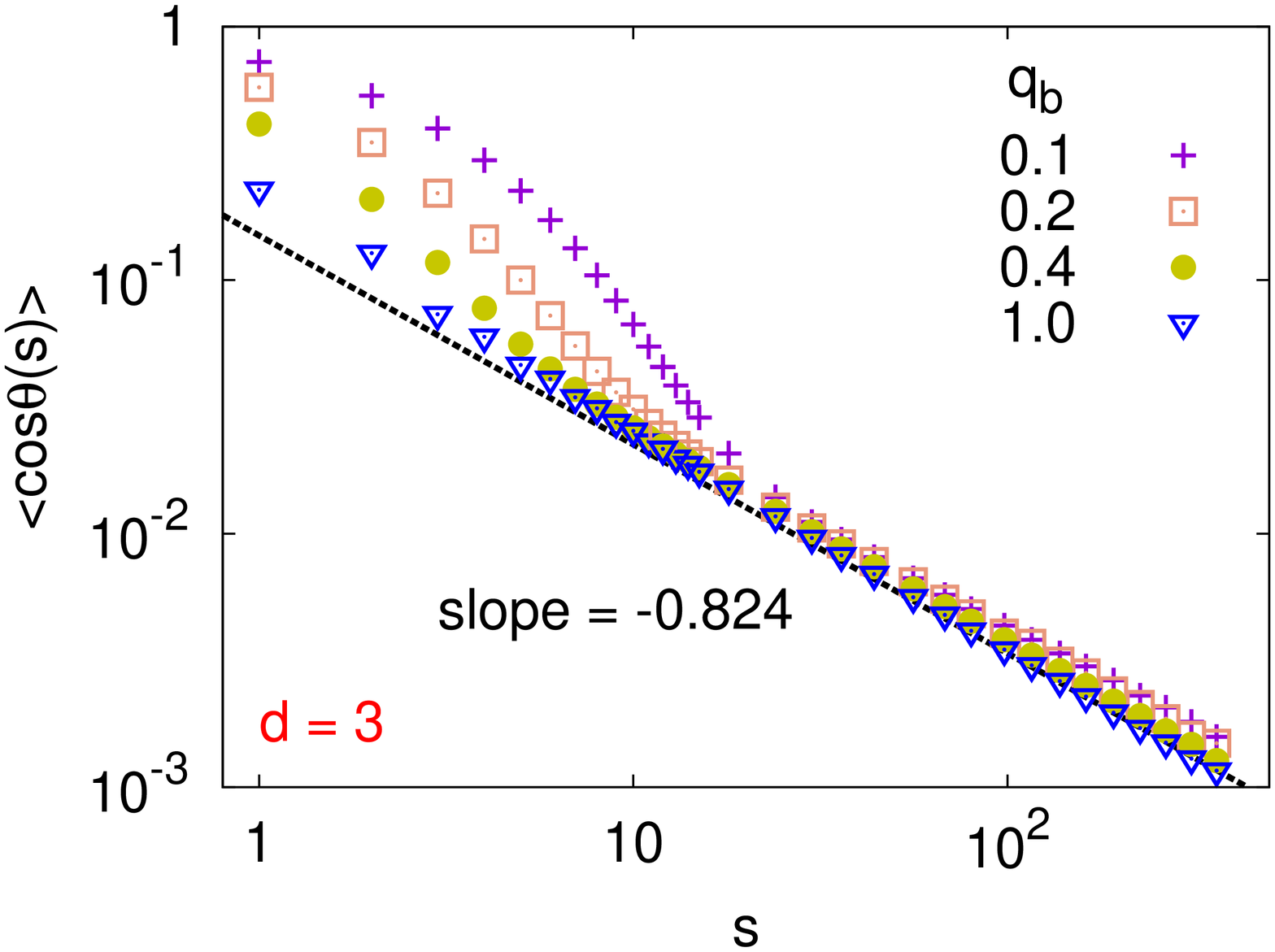}\\
\caption{Log-log plot of $\langle \cos \theta (s) \rangle$ versus $s$,
for $q_b=0.1,0.2,0.4$ and 1.0, including only data for $N_b=25600$ in
$d=2$ (a) and for $N_b = 50000$ in $d=3$ (b). The straight line indicates a
fit of the power law, Eq.~(\ref{eq49}), to the data for $q_b=1.0$, including
only data for $s \ell_b \geq 10$ in the fit, and requesting the theoretical
exponent, $\beta = 2-2\nu$, with $\nu = 3/4$ in $d=2$ (a), and
$\nu= 0.588$ in $d=3$ (b).}
\label{fig3}
\end{center}
\end{figure*}

\begin{figure*}[htb]
\begin{center}
(a)\includegraphics[scale=0.29,angle=270]{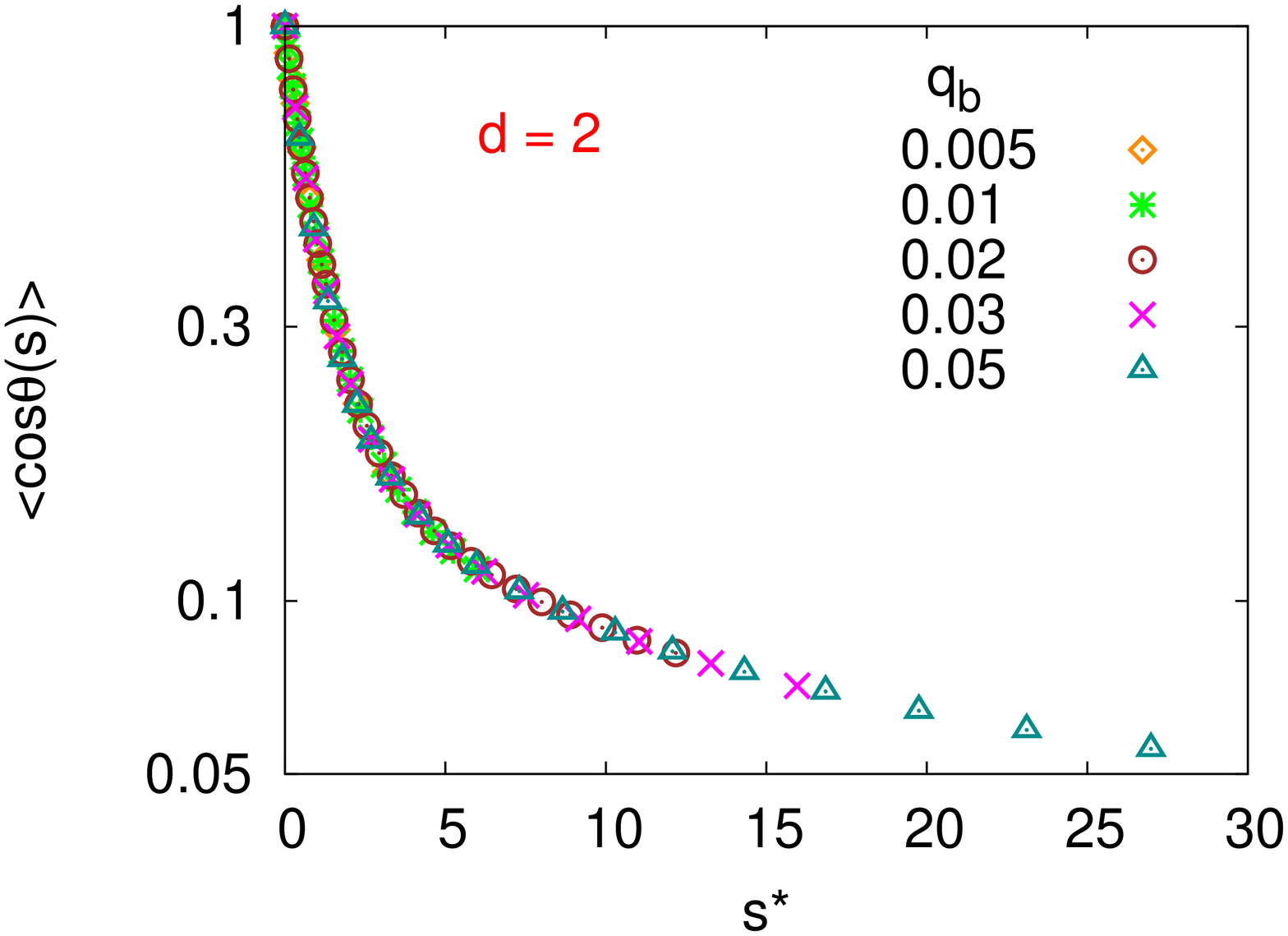}\hspace{0.4cm}
(b)\includegraphics[scale=0.29,angle=270]{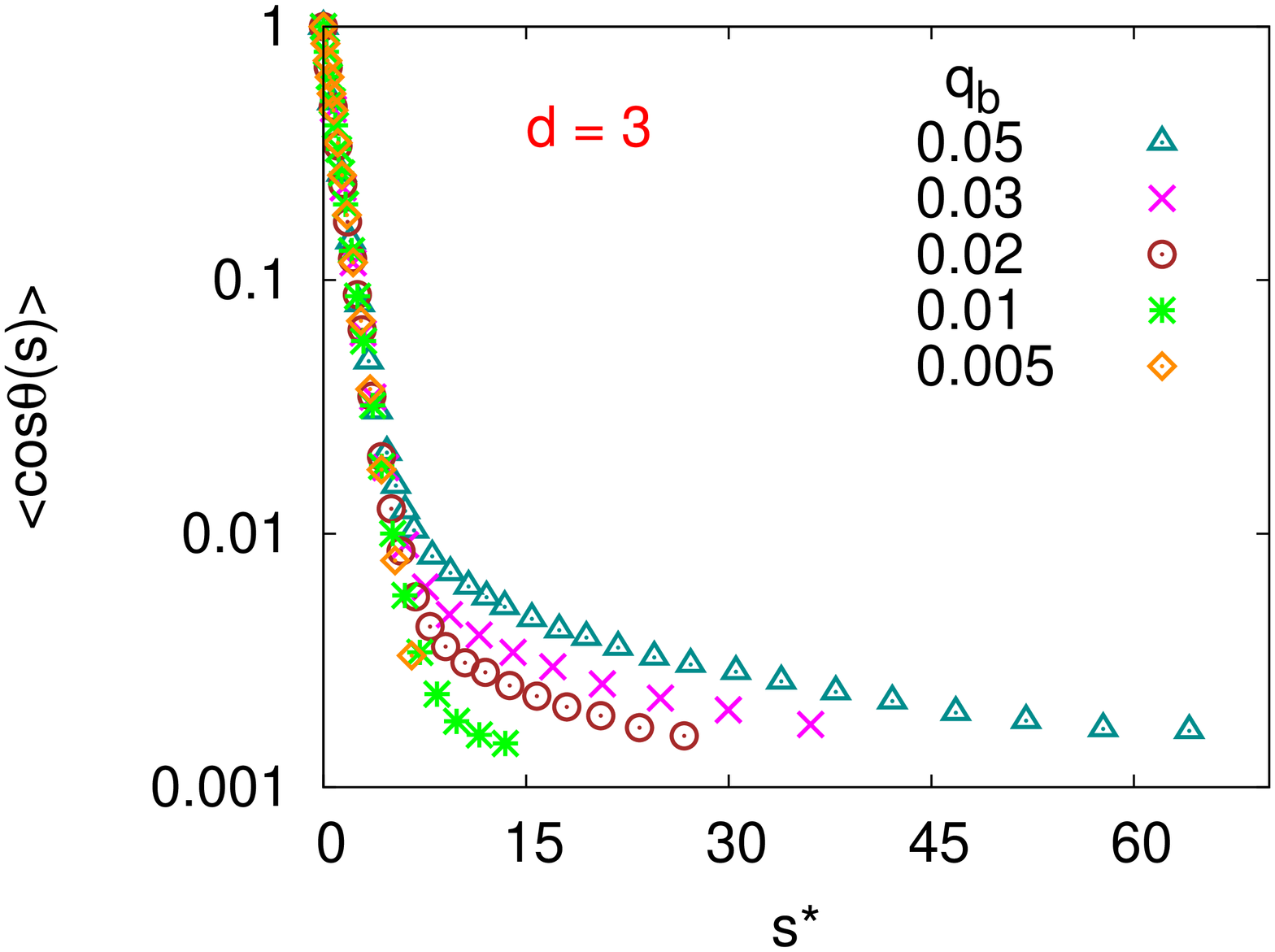}\\
\caption{Semi-log plot of $\langle \cos \theta (s) \rangle$ versus the
scaled distance $s^*= s\ell_b/\ell_p$ along the chain, for $d=2$ (a)
and $d=3$ (b). Data for $\ell_p/\ell_b$ extracted from Fig.~\ref{fig2},
as described above and listed in Table~\ref{table1} and \ref{table2},
were used.}
\label{fig4}
\end{center}
\end{figure*}

\section{Semiflexible polymers in the absence of stretching forces}
In this section, we summarize our Monte Carlo results for bond orientational
correlations and chain linear dimensions obtained for the model described
in the previous section. While some of these results have recently been
described in our earlier work~\cite{35,36,37}, the information provided
will be crucial for the understanding of our results for the extension
versus force curves as well.

We start with the bond orientational correlation function
$\langle \cos \theta (s) \rangle$, Figs.~\ref{fig2}-\ref{fig4},
since the decay of this function with
$s$ is traditionally used to extract ``the'' persistence length
$\ell_p$, using Eq.~(\ref{eq29}). As expected from Sec.~II B, however, one must
not rely on Eq.~(\ref{eq29}) to describe the asymptotic decay of
$\langle \cos \theta (s) \rangle$ (in the limit where first
$N_b \rightarrow \infty$ has been taken, so that one can study large
$s$ without being affected by the finite size of the chain) for large
$s$, but rather one must consider the initial decrease of
$\langle \cos \theta (s) \rangle$ with $s$, cf. Eq.~(\ref{eq50}).
Since $s=0,1,2,3,\ldots$ is a discrete variable, such a fit becomes
ill-defined for flexible chains; then the only possible procedure is to
use Eq.~(\ref{eq30}) as a definition of the persistence length,
$\ell_{p, \theta} = - \ell_b/\ln [\langle \cos \theta (s=1)\rangle ]$.
Both estimates for $\ell_p$ (from an extended fit over a range of $s$,
and from the latter formula) are collected in Table~\ref{table1},
together with the
prediction based on Eq.~(\ref{eq36}), where excluded volume is neglected.
One recognizes that Eq.~(\ref{eq36}) becomes accurate for $d=3$ as the chains
become very stiff, $q_b\rightarrow 0$ while in $d=2$ Eq.~(\ref{eq36})
\{predicting $\ell_p/\ell_b=0.5 q^{-1}_b$ in this case\}
{never becomes valid}.
As a consequence, we emphasize that the rule (based on the Kratky-Porod model)
that for the same bending stiffness $\kappa/k_BT$ (the continuum analog of
our parameter $q_b$) the persistence length in $d=2$ is twice as large as
in $d=3$ is not accurate (since this rule fails only by about 24 \%,
in experimental work where it was tried to extract estimates of $\ell_p$
from adsorbed semi-flexible chains on two-dimensional substrates this
problem was not noticed, due to other uncertainties in the data analysis).

\begin{table*}[!htb]
\caption{Various possible estimates for persistence lengths,
$\ell_p/\ell_b$ from Eq.~(\ref{eq29}), $\ell_{p,\theta}/\ell_b$ from Eq.~(\ref{eq30}),
$n_p$ and $\langle n_{\rm str} \rangle$ from Eq.~(\ref{eq71}) including the
fitting parameter $a_p$, and $\ell_{p,R}/\ell_b$ from Fig.~\ref{fig6}, for
semiflexible chains in $d=2$ with various values of $q_b$.}
\begin{center}
\begin{tabular}{|c|ccccccccc|}
\hline
$q_b$     & 0.005 & 0.01 & 0.02 & 0.03 & 0.05 & 0.10 & 0.20 & 0.40 & 1.0 \\
\hline
$\ell_p/\ell_b$ & 123.78 & 62.20 & 31.36 & 21.55 & 13.41 & 7.53 & 4.16 & 2.67 & - \\
$\ell_{p,\theta}/\ell_b$ & 118.22 & 59.44 & 30.02 & 20.21 & 12.35 & 6.46 & 3.50 & 2.00 & 1.06 \\
$a_p$ & 0.009 & 0.017 & 0.034 & 0.051 & 0.085 & 0.168 & 0.331 & 0.646 & 1.539 \\
$n_p$ & 116.34 & 58.81 & 29.78 & 20.07 & 12.28 & 6.44 & 3.50 & 2.01 & 1.08 \\
$\langle n_{\rm str} \rangle $ & 118.06 & 59.78 & 30.48 & 20.69 & 12.85 & 6.97 & 4.02 & 2.54 & 1.64 \\
$\ell_{p,R}/\ell_b$ & 3.34 & 2.37 & 1.70 & 1.39 & 1.09 & 0.81 & 0.61 & 0.48 & 0.39 \\
\hline
\end{tabular}
\end{center}
\label{table1}
\end{table*}

\begin{table*}[!htb]
\caption{Various possible estimates for persistence lengths,
$\ell_p/\ell_b$ from Eq.~(\ref{eq29}), $\ell_{p,\theta}/\ell_b$ from Eq.~(\ref{eq30}),
$n_p$ and $\langle n_{\rm str} \rangle$ from Eq.~(\ref{eq71}) including the
fitting parameter $a_p$, and $\ell_{p,R}/\ell_b$ from Ref.~\cite{36}, for
semiflexible chains in $d=3$ with various values of $q_b$.}
\begin{center}
\begin{tabular}{|c|ccccccccc|}
\hline
$q_b$     & 0.005 & 0.01 & 0.02 & 0.03 & 0.05 & 0.10 & 0.20 & 0.40 & 1.0 \\
\hline
$\ell_p/\ell_b$ &  52.61 & 26.87 & 13.93 & 9.54 & 5.96 & 3.35 & 2.05 & - & -\\
$\ell_{p,\theta}/\ell_b$ & 51.52 & 26.08 & 13.35 & 9.10 & 5.70 & 3.12 & 1.18 & 1.12 & 0.67  \\
$a_p$ &  0.02 & 0.04 & 0.08 & 0.12 & 0.19 & 0.38 & 0.73 & 1.42 & 3.37 \\
$n_p$ & 51.17 & 25.95 & 13.30 & 9.07 & 5.68 & 3.12 & 1.82 & 1.13 & 0.68 \\
$\langle n_{\rm str} \rangle $ &  51.72 & 26.50 & 13.83 & 9.60 & 6.20 & 3.65 & 2.36 & 1.70 & 1.29 \\
$\ell_{p,R}/\ell_b$ & 5.35  & 3.49 & 2.39 & 1.94 & 1.54 & 1.12 & 0.87 & 0.71 & 0.61 \\
\hline
\end{tabular}
\end{center}
\label{table2}
\end{table*}

\begin{figure*}
\begin{center}
(a)\includegraphics[scale=0.29,angle=270]{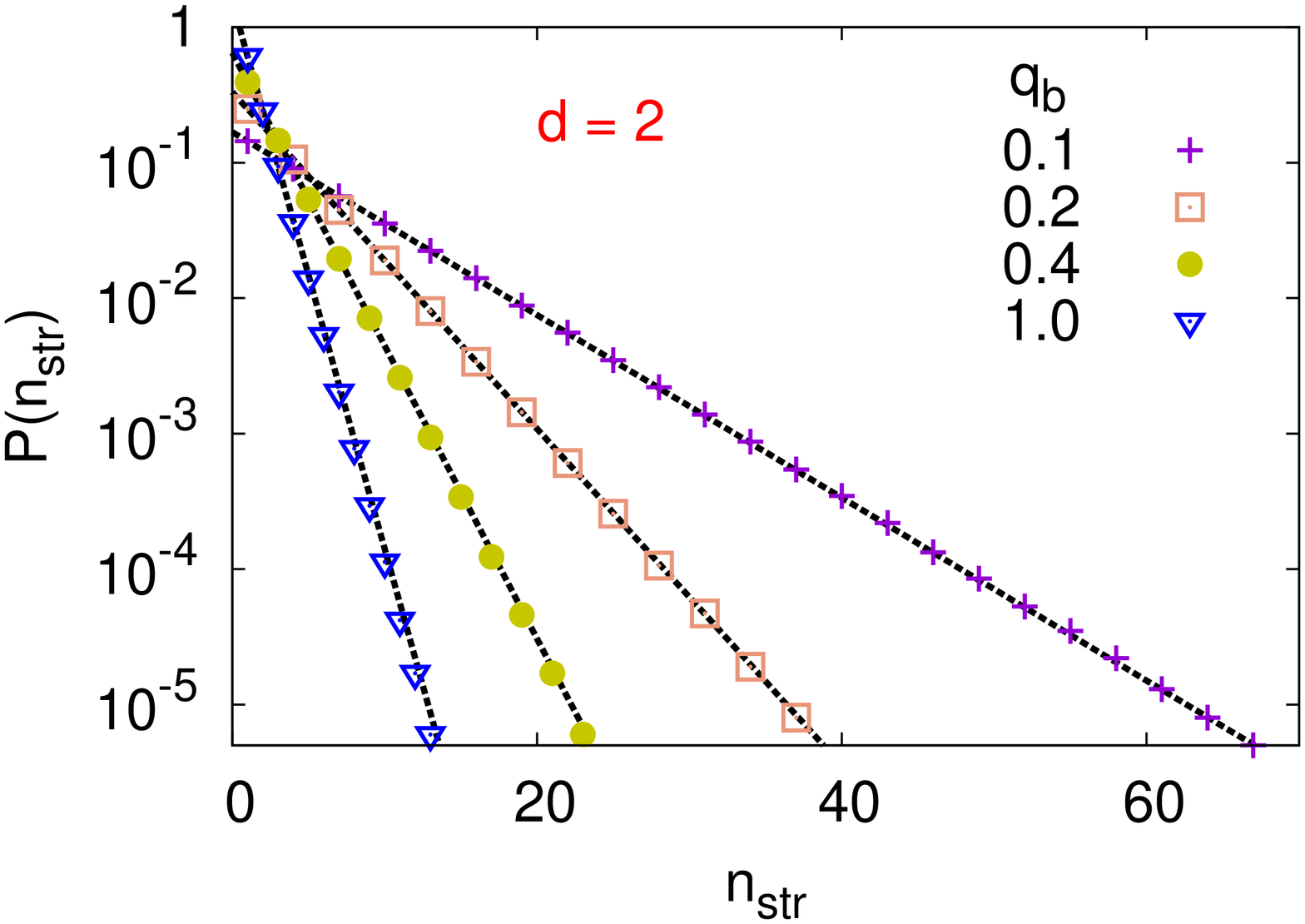}\hspace{0.4cm}
(b)\includegraphics[scale=0.29,angle=270]{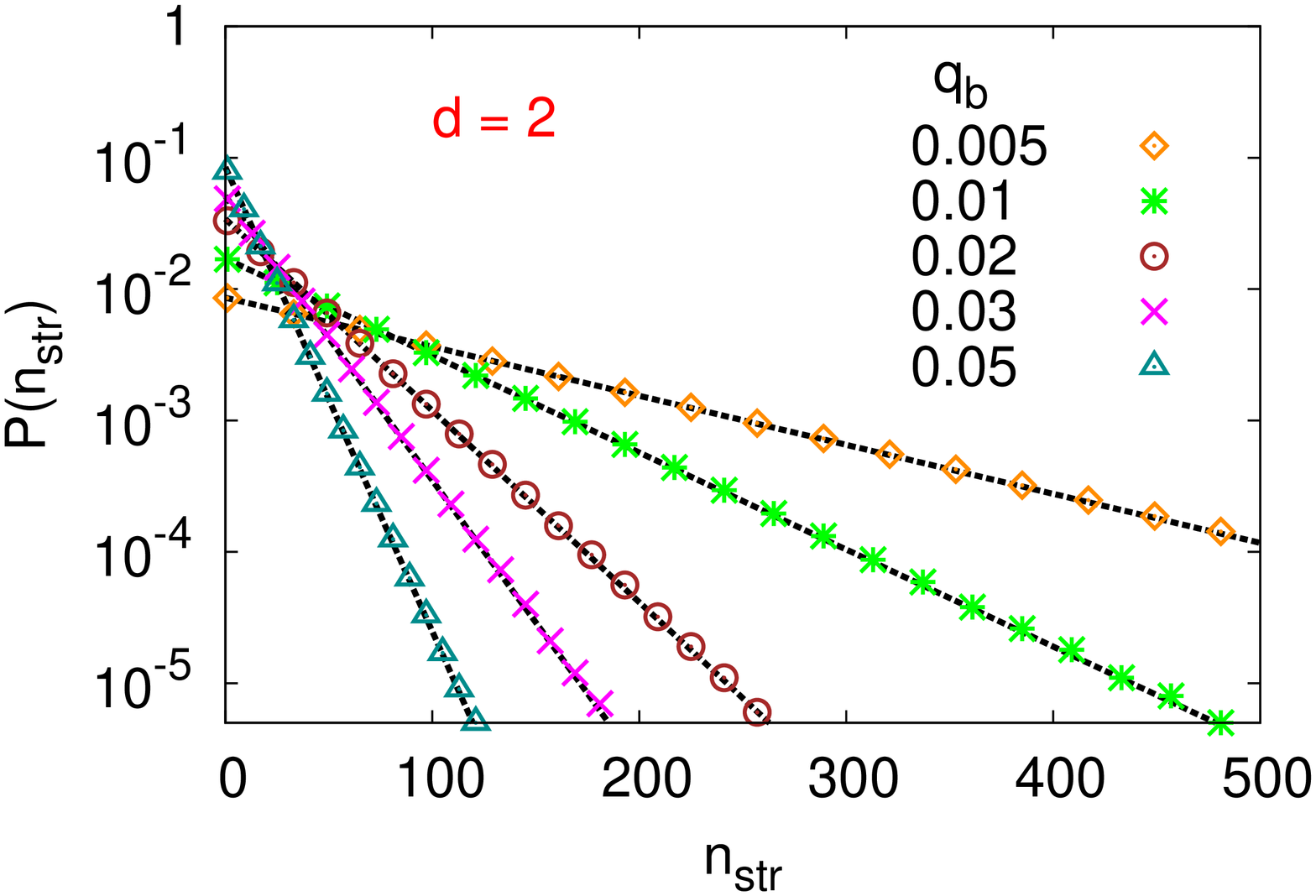}\\
(c)\includegraphics[scale=0.29,angle=270]{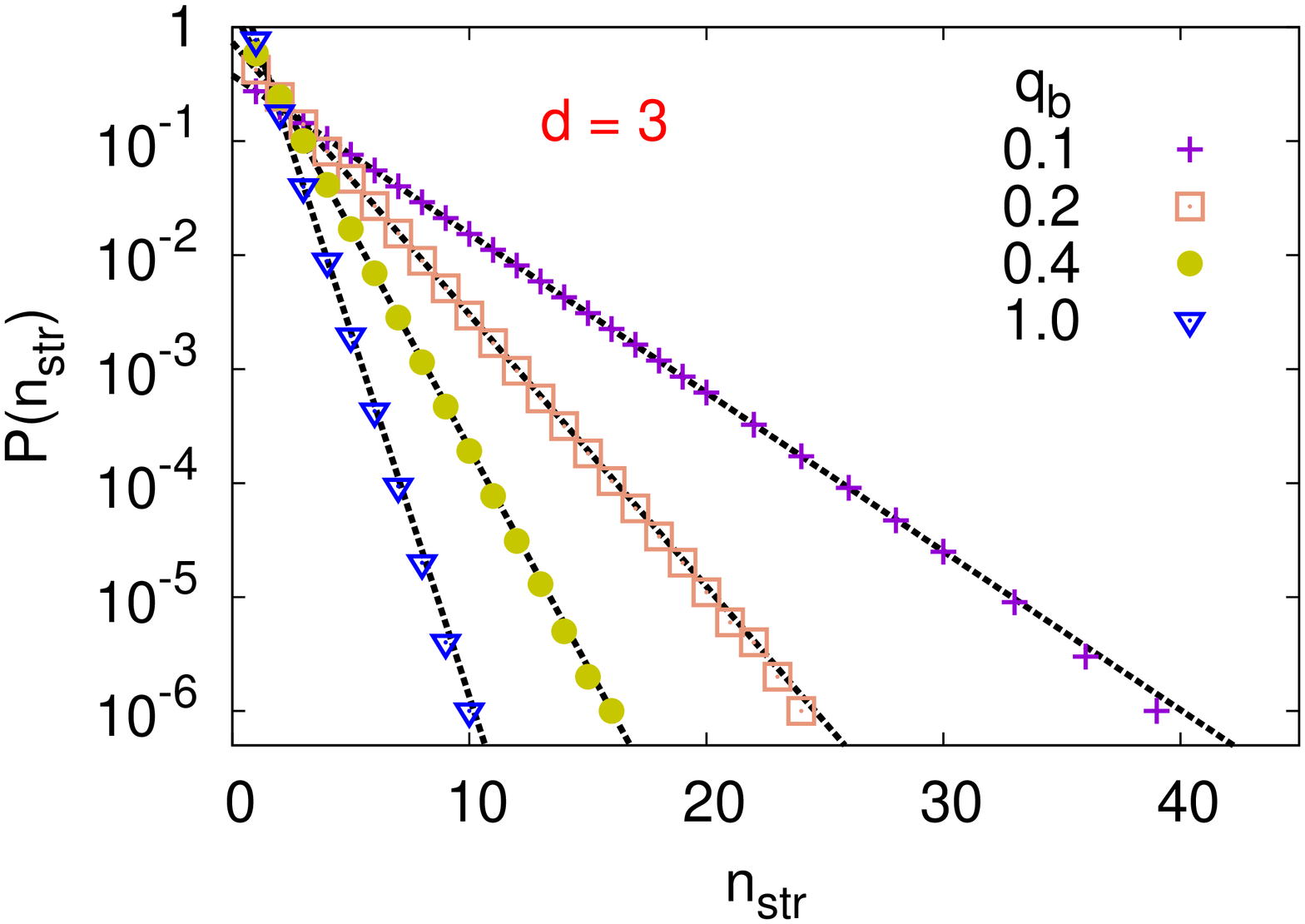}\hspace{0.4cm}
(d)\includegraphics[scale=0.29,angle=270]{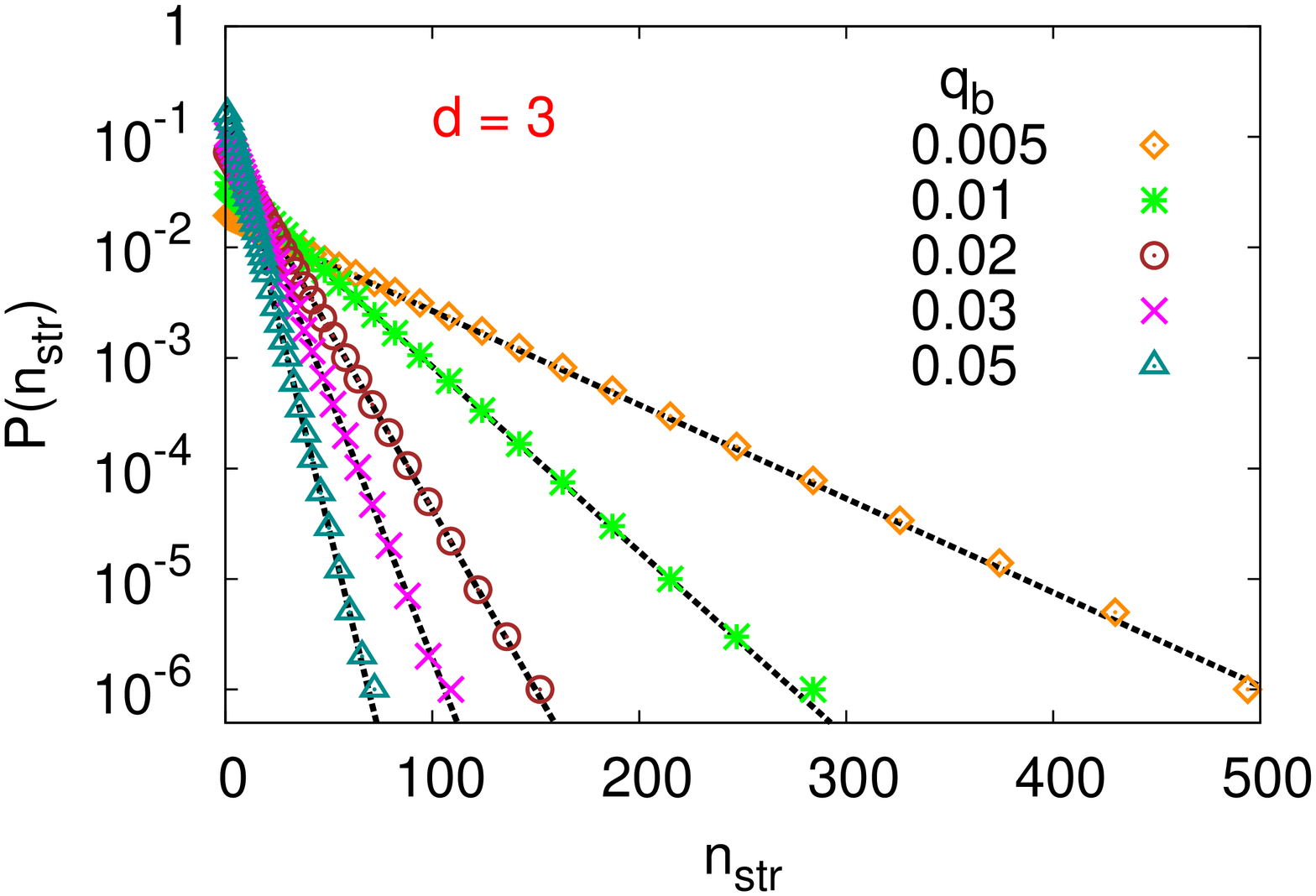}\\
\caption{Semi-log plot of the distribution $P(n_{\rm str})$ of $n_{\rm str}$
successive bonds along the chain which continue straight along a lattice
direction until a kink appears, versus $n_{\rm str}$ for rather
flexible chains,
i.e. $0.1 \leq q_b\leq 1.0$ in $d=2$ (a) and $d=3$ (c), as well as for rather
stiff chains, i.e. $0.005 \leq q_b \leq 0.05$, in $d=2$ (b) and $d=3$ (d).
The straight lines indicate fits to simple exponential functions,
Eq.~(\ref{eq71}), $P(n_{\rm str})= a_ p \exp(-n_{\rm str}/n_p)$,
with constants $a_p$
and $n_p$ quoted in Table~\ref{table1} and ~\ref{table2}.
All data are taken for $N_b= 25600$ in $d=2$
and $N_b= 50000$ in $d=3$.}
\label{fig5}
\end{center}
\end{figure*}

Fig.~\ref{fig3} plots our data for the bond orientational correlations in
a log-log form, to clearly demonstrate that the asymptotic decay is a power
law \{Eq.~(\ref{eq49}\} rather than exponential \{Eq.~(\ref{eq29})\}.
Eqs.~(\ref{eq50}), (\ref{eq52}) in fact suggest to study 
$\langle \vec{a}_i \cdot \vec{a}_{i+s}\rangle$ not simply as a function
of $s$ but as a function of the rescaled variable $s^*=s\ell_b/\ell_p$.
In $d=2$, one expects a data collapse on a universal master curve and
this is indeed found (Fig.~\ref{fig4}(a)). No such simple scaling
is possible in $d=3$, however, as expected from Eq.~(\ref{eq51}).

Another measure of a persistence length in our model is the average number 
$\langle n_{\rm str}\rangle$ of successive bonds along the chain that have the 
same orientation without any kink. The distribution $P(n_{\rm str})$ of such 
straight sequences along the chain is plotted in Fig.~\ref{fig5}. We find 
that irrespective of dimensionality and for all values of $q_b$ the 
distribution shows a simple exponential decay
\begin{equation}\label{eq71}
P(n_{\rm str}) = a_p \exp(-n_{\rm str}/n_p),
\end{equation}
and either the average 
$\langle n_{\rm str}\rangle = \sum \limits _{n_{\rm str}=1} ^\infty 
P(n_{\rm str})n_{\rm str}$ 
or the decay constant $n_p$ can be taken as a characteristic (similar but not 
identical to the persistence length) of local intrinsic chain stiffness. 
Clearly, fitting the data shown in Fig.~\ref{fig5} is less ambiguous than 
fitting the data for the bond orientational correlations. We also note that 
$\langle n_{\rm str}\rangle$ has an obvious physical correspondence in real 
macromolecules: in alkane-type chains, where the torsional potential has one 
deeper minimum (the ``trans'' state, torsional angle $\varphi =0^\circ$) and 
two less deep minima (gauche $\pm$, $\varphi = \pm 120^\circ$) separated 
from the trans state by high energy barriers, $n_{\rm str}$ simply is the 
number of successive carbon-carbon bonds in an all-trans configuration. 
Of course, in this case successive bonds in this state are not oriented 
along the same direction, since the ground state configuration of the alkanes 
is a zigzag-configuration, and so Eqs.~(\ref{eq29}), (\ref{eq50}) need to be 
generalized (bond angles need to be measured relative to their values in 
the ``all-trans'' configuration). Similarly in biopolymers a sequence of 
$n_{\rm str}$ bonds in an $\alpha$-helix configuration can be the right object 
to characterize stiffness. Such considerations will be needed when one
 wants to adapt our findings to real polymer chains.
{Finding an analogue of Eq.~(\ref{eq71}) that is generally
valid for off-lattice models is in interesting problem but beyond
the scope of the present study.}

\begin{figure*}
\begin{center}
(a)\includegraphics[scale=0.29,angle=270]{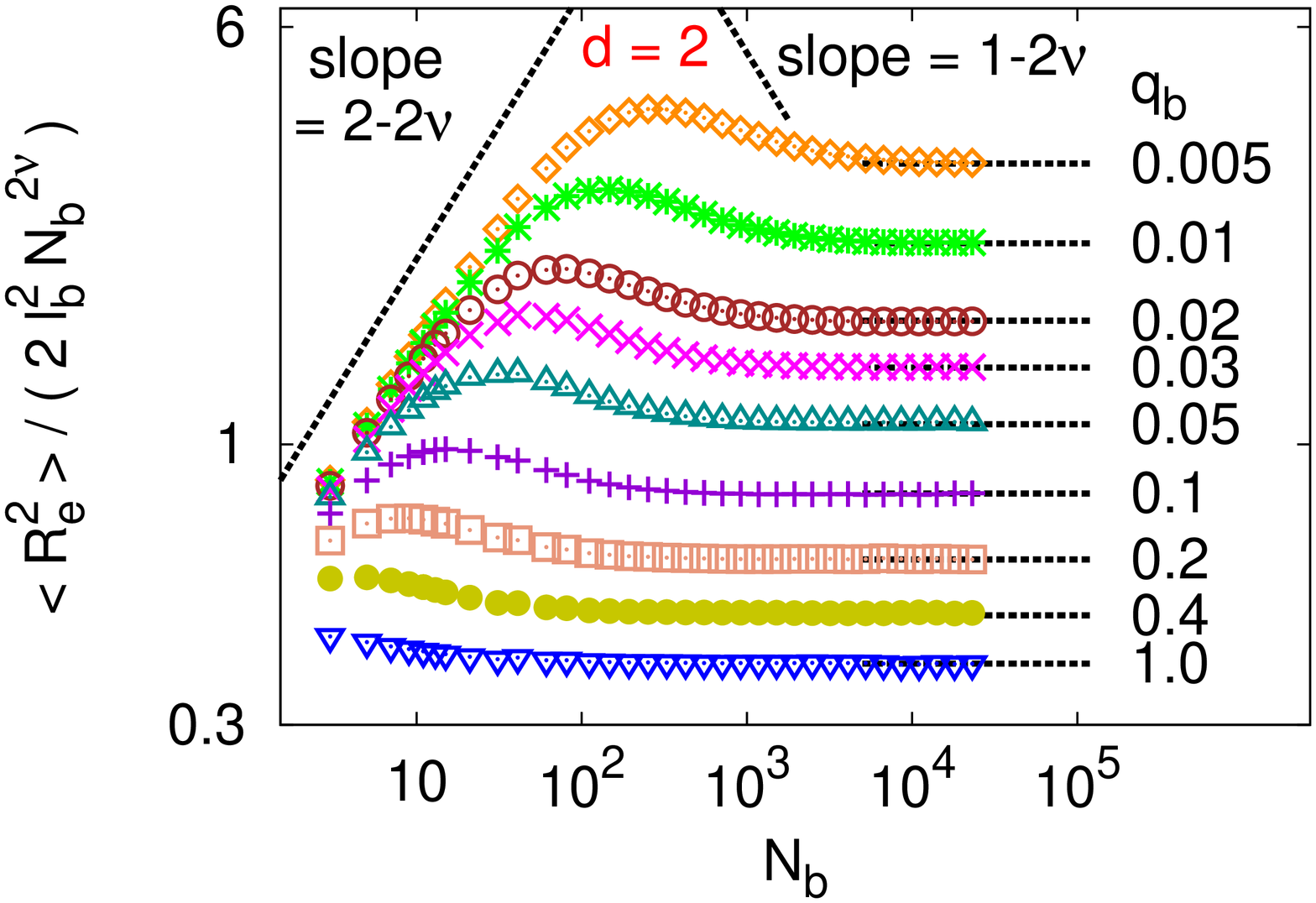}\hspace{0.4cm}
(b)\includegraphics[scale=0.29,angle=270]{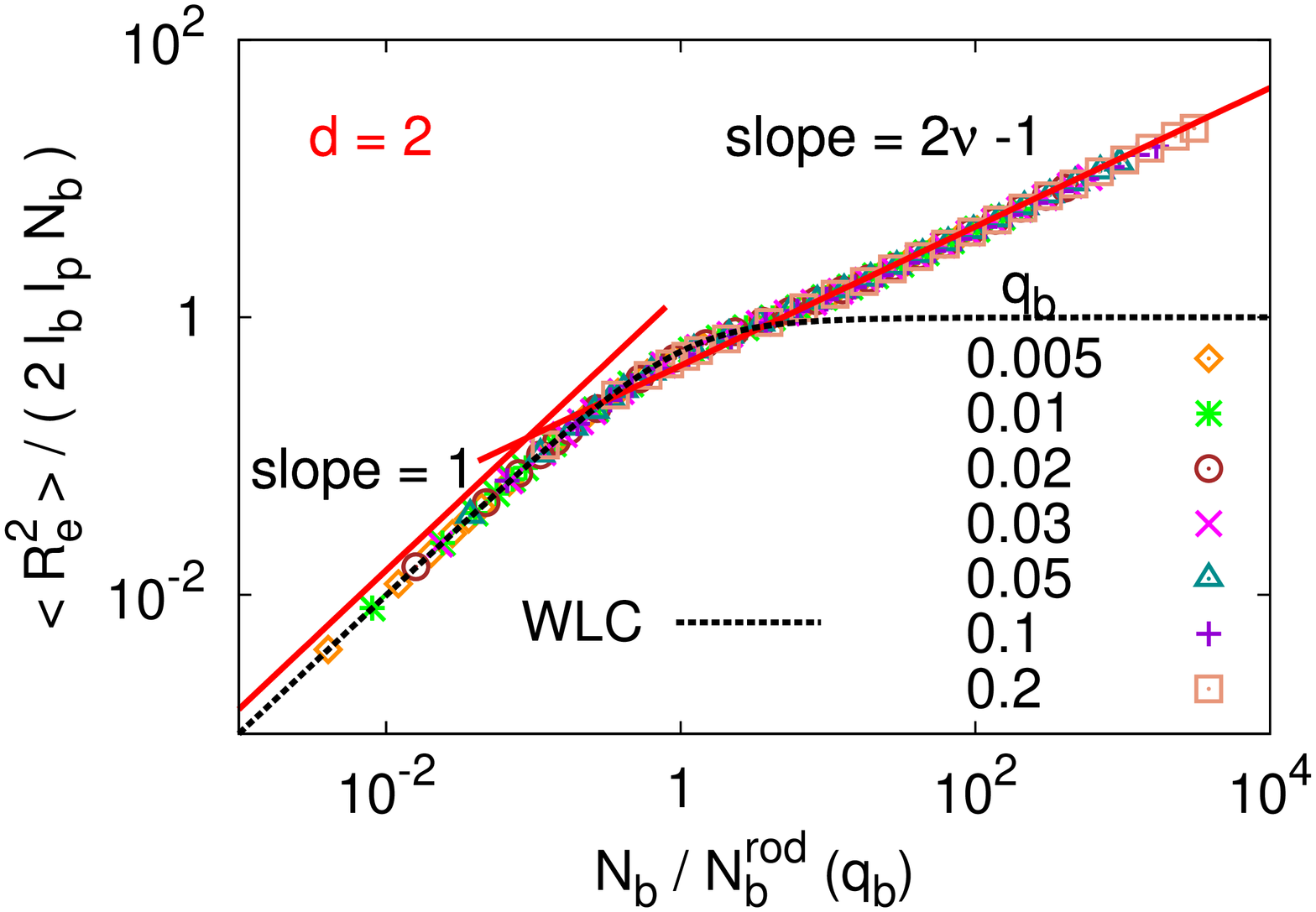}
\caption{Log-log plot of the rescaled mean square end-to-end distance in
$d=2$ dimensions, $\langle R_e^2 \rangle /(2 \ell_b^2N^{2\nu}_b)$
with $\nu=3/4$, versus $N_b$ (a) and {
the normalized rescaled mean square
end-to-end distance $\langle R_e ^2\rangle /(2 \ell_{p} \ell_b N_b)$
versus rescaled chain length $N_b/N_b^{\rm rod}$ (b). These data include chain
lengths $N_b$ up to $N_b=25600$, and all values of the stiffness parameter
$q_b$, as indicated. Straight lines in (a) show the slope $2-2\nu=0.5$
describing the rod-like regime (that occurs for small $N_b)$ and the
slope $1-2 \nu=-0.5$ that would occur if a Gaussian-like regime was
present (which is not). Dotted horizontal plateaus for large $N_b$ in (a) show
estimates for $\ell_{p,R}(q_b)/\ell_b$ (Table~\ref{table1}).
Part (b) shows that all data collapse to a single master curve which
describes a crossover from a rod-like regime to a self-avoiding walk regime.
The Kratky-Porod function, Eq.~(\ref{eq32}), indicated by the dotted curve
(WLC) is also shown for comparison. The chain length
$N_b^*=N_b^{\rm rod}=\ell_k/\ell_b=2\ell_p/\ell_b$ describing the number
of bonds per effective Kuhn segment $\ell_k$ is extracted from the
persistence length estimates (Table~\ref{table1}).}
}
\label{fig6}
\end{center}
\end{figure*}

\begin{figure*}
\begin{center}
(a)\includegraphics[scale=0.29,angle=270]{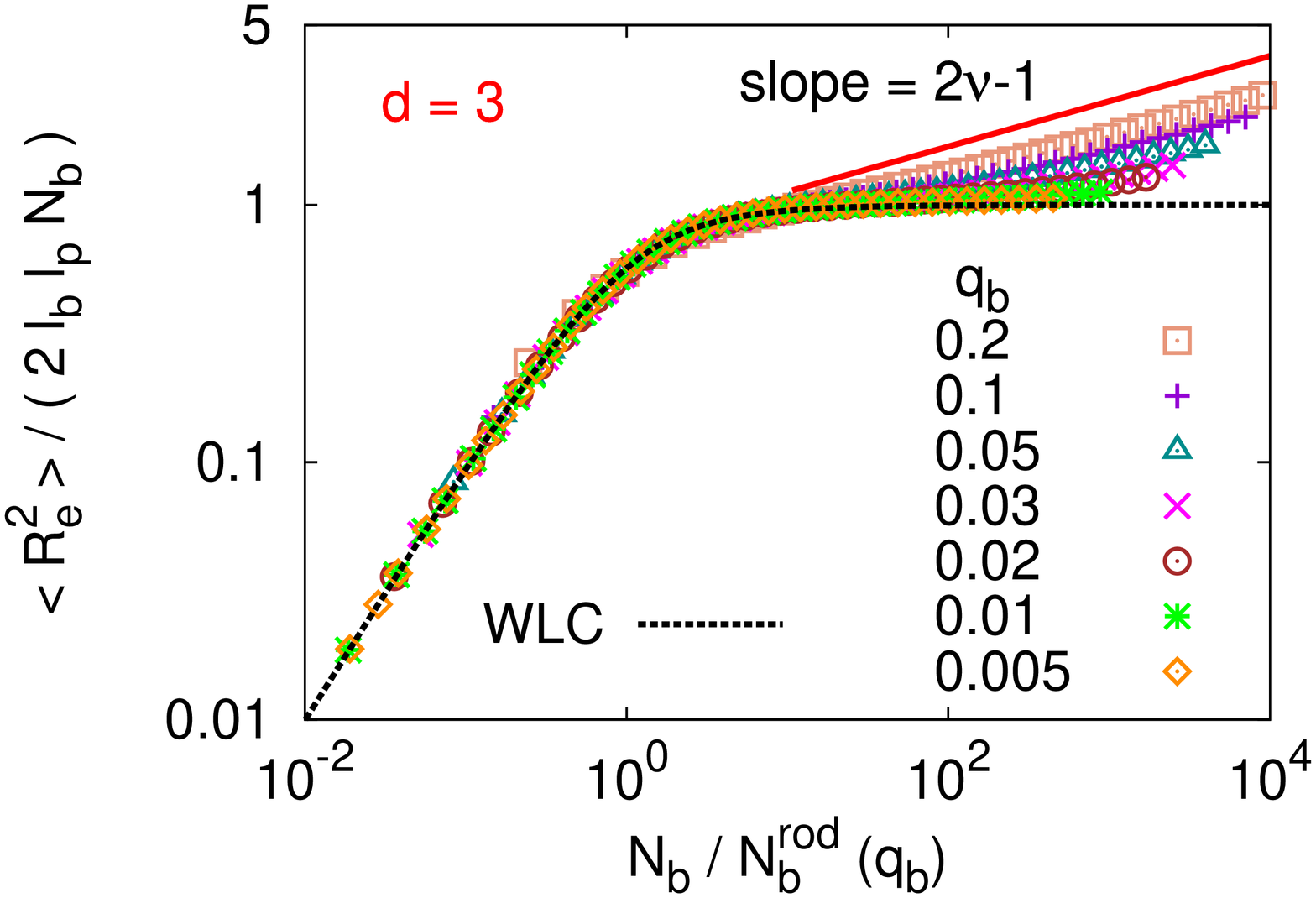}\hspace{0.4cm}
(b)\includegraphics[scale=0.29,angle=270]{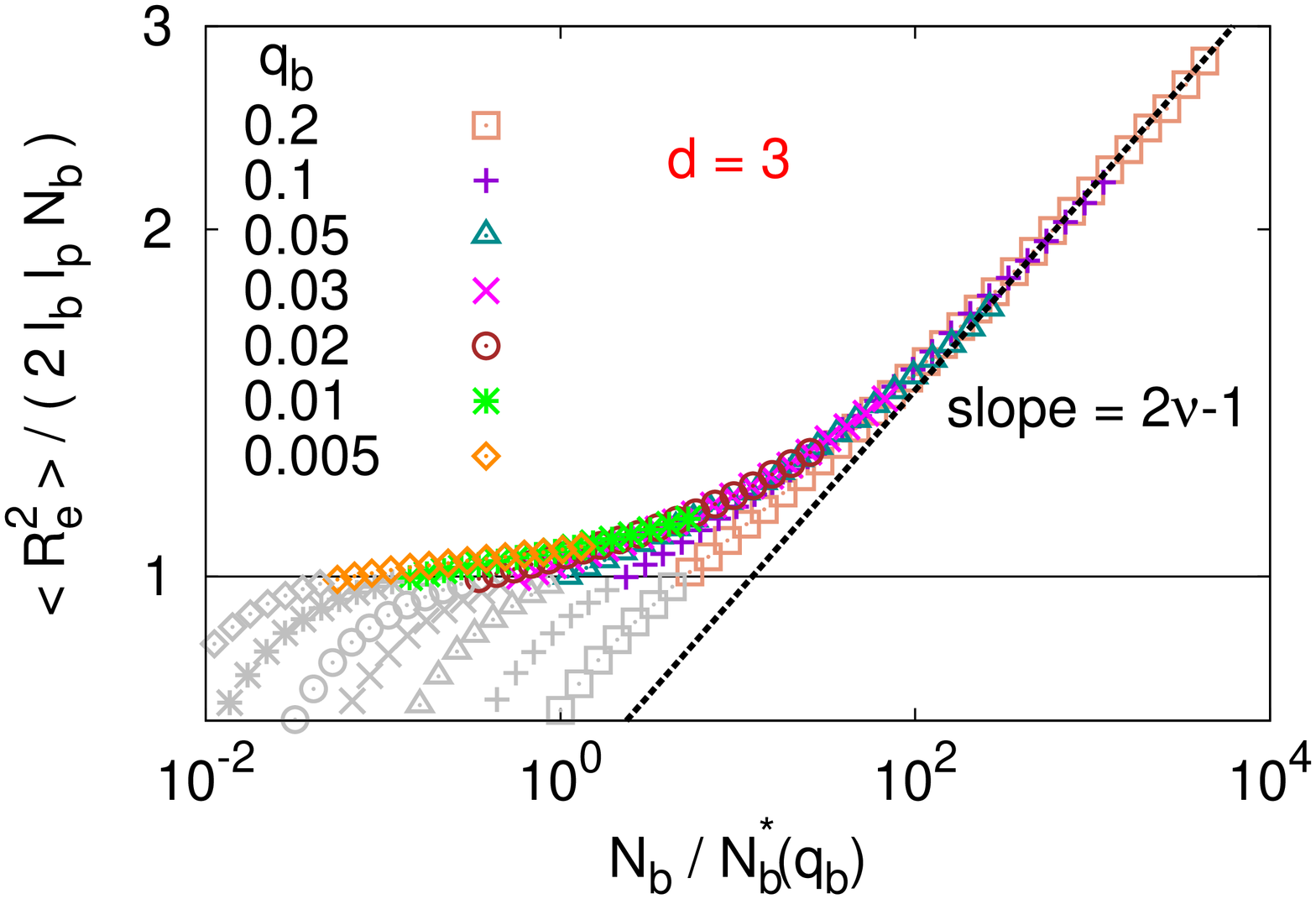}\\
\caption{{
Log-log plot of $\langle R_e^2 \rangle/(2 \ell_b \ell_p(q_b)N_b)$
versus
$N_b/N_b^{\rm rod}(q_b)$ with $N_b^{\rm rod}=2\ell_p(q_b)/\ell_b$. As always,
$\ell_p(q_b)/\ell_b$ is extracted from the initial decay of the bond vector
correlation function (Fig.~\ref{fig2} and Table~\ref{table2}).
Now the initial part of the data
scale, and the smaller $q_b$ (increasing $\ell_p$) the more the data follow
the Kratky-Porod function, Eq.~(\ref{eq32}), indicated by the dotted curve
labeled WLC. Part(b) shows the same data but plotted versus
$N_b/N_b^*(q_b)$ (a) where $N_b^*(q_b)$ is defined such
that for large $N_b$ an optimal data collapse on the straight line representing
the three-dimensional self-avoiding walk behavior
$\langle R_e^2\rangle \propto N_b^{2 \nu}$ is obtained. Note that data that
fall below the Kratky-Porod plateau (horizontal straight line) for different
stiffness parameters systematically splay out, there is no scaling over the
full parameter range. }}
\label{fig8}
\end{center}
\end{figure*}

{
We now turn to the end-to-end distance of the chains
(Figs.~\ref{fig6}, \ref{fig8}). As predicted by the theoretical
considerations of Sec.~II, we find in $d=2$ and $d=3$ dimensions very
different behaviors. In $d=2$ (Fig.~\ref{fig6}) the data for small enough
$N_b$ show the rod-like behavior, $\langle R_e^2 \rangle \propto N_b^2$,
indicated by the slope of the straight line in the left of
Fig.~\ref{fig6}(a), (b), and then $\langle R_e^2\rangle \propto N_b^{2 \nu}$
with $\nu=3/4$ reaches a broad maximum, and thereafter decreases only a
little bit and then settles down at the limiting value expected for
two-dimensional self-avoiding walks. As the rescaled plot (Fig.~\ref{fig6}(b))
shows, there is a single crossover from rods to self-avoiding walks,
and irrespective of stiffness there is never a regime where the Gaussian
plateau predicted by the Kratky-Porod model \{Eq.~(\ref{eq32})\} describes
part of the data approximately. Of course, the latter can describe the
initial rod-like behavior~\cite{41} but this is of little interest and
clearly from this regime one cannot estimate $\ell_p$ reliably at all.
Interestingly we find from the rescaled plot (Fig.~\ref{fig6}(b)) that the
maximum which appear at $N_b=N_b^{\rm max}$ in Fig.~\ref{fig6}(a)
rather accurately coincides with the value $N_b=N_b^*$, the chain
length corresponding to the effective Kuhn length $\ell_k=2 \ell_p$. As an
immediate consequence of this finding we can suggest as a recipe for
experimentalists who analyze end-to-end-distances of two-dimensional
adsorbed chains to plot their data in analogy to Fig.~\ref{fig6}(a)):
if their chain lengths $N_b$ are long enough to reach the region where
the maximum $N_b^{\rm max}$ in such a plot occurs, they can immediately
estimate the persistence length as}
\begin{equation}\label{eq72}
\ell_p=\ell_bN_b^{\rm max}/2 \;.
\end{equation}

{
For this method to work, it is not necessary at all to have chains long
enough to see the asymptotic $d=2$ SAW behavior,
$\langle R_e^2 \rangle \propto N_b^{3/2}$. Since the Kratky-Porod model
\{Eq.~(\ref{eq32})\} has so widely been used by experimentalists to fit their
data and by theorists to build more sophisticated extensions on it, we
emphasize again that Eq.~(\ref{eq32}) is accurate only in the rod-like
regime and in the initial part of the crossover towards the self-avoiding
walk regime as shown in Fig.~\ref{fig6}(b).
Note that unlike
experimental work, this comparison does not involve any adjustable
parameter whatsoever. We see that for small $N_b$ and small $q_b$ theory
and simulation agree qualitatively, but in this regime, where curves for
small $N_b$ collapse on the straight line
$\langle R_e^2\rangle/\ell_bN_b = \ell_bN_b$, and then gradually bend over
to a slower increase, the data are not very sensitive to the actual value
of $\ell_p$. For $N_b <2N_b^*$ the
Kratky-Porod result slightly overestimates the actual data, while for
$N_b\gg 2N_b^*$ it strongly underestimates them, since the increase
proportional to $\langle R_e^2\rangle \propto N_b^{3/2}$ cannot be
described. Clearly, the plateaus predicted by Eq.~(\ref{eq32})
for $N_b>2N_b^*$, as displayed in Fig.~\ref{fig6}(b), do not have any
correspondence to the actual data.}

{However, in the three-dimensional case the situation is clearly
different~\cite{35,36}. We shall not reproduce in full detail the data
published already elsewhere~\cite{35,36} but only show as a summary
of the scaling plots in the Kratky-Porod representation for
the three-dimensional cases (Fig.~\ref{fig8}). Now there
is clear evidence from the data (Fig.~\ref{fig8}(a)) that with increasing
stiffness (decreasing $q_b$) a Gaussian plateau in the plots of
$\langle R_e^2\rangle /(2 \ell_b N_b \ell_p)$ versus $N_b/N_b^{\rm rod}(q_b)$
develops, before the regime ruled
by excluded volume interactions sets in. Here we rescale the chain length
$N_b$ such that data collapse occurs in the rod-like regime,
i.e. $N_b$ is rescaled with $N_b^{\rm rod}=2\ell_p/\ell_b$.
We see that with increasing $\ell_p$ the data gradually approach the
Kratky-Porod result \{Eq.~(\ref{eq32})\} over an increasing range of
$N_b/N_b^{\rm rod}$, while ultimately the data increase beyond the
Kratky-Porod plateau values, to cross over to the asymptotic relation
$\langle R_e^2\rangle/N_b \propto N_b^{2 \nu-1}$ as it should be
\{compare Eqs.~(\ref{eq39})-(\ref{eq43})\}. Alternatively, we can estimate
another crossover chain length $N_b^*(q_b)$ such that the curves collapse
in the regime of large $N_b$, so that the asymptotic regime where excluded
volume interactions dominate, shows proper scaling behavior (Fig.~\ref{fig8}b).
Obviously, while in $d=2$ dimensions $N_b^*(q_b)=N_b^{\rm rod}(q_b)$, so there is
no need to distinguish these crossover chain lengths at all,
(Fig.~\ref{fig6}b), and there is a single crossover from rods to
self-avoiding walks described by one universal crossover scaling function,
this is not true in $d=3$ dimensions: there occur two successive crossovers,
from rods to Gaussian coils at $N_b=N_b^{\rm rod}$, and from Gaussian coils to
three-dimensional self-avoiding walks, at $N_b= N_b^*$. Of course,
these crossovers are rather gradual and not sharp: therefore a well-defined
Gaussian plateau comes into existence only for $N_b^* \gg N_b^{\rm rod}$,
which requires extremely stiff chains.
These findings are in beautiful qualitative agreement with the theoretical
considerations of Sec.~II B and with available experiments in $d=3$ that
did show two successive crossovers~\cite{51}. Surprisingly, some
authors~\cite{Yoshinaga} claim to have observed two successive crossovers
(with an
intermediate Gaussian regime) for two-dimensional adsorbed chains. We suspect
that the observations may be due to incomplete equilibration of the chains,
and we feel that the theoretical interpretation given there is inappropriate,
however.}

\begin{figure*}
\begin{center}
(a)\includegraphics[scale=0.29,angle=270]{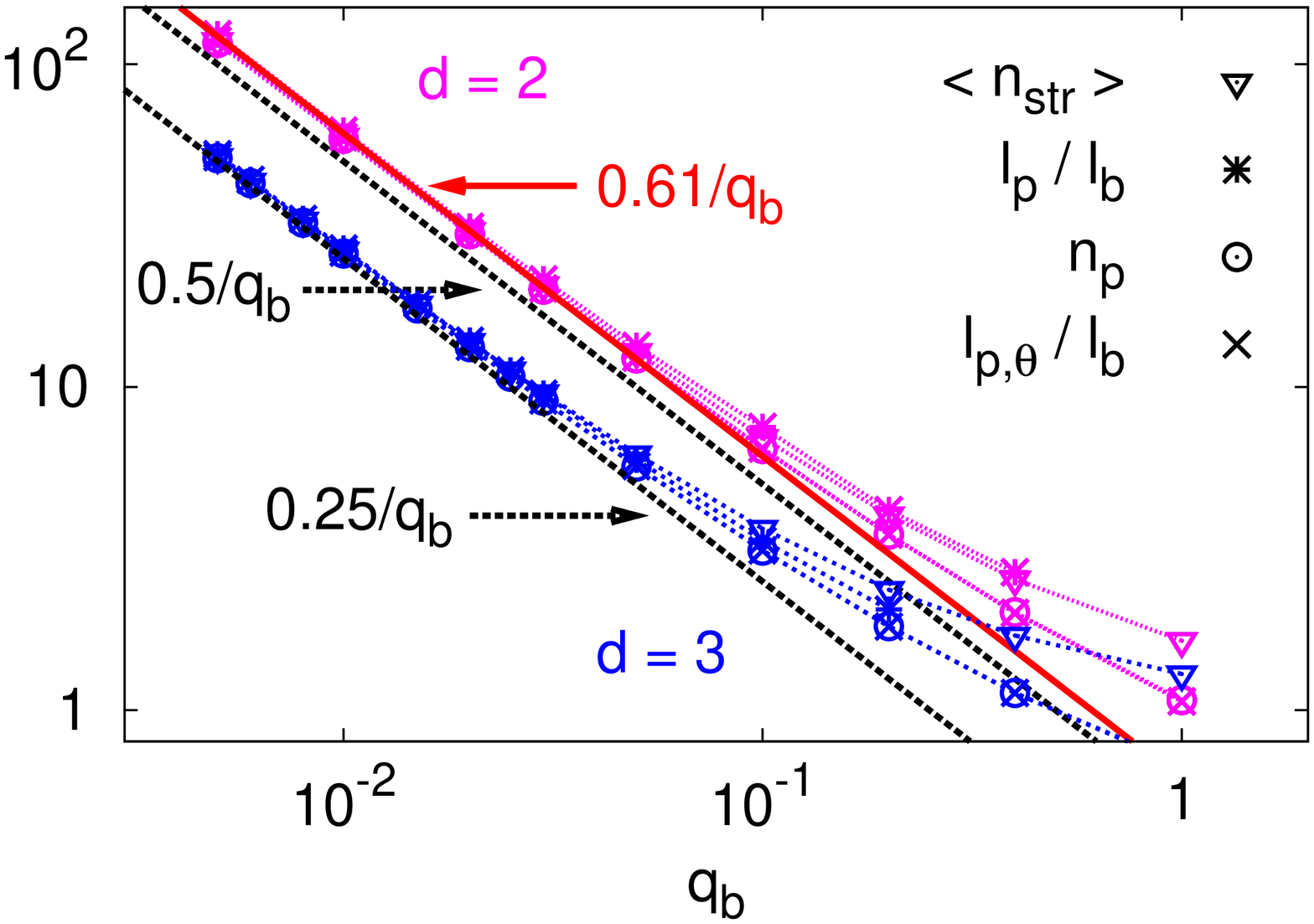}\hspace{0.4cm}
(b)\includegraphics[scale=0.29,angle=270]{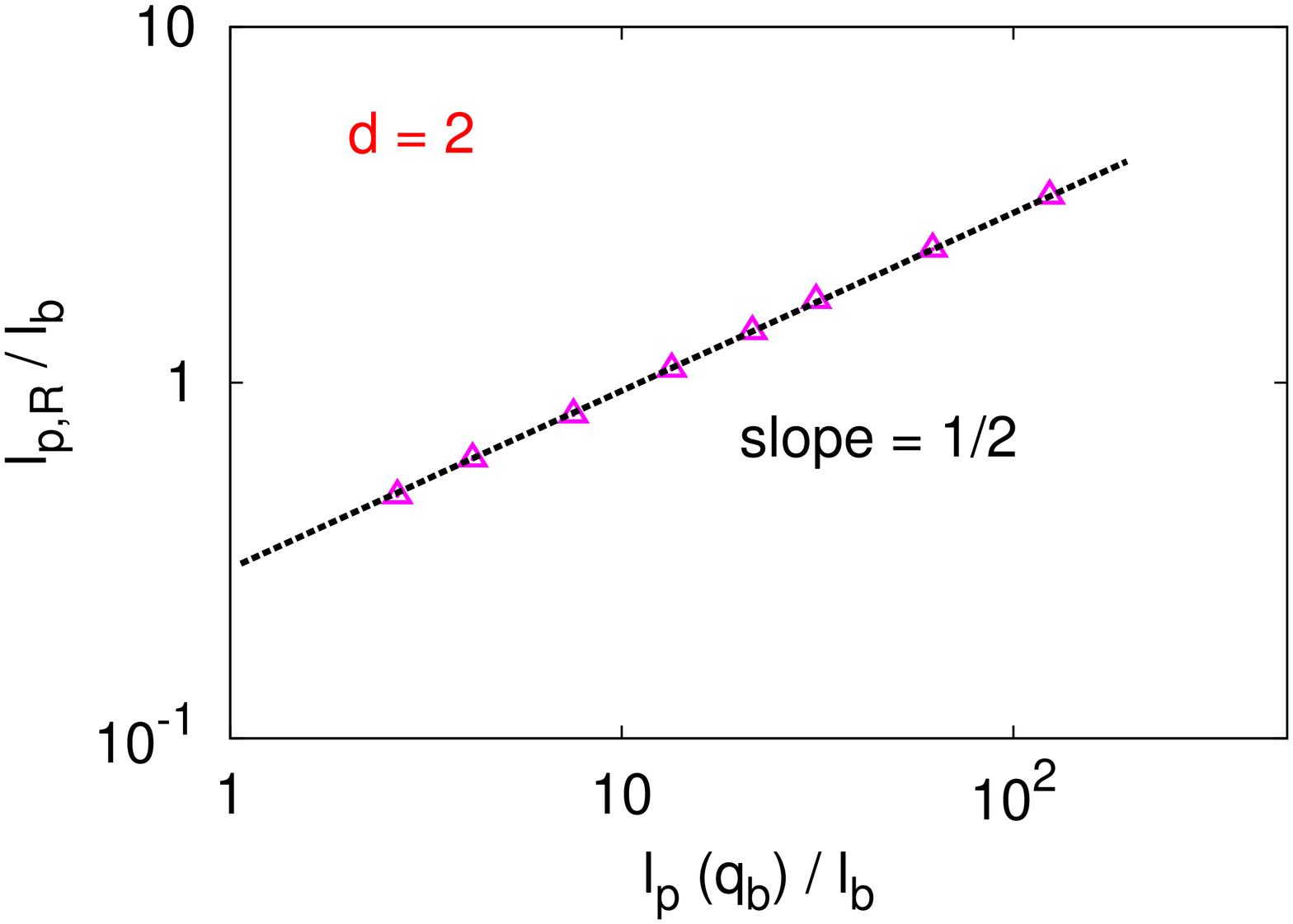}\\
(c)\includegraphics[scale=0.29,angle=270]{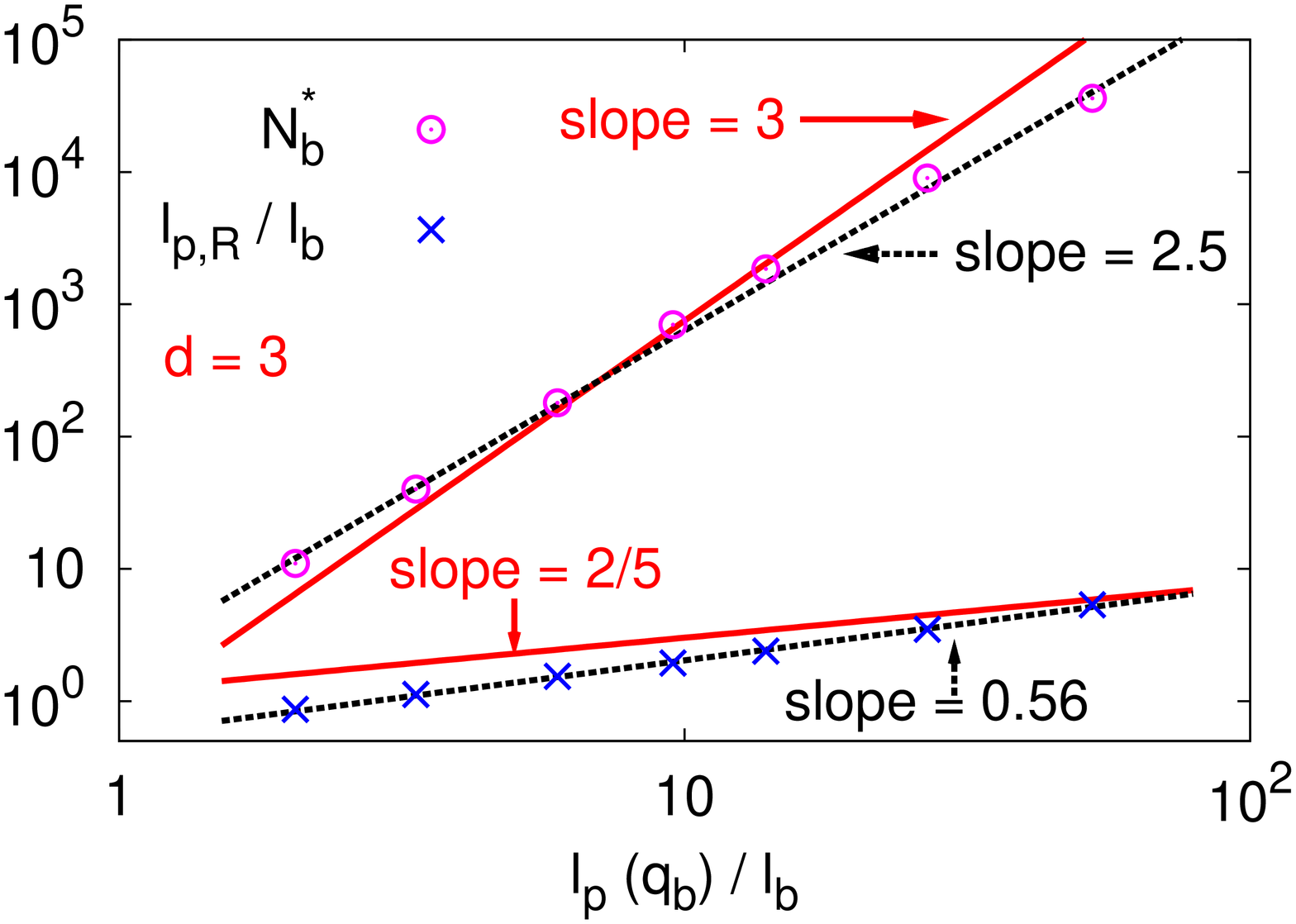}
\caption{Log-log plot of various possible estimates for a persistence length
$\ell_p$ plotted vs.~$q_b$ in both $d=2$ and $d=3$, as indicated. Here $n_p$
and $\langle n_{\rm str} \rangle$ are extracted from the use of
Eq.~(\ref{eq71}),
cf.~Fig.~\ref{fig5}, while $\ell_p/\ell_b$ is taken from the fit to the initial
decay of $\langle \cos \theta (s) \rangle $ with $s$, and
$\ell_{p,\theta}/\ell_b$ is
taken directly from $\langle \cos \theta (s=1)\rangle$
($\ell_{p,\theta}=-\ell_b/\ln$ ($[\langle \cos \theta (s=1)\rangle ]$).
(b) Log-log plot of the amplitude $\ell_{p,R}/\ell_b$ characterizing
the prefactor
of the asymptotic excluded volume region,
$\langle R_e^2\rangle =2\ell_b\ell_{p,R}N^{2\nu}$ with $\nu=3/4$, in $d=2$,
versus
$\ell_p(q_b)/\ell_b$.
Straight line shows a fit to $\ell_{p,R}/\ell_b=0.3 (\ell_p/\ell_b)^{1/2}$.
(c) Log-log plot of $\ell_{p,R}/\ell_b$ and $N_b^*$ in $d=3$ dimensions versus
$\ell_p/\ell_b$, as extracted from the fit shown in Fig.~\ref{fig8}.
Both effective
exponents and theoretical power laws are indicated (cf.~text).}
\label{fig9}
\end{center}
\end{figure*}

It remains to test to what extent the predictions given in Sec.~II B for 
the crossover chain lengths $N_b^{\rm rod}, N_b^*$ are actually compatible 
with our data. First of all, Fig.~\ref{fig9}(a) illustrates that both in 
$d=2$ and $d=3$ the region where $\ell_b$ follows the simple asymptotic 
power law $\ell_b \propto q_b^{-1}$ is quickly reached, and we have ample 
data where $\ell_p$ exceeds $\ell_b$ by at least an order of magnitude.

\begin{figure*}
\begin{center}
(a)\includegraphics[scale=0.29,angle=270]{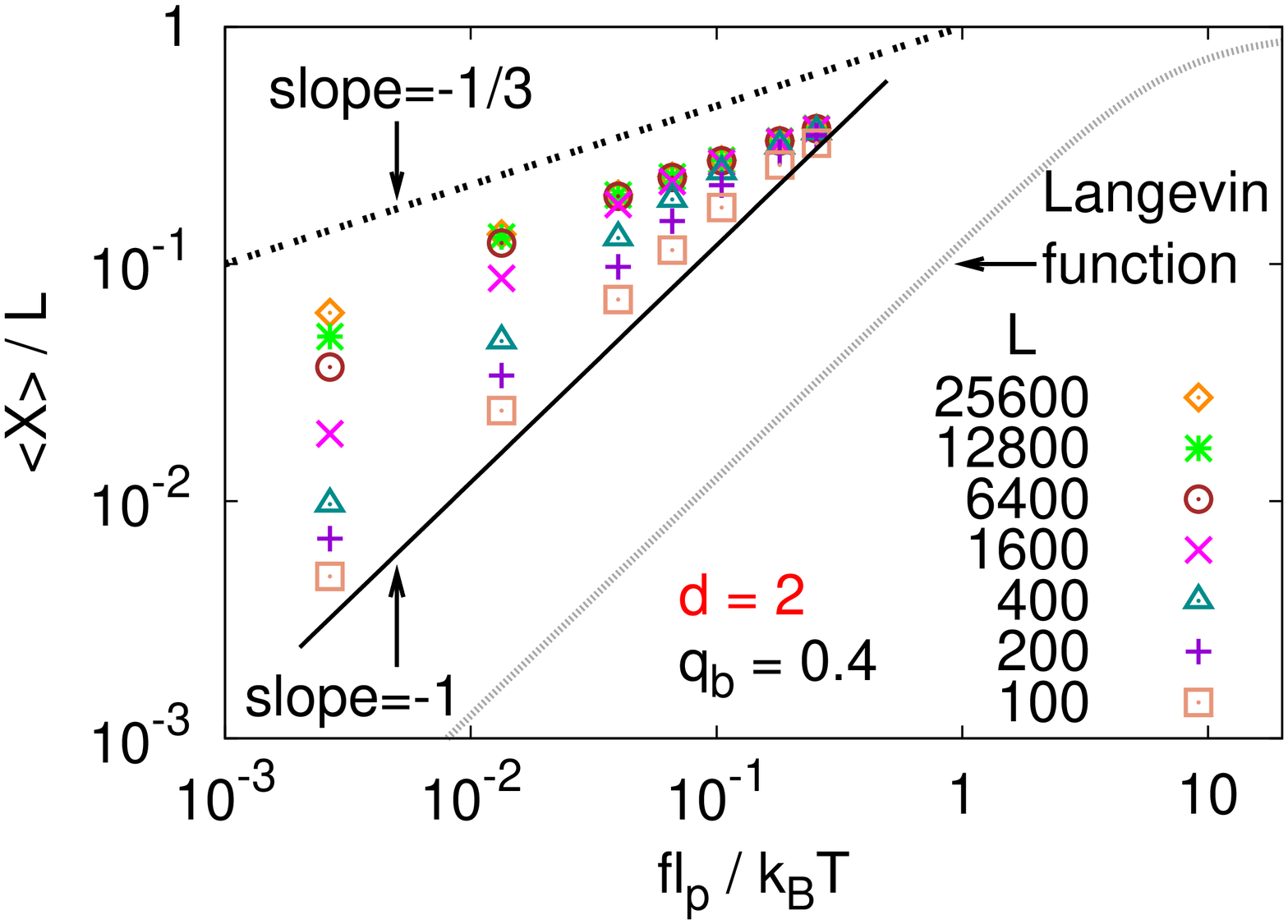}\hspace{0.4cm}
(b)\includegraphics[scale=0.29,angle=270]{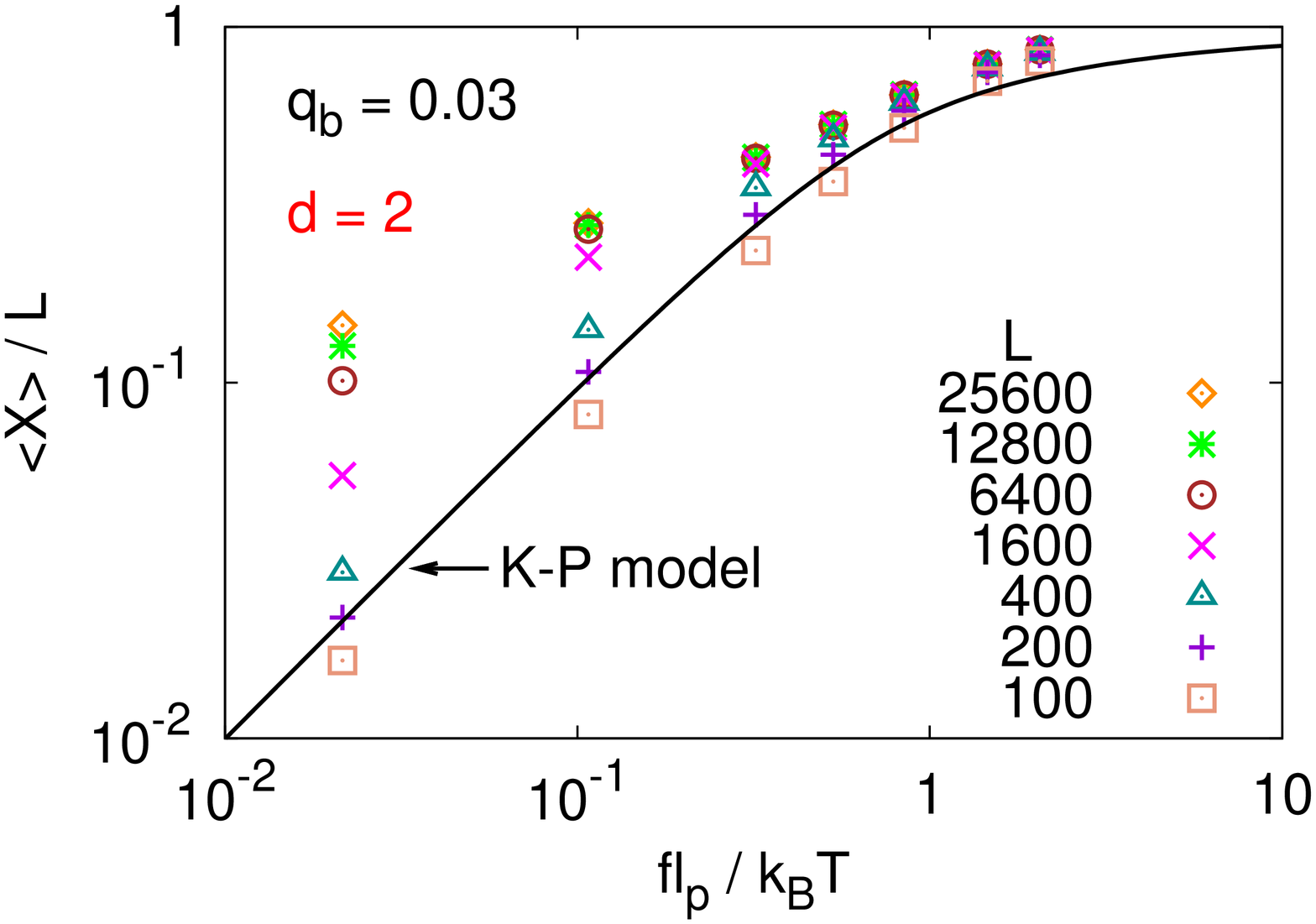}\\
(c)\includegraphics[scale=0.29,angle=270]{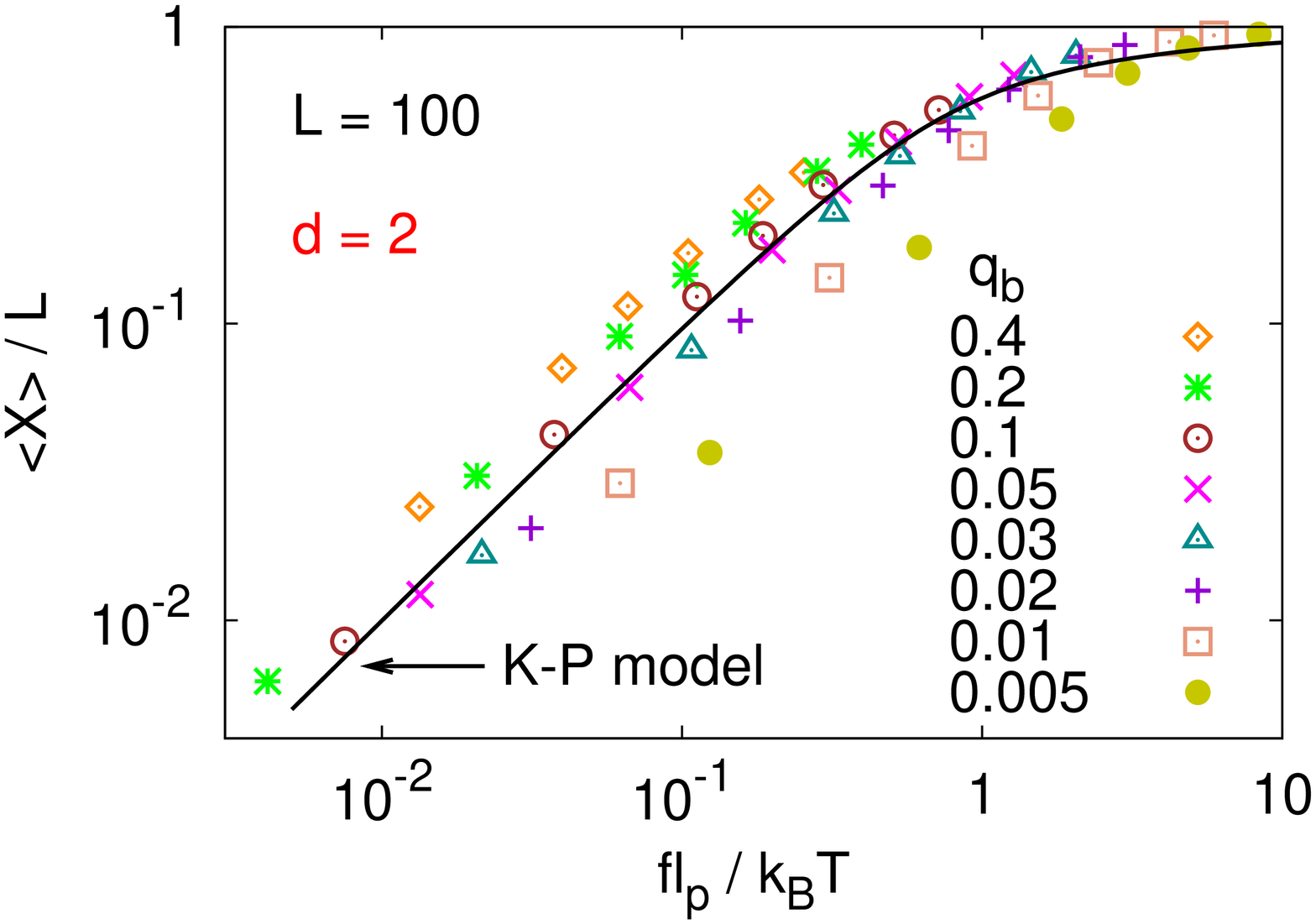}\hspace{0.4cm}
(d)\includegraphics[scale=0.29,angle=270]{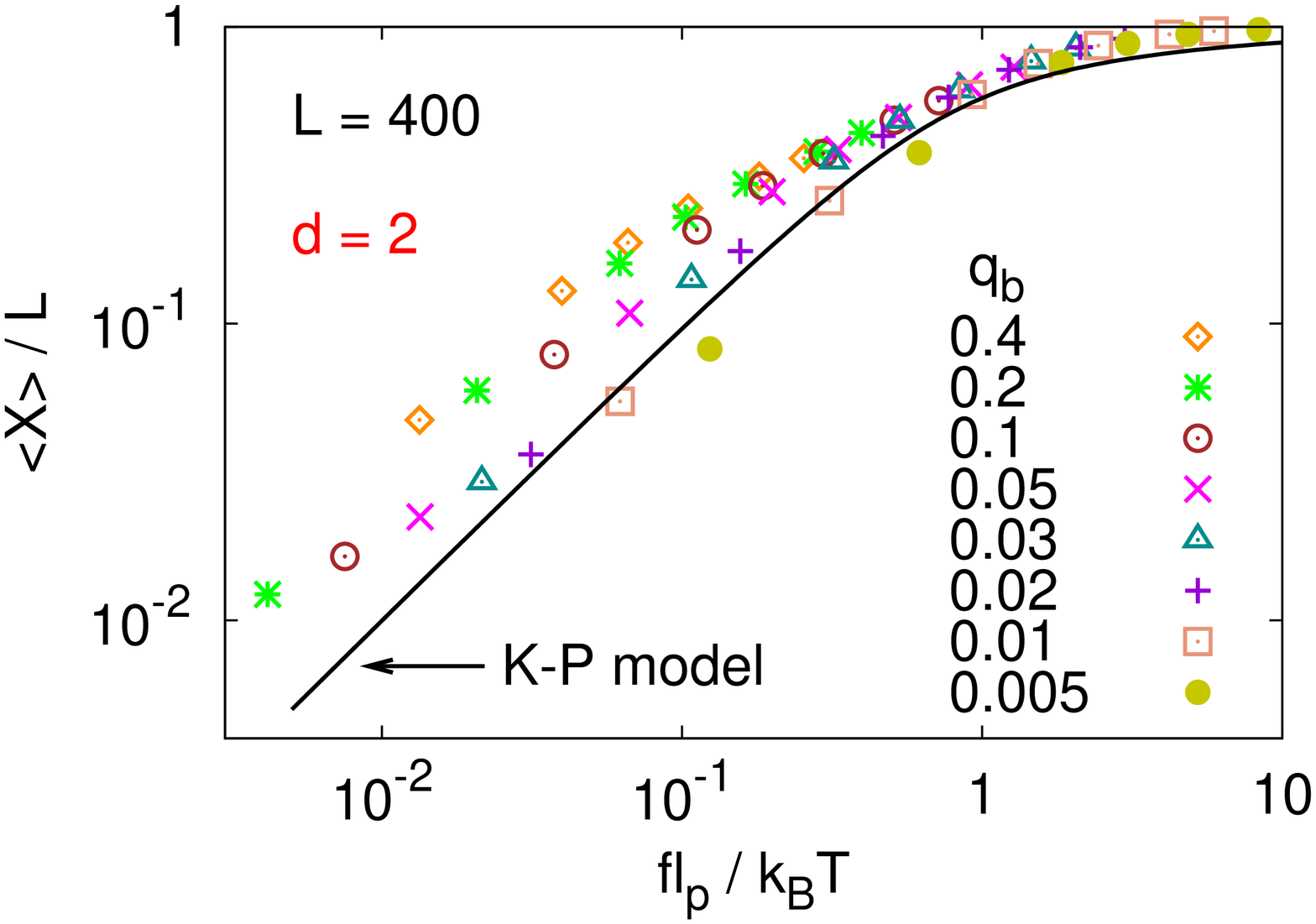}\\
\caption{Relative extension $\langle X \rangle/L$ plotted versus scaled
force $f\ell_p/k_BT$ for rather flexible chains ($q_b=0.4$, case (a)) and
for rather stiff chains ($q_b=0.03$, case (b)) in $d=2$,
including several different contour lengths $L=N_b\ell_b$,
as indicated. In (a), the prediction Eq.~(\ref{eq4}),
$\langle X \rangle/L \propto \mathcal{L}((f\ell_p/k_BT)(\ell_b/\ell_p))$,
for the freely jointed chain and Eq.~(\ref{eq64}) for the Pincus blob
prediction are included for comparison, while in (b)
the result Eq.~(\ref{eq55}) for the Kratky-Porod model is included.
Note that in (a) the result $\ell_p=2.67\ell_b$ (Table~\ref{table1})
was used to convert the scale from $f\ell_p/k_BT$ to $f\ell_b/k_BT$.
Case (c) plots $\langle X \rangle /L$ versus $f\ell_p/k_BT$ and variable
$q_b$ (and hence variable $\ell_p$, cf. Table~\ref{table1}) for
$L=100$ and case (d) for $L=400$, respectively. Eq.~(\ref{eq55})
is again included for comparison.}
\label{fig10}
\end{center}
\end{figure*}

Fig.~\ref{fig9}b also illustrates that in the case of a single crossover 
Eq.~(\ref{eq47}) is quantitatively verified, since Eq.~(\ref{eq47}) says 
$R^2\propto \ell_k^{1/2} L^{3/2} = \ell_b^{3/2} \ell_k^{1/2}N_b^{3/2}$, and 
hence using $\langle R_e^2\rangle = 2 \ell_{p,R}\ell_bN_b$ we would 
conclude $\ell_{p,R}=(\ell_p\ell_b)^{1/2}/\sqrt{2}$, if the proportionally 
constant in Eq.~(\ref{eq47}) is taken to be unity. Of course, only the 
exponent in this relation and not the prefactor can be taken seriously. 
However, in $d=3$ the theoretical relations $N_b^*\propto \ell_p^3$ and
$\ell_{p,R} \propto \ell_p^{2/5}$ 
\{Eqs.~(\ref{eq40}), (\ref{eq43})\} are not 
quantitatively verified: rather we found effective exponents 
$\ell_{p,R}\propto \ell_p^{0.56}$ and $N_b^*\propto \ell_p^{2.5}$ 
(Fig.~\ref{fig9}c). We cannot rule out that this result is due to a still 
somewhat slower crossover to the asymptotic excluded volume dominated regime 
than assumed in our fit in Fig.~\ref{fig8}; still much longer chains than 
$N_b=50000$ would be needed to check this, but this is a very tough task 
even for the PERM algorithm. On the other hand, we note that 
Eqs.~(\ref{eq38})-(\ref{eq43}) clearly are not exact, the Flory argument 
invariably implies that $\nu=3/5 = 0.60$ instead of $\nu=0.588$~\cite{44} 
and it is unclear to us to what extent these exponents describing the 
variation of $\ell_{p,R}$ and $N_b^*$ with $\ell_p$ are modified. 
This problem could possibly be addressed with the renormalization group 
approach.

\begin{figure*}
\begin{center}
(a)\includegraphics[scale=0.29,angle=270]{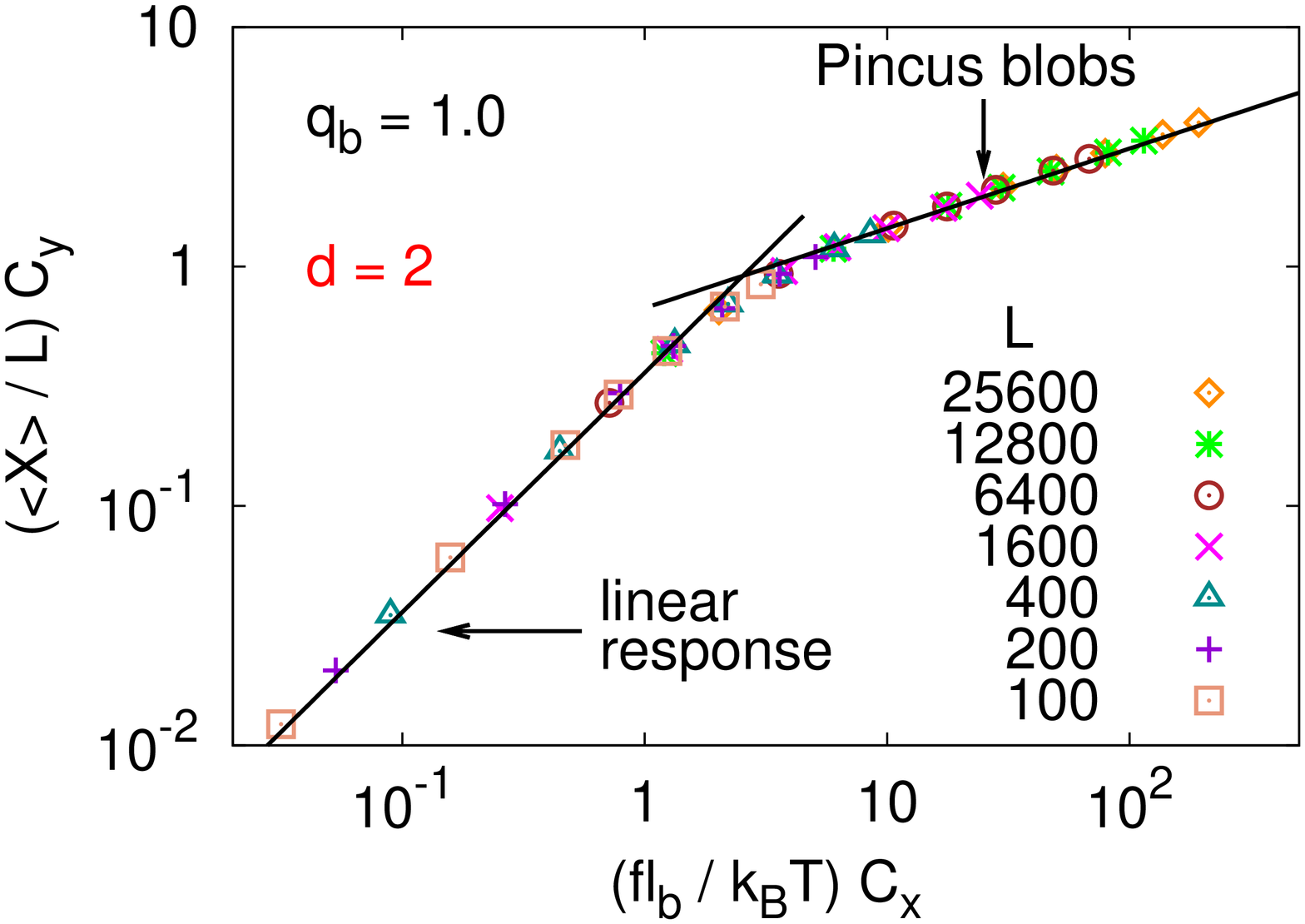}\hspace{0.4cm}
(b)\includegraphics[scale=0.29,angle=270]{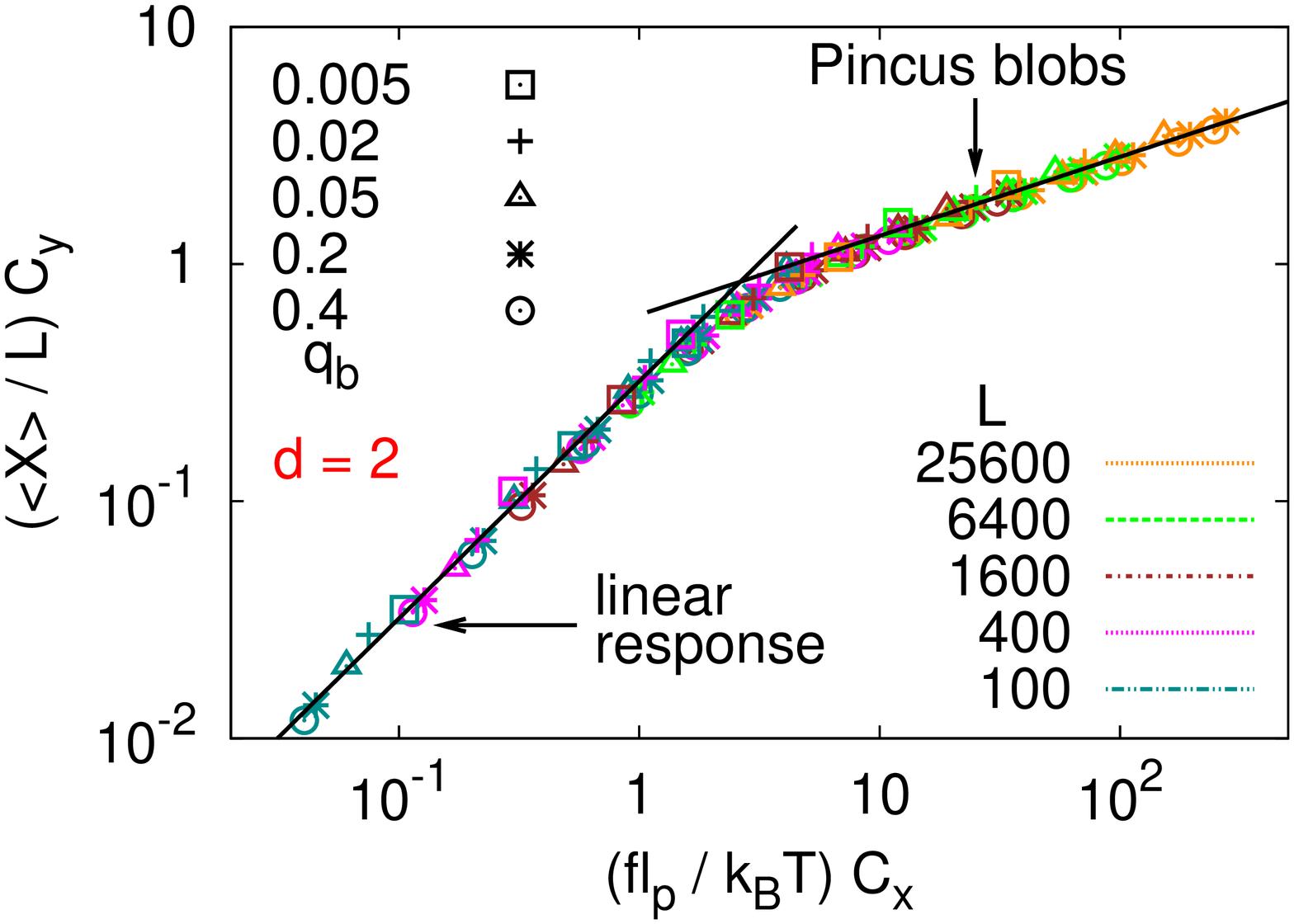}\\
(c)\includegraphics[scale=0.29,angle=270]{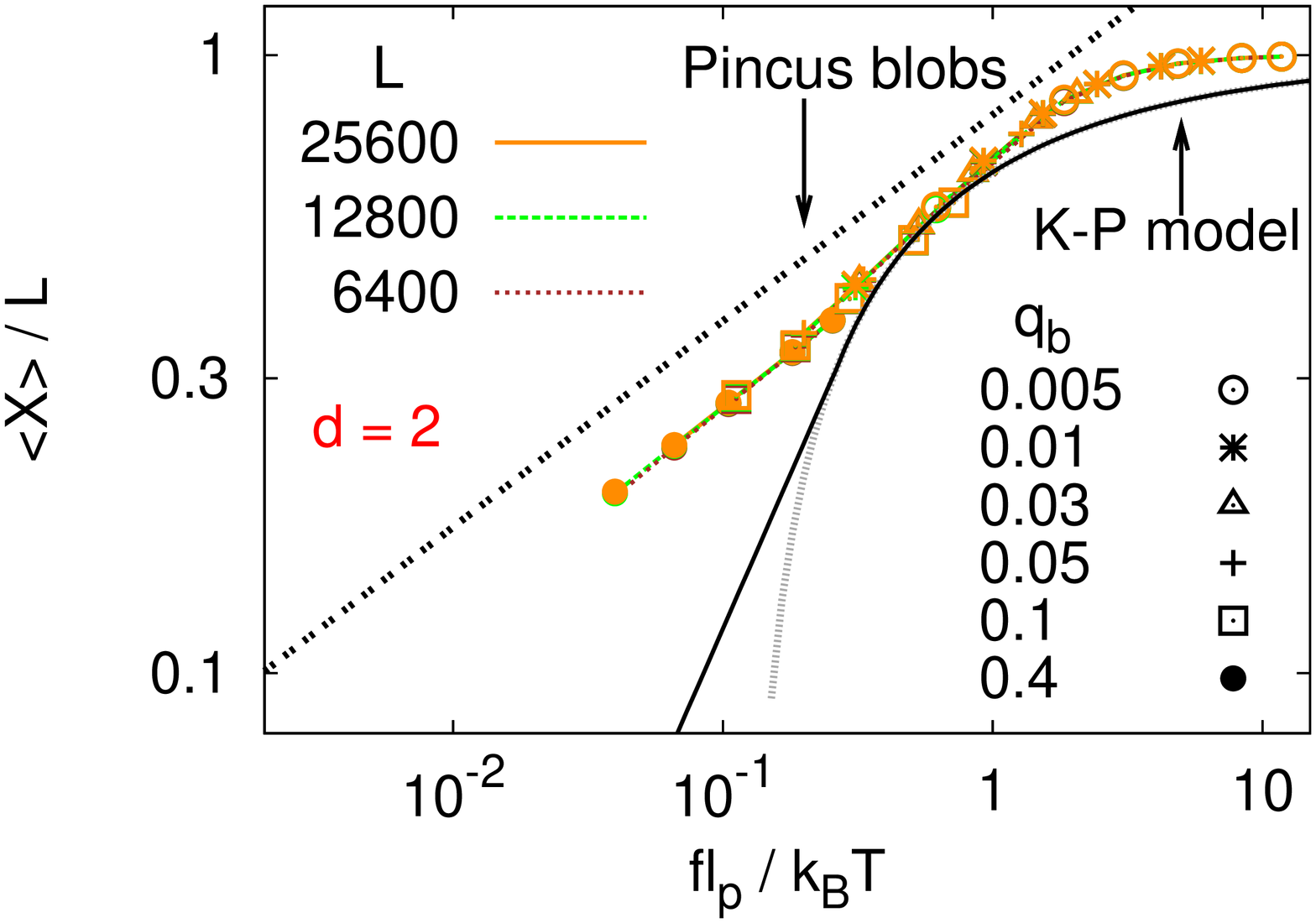}\\
\caption{Plot of $y=(\langle X \rangle/L)C_y$ versus $x=(f\ell_b/k_BT)C_x$
for flexible SAW's in $d=2$ ($q_b=1$) where the scaling factors
$C_x$, $C_y$ for abscissa and ordinate have been chosen $C_x=N_b^{3/4}$,
$C_y=N_b^{1/4}$, so that the coordinates ($x_{\rm cr}, y_{\rm cr}$)
of the crossover point from the linear response regime to the Pincus
blob regime in the (x,y) plane are of order unity. Several choices
of $L$ are included,
as indicated. Part (b) is similar as (a), but for semiflexible chains
with several
choices of $q_b$ and $L$, as indicated. Now $x=(f\ell_p/k_BT)C_x$ and
the scaling factors are
chosen as $C_x=(L/\ell_p)^{3/4}$, $C_y=(L/\ell_p)^{1/4}$. Straight
lines in both parts indicate the theoretical power law
$\langle X \rangle \propto f$ (linear response regime) and
$\langle X \rangle \propto f^{1/3}$ (Pincus blobs regime), respectively.
Part (c) is the same as (b), but choosing $C_x=C_y=1$ including only data for
$L=25600$, $12800$, and $6400$, to show the crossover from the Pincus
blob regime to a Kratky-Porod (K-P) like regime.}
\label{fig11}
\end{center}
\end{figure*}

\begin{figure*}
\begin{center}
(a)\includegraphics[scale=0.29,angle=270]{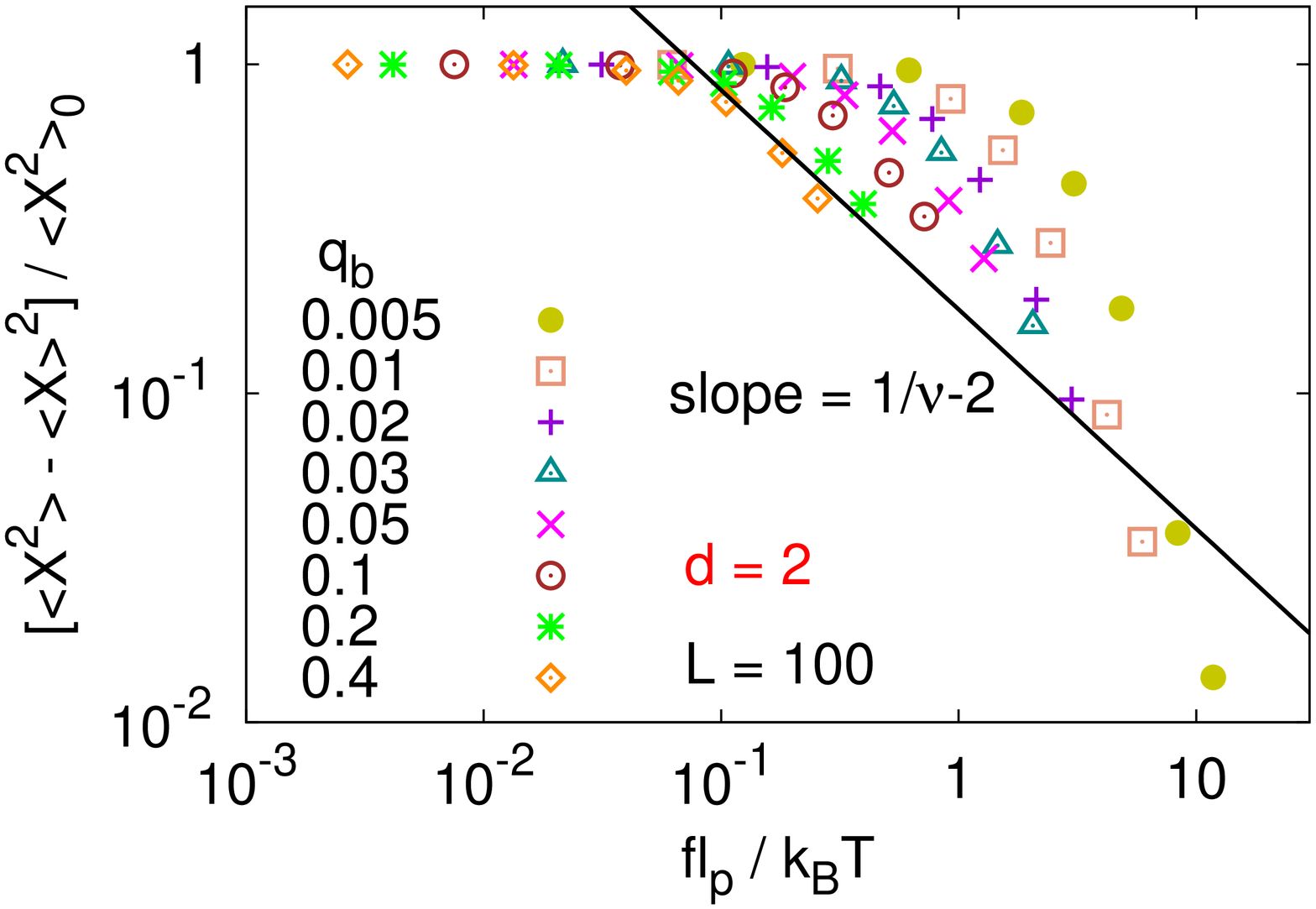}\hspace{0.4cm}
(b)\includegraphics[scale=0.29,angle=270]{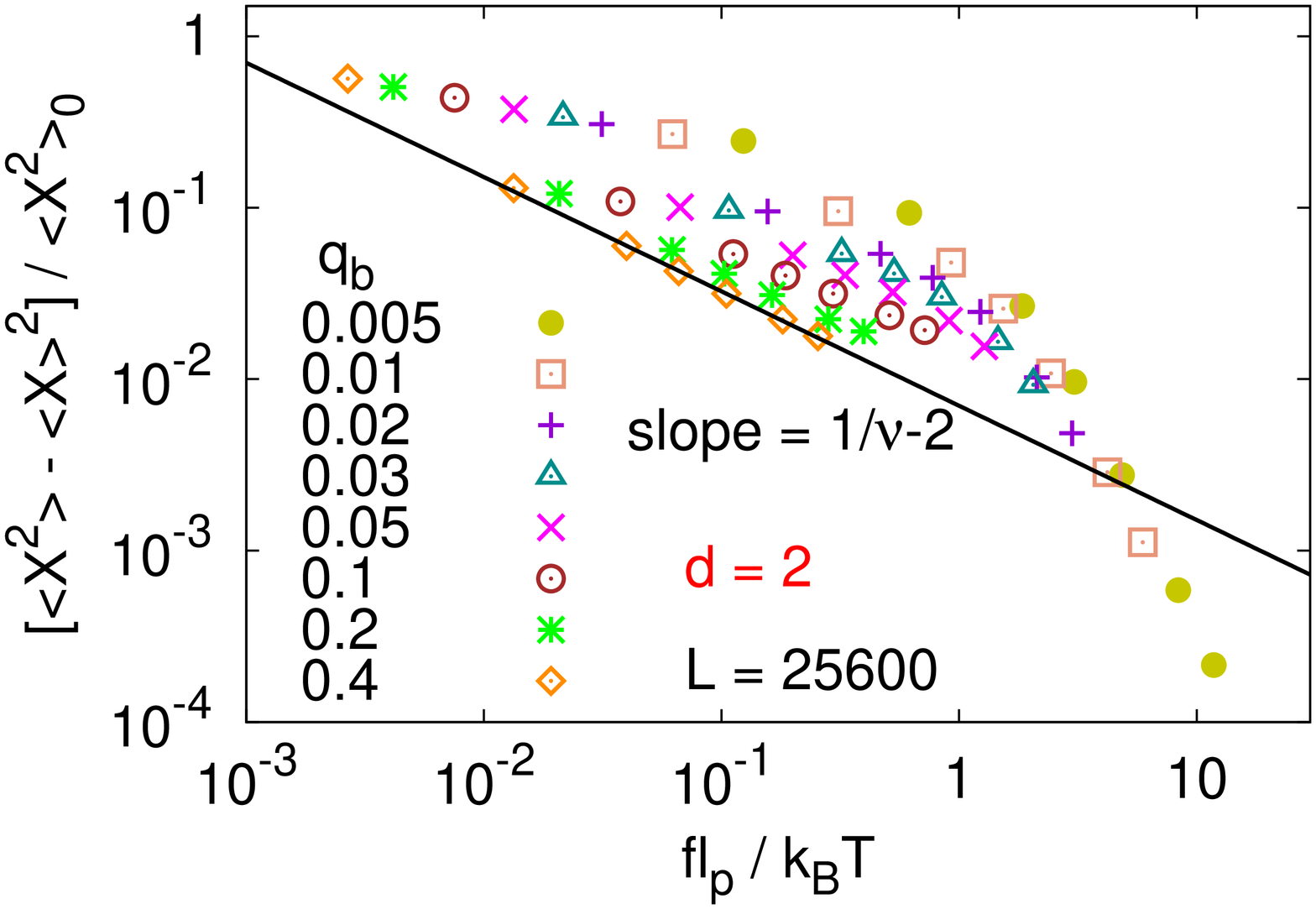}\\
(c)\includegraphics[scale=0.29,angle=270]{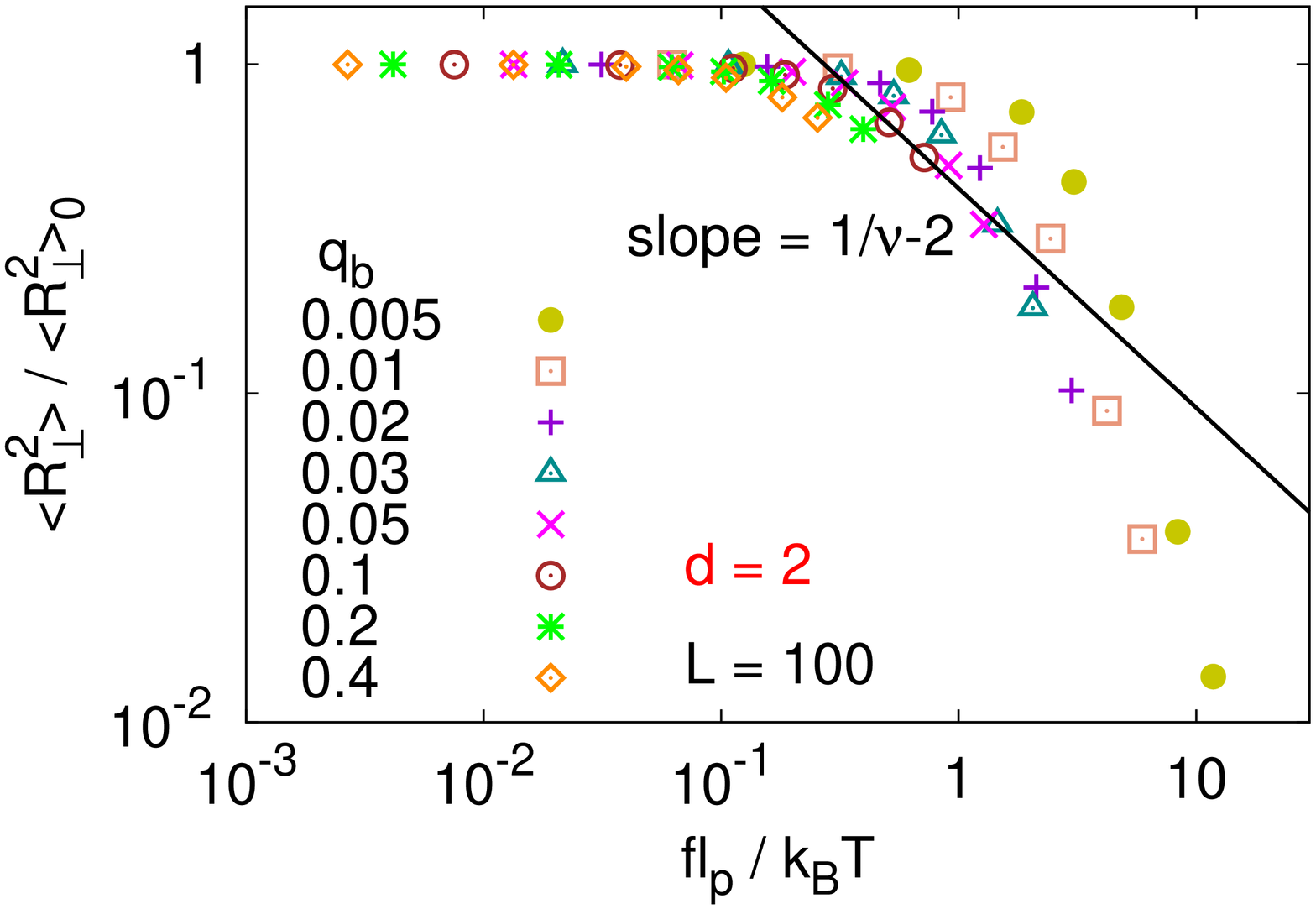}\hspace{0.4cm}
(d)\includegraphics[scale=0.29,angle=270]{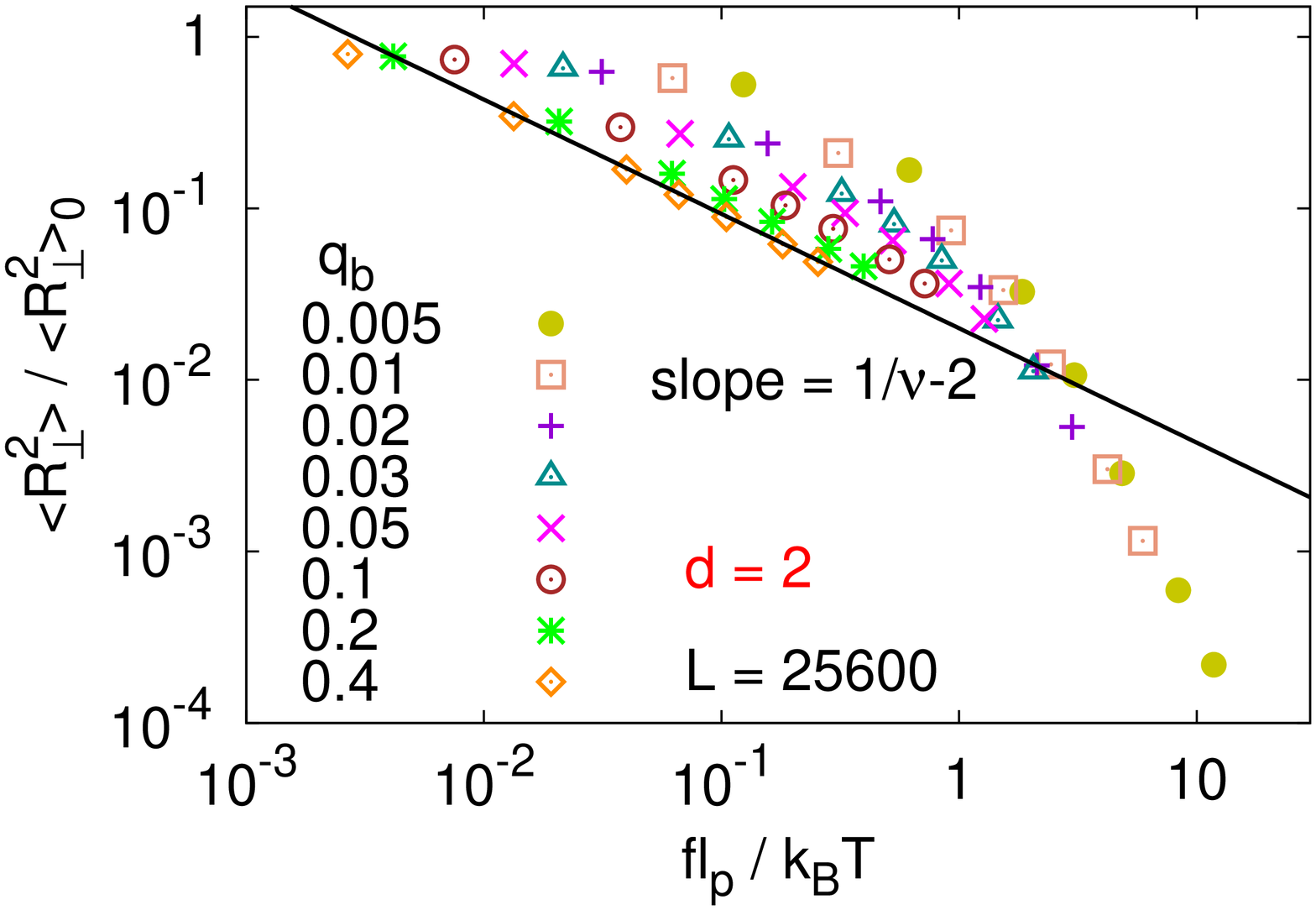}\\
\caption{Log-log plot of $[\langle X^2 \rangle -\langle X \rangle^2 ]/
\langle X^2 \rangle_0$ vs. $f\ell_p/k_BT$ for $L=100$ (a),
and $L=25600$ (b), including several choices for $q_b$ as indicated.
Log-log plot of $\langle R_\perp^2 \rangle /\langle R_{\perp}^2 \rangle_0$ vs.
$f\ell_p/k_BT$ for $L=100$ (c) and $L=25600$ (d), including several
choices for $q_b$ as indicated.
A straight line with slope $1/\nu-2$ ($\nu=3/4$) is shown for comparison.
Data are for semiflexible chains in $d=2$.}
\label{fig12}
\end{center}
\end{figure*}

\section{Stretching semiflexible polymers in $d=2$ dimensions}
Fig.~\ref{fig10} presents now a selection of our results for extension versus
force curves for the two-dimensional SAW's of variable stiffness on a square
lattice. As expected, neither the simple result for freely jointed
chains \{Eq.~(\ref{eq4})\} nor the Kratky-Porod result \{Eq.~(\ref{eq55})\}
are compatible with the data. For very short chains ($N_b=100$) and
intermediate values of the stiffness ($q_b=0.05$ which corresponds to
$\ell_p \approx 13 \ell_b$, cf. Table~\ref{table1}) we note that
$\langle X \rangle /L$ roughly agrees with Eq.~(\ref{eq55}), however:
this agreement probably is not accidental, since also in the absence of a force
for such short chains and this choice of $q_b$ the Kratky-Porod
prediction for the mean square end-to-end distance (Fig.~\ref{fig6}(b))
still is rather close to the actual result for
$\langle R_e^2 \rangle /2\ell_pL$.
Similar agreement was also noted in our earlier work~\cite{37} for somewhat
longer and stiffer chains ($N_b=200$ and $q_b=0.03$ and $0.02$, respectively)
for exactly the same reason: as long as $\langle R_e^2 \rangle$ in the
absence of forces is still more or less correctly predicted,
and this can be judged from the data presented in the previous section,
the general linear response relation, Eq.~(\ref{eq9}), which holds not only
for flexible SAW's but also for stiff chains, implies that the K-P model
still provides an accurate description of the initial linear part of the
extension versus force curve. Since in such a case where $L$ is larger then
$\ell_p$ by only a small factor, beyond the linear response regime there
is no regime of Pincus blobs possible, since $\sqrt{\langle R^2 \rangle}_0$
in Fig.~\ref{fig1}(a) and $\ell_p$ then are of the same order (each
Pincus blob needs to be formed from many subunits of size $\ell_p$, in order
that the power law regime $\langle X \rangle /L \propto (f\ell_p/k_BT)^{1/3}$
can develop!) Thus, we arrive at the general conclusion that in $d=2$
the K-P result Eq.~(\ref{eq55}), is applicable only for such short chains that
$L$ is larger than $\ell_p$ by only a small factor ($L \le 10 \ell_p$, say),
so that in Fig.~\ref{fig1}(a) the Pincus blob regime is essentially absent,
and the K-P model also achieves an approximate description of the linear
response regime.

Of course, it is of great interest to clarify what happens when $L\gg \ell_p$.
Fig.~\ref{fig11} hence presents a log-log plot of the data for the
extension versus force curves including long chains and rescaling the data
such that a scaling description for the crossover from the linear response
to the regime of Pincus blobs is obtained. One sees that both for flexible
chains (Fig.~\ref{fig11}(a)) and for rather stiff chains (Fig.~\ref{fig11}(b))
a reasonable data collapse on a master curve is obtained, consistent with
the predicted exponents. As expected, the crossover between both power laws
is gradual and not sharp. If one includes data for too large forces, one can
see that the data fall systematically below the Pincus power law. Similarly,
when one puts the focus on the crossover from the Pincus blob regime to
the saturation behavior, Fig.~\ref{fig11}(c), one finds that the data
fall systematically below the Pincus power law for small forces (due to
the crossover towards the linear response regime). As expected from the
theoretical considerations of Sec.~II, there cannot exist a scaling
representation which brings both crossovers of Fig.~\ref{fig1}(a)
to a data collapse on a master curve together. Note also that for the
long chains the K-P model does not fit our data at large relative
extensions $\langle X \rangle /L$ either, since our simulations are
based on a discrete chain model. Since our choices of $q_b$ do not yield
extremely large persistence lengths, the crossover from the saturation
behavior predicted by the K-P model \{Eq.~(\ref{eq59})\}
to that of the FJC model \{Eq.~(\ref{eq7})\} is not clearly
resolved either.
Actually, for very large forces one must consider that our model is
a lattice model, not a model of rigid bonds in the continuum
where arbitrary bond angles occurs such as the FJC model: hence we expect
that for $f\rightarrow \infty$ the saturation behavior is
$1-\langle X \rangle/L \propto \exp(-f\ell_b/k_BT)$ rather than
$k_BT/f\ell_b$.

\begin{figure*}
\begin{center}
(a)\includegraphics[scale=0.29,angle=270]{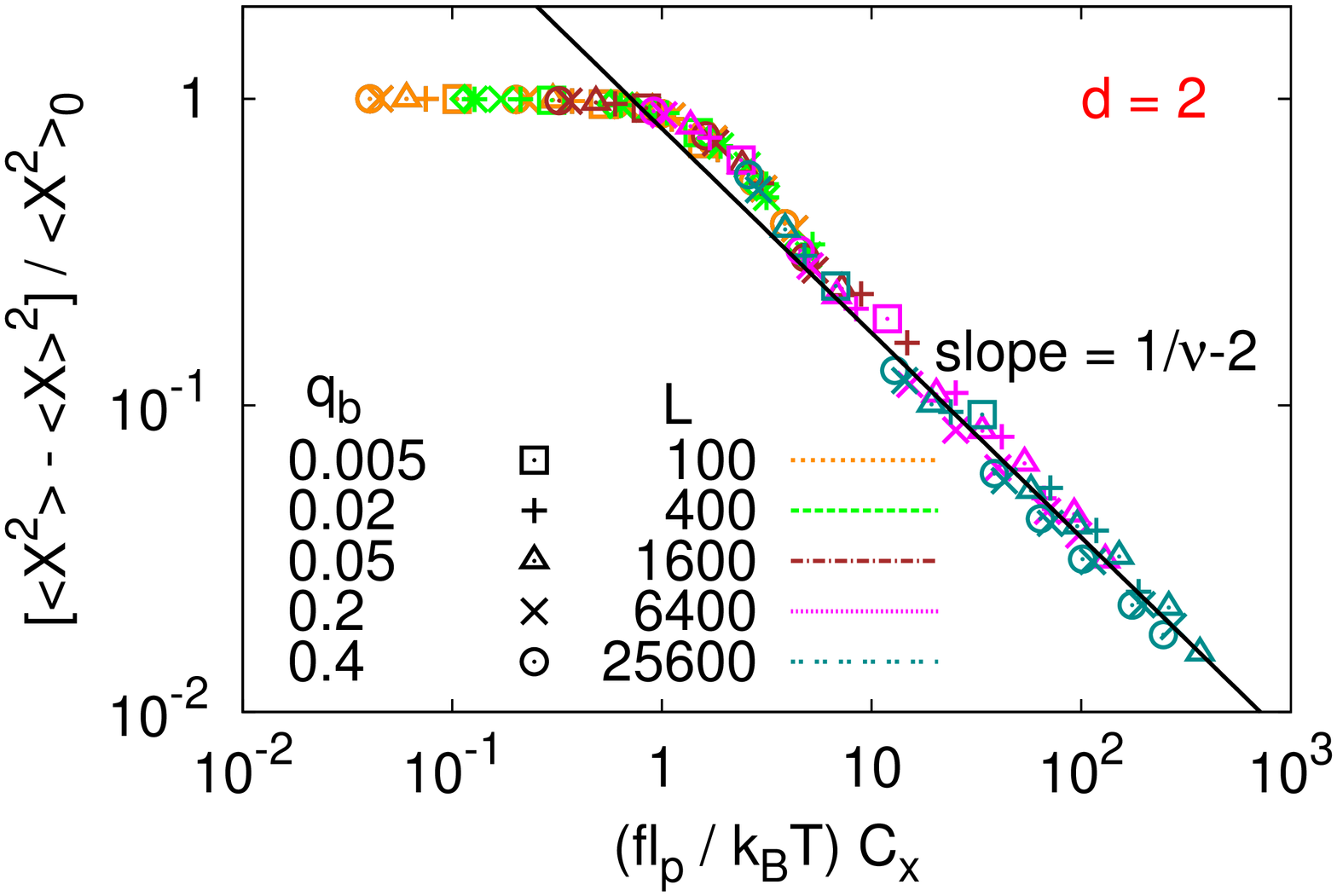}\hspace{0.4cm}
(b)\includegraphics[scale=0.29,angle=270]{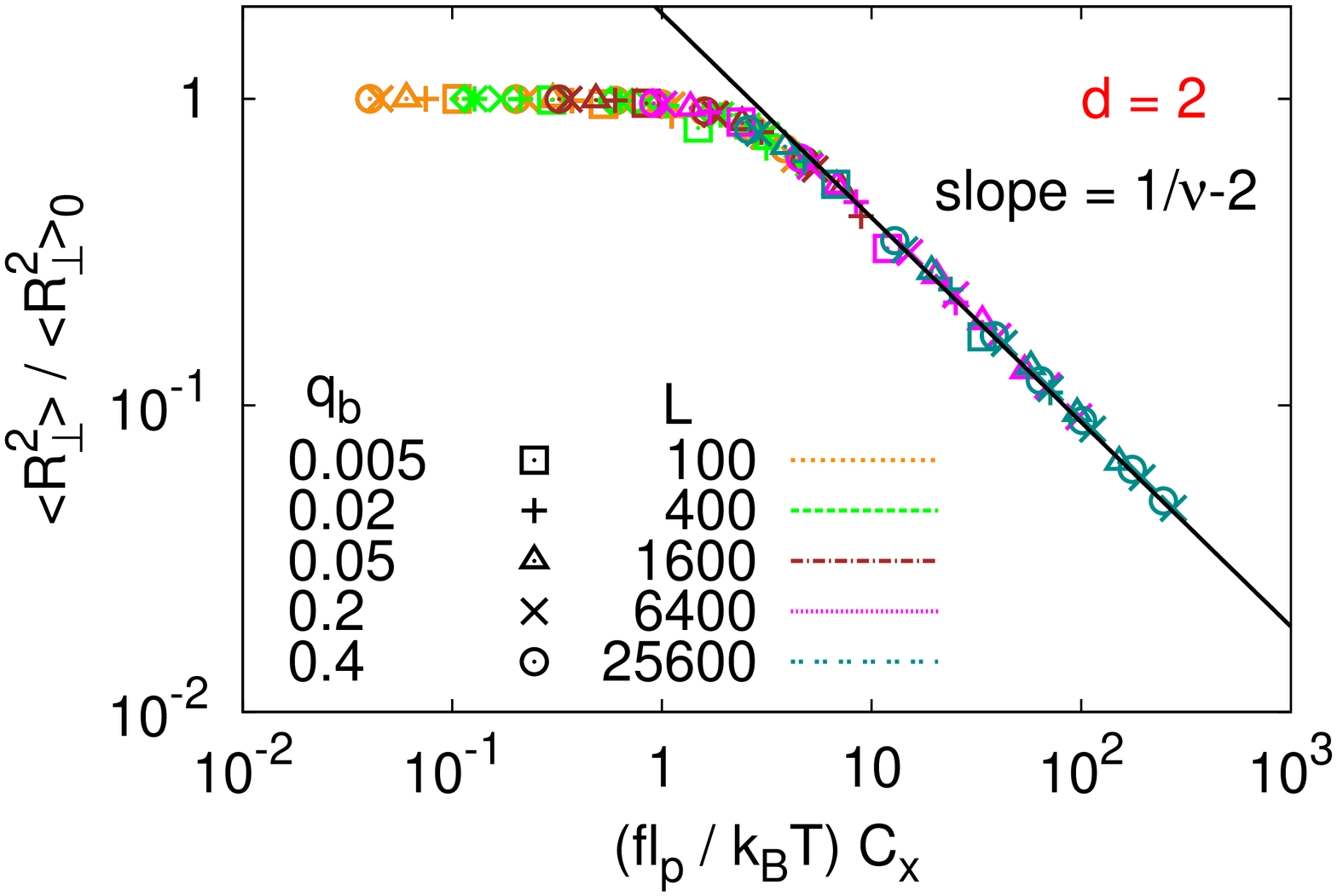}\\
\caption{Log-log plot of $[\langle X^2 \rangle -\langle X \rangle^2 ]/
\langle X^2 \rangle_0$ (a) and
$\langle R_\perp^2 \rangle /\langle R_{\perp}^2 \rangle_0$ (b) vs.
$(f\ell_p/k_BT)C_x$ with $C_x=(L/\ell_p)^{3/4}$ (see Fig.~\ref{fig11}(b))
for semiflexible chains in $d=2$ including several choices of $L$ and $q_b$,
as indicated.}
\label{fig13}
\end{center}
\end{figure*}

\begin{figure*}
\begin{center}
(a)\includegraphics[scale=0.29,angle=270]{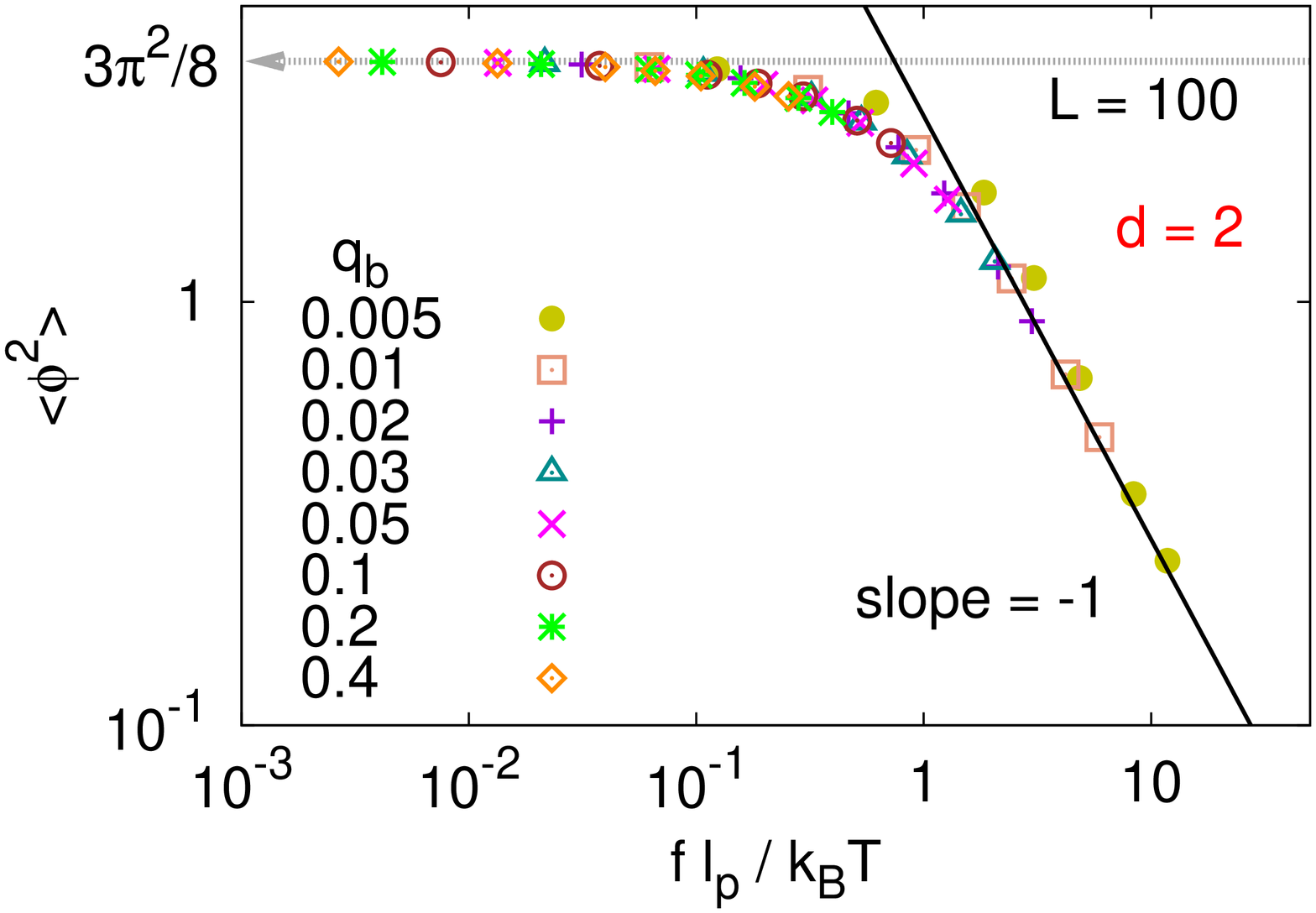}\hspace{0.4cm}
(b)\includegraphics[scale=0.29,angle=270]{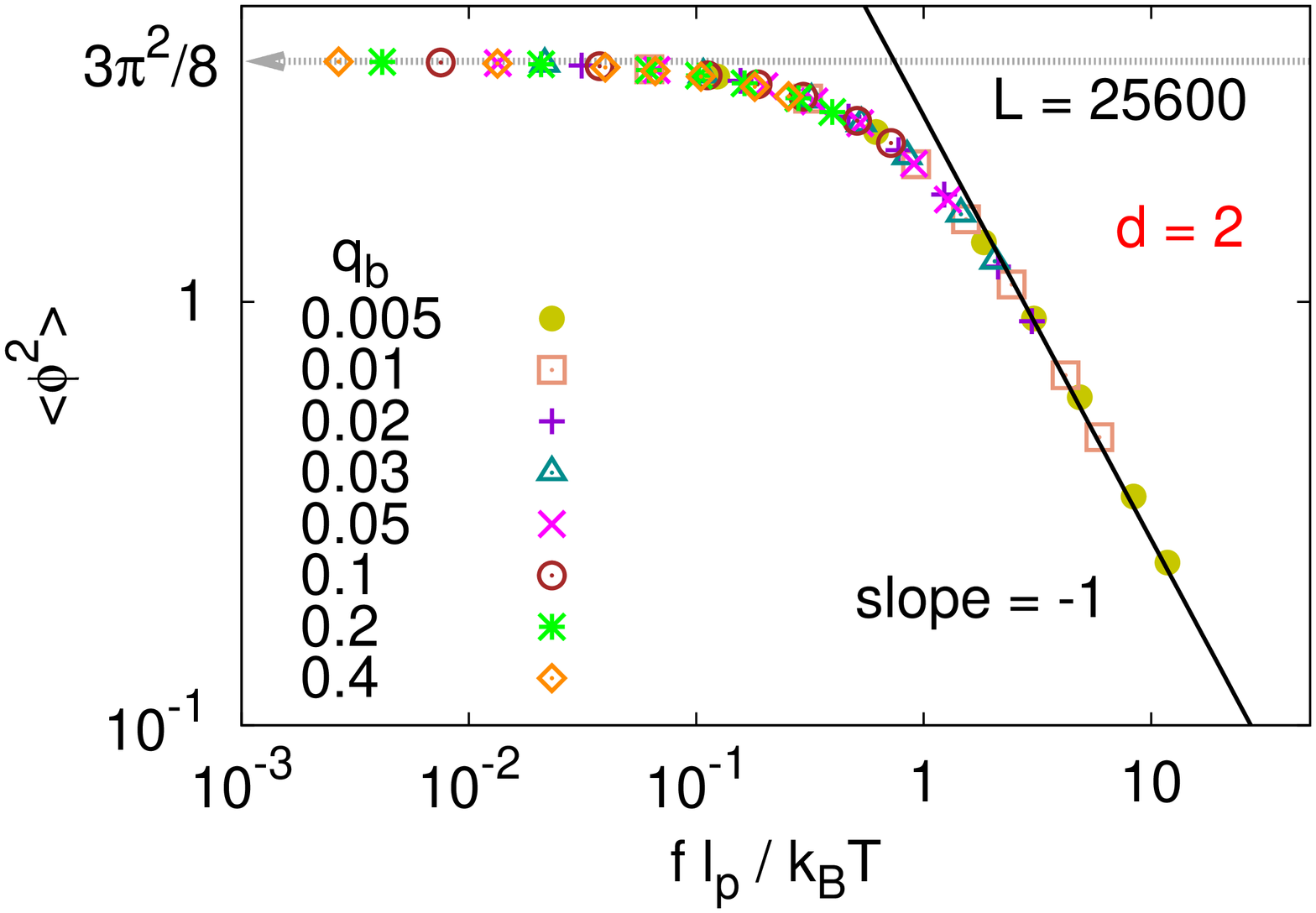}\\
\caption{Log-log plot of $\langle \phi^2 \rangle$ vs. $f\ell_p/k_BT$
for $L=100$ (a),
and $L=25600$ (b), including several choices for $q_b$ as indicated.
$\langle \phi^2 \rangle =3\pi^2/8$ as $f\rightarrow 0$. Data are
for semiflexible chains in $d=2$.}
\label{fig14}
\end{center}
\end{figure*}

  As a last point of this section, we consider both longitudinal 
(Fig.~\ref{fig12}(a)(b)) and transverse (Fig.~\ref{fig12}(c)(d)) 
fluctuations of the chain dimensions. These fluctuations have been 
normalized such that they are of order unity (and independent both
$L$ and $\ell_p$) in the linear response regime, while in the Pincus
blob regime ($0 \ll f\ell_p/k_BT<1$) a crossover to a simple 
power law proportional to $(f\ell_p/k_BT)^{1/\nu-2}$ occurs. 
Using the same scaling factors $C_x=(L/\ell_p)^{3/4}$ (Fig.~\ref{fig11}(b))
for abscissa, a nice data collapse on the master curve is seen
in Fig.~\ref{fig13} for both longitudinal and transverse fluctuations.
Note that our scaling description for the crossover from the linear
response regime to the Pincus blob regime, exemplified in Fig.~\ref{fig11}
and \ref{fig13}, does not invoke any adjustable parameters whatsoever
(unlike the case of experiments, where often both $L$ and $\ell_p$ are fit
parameters). However, the behavior at larger forces (beyond Pincus blob 
regime) is more subtle.
While in the Kratky-Porod regime a power law proportional to 
$(f\ell_p/k_BT)^{-2}$ is expected for
large enough $f$, where a behavior similar as that has been found for freely
jointed chains \{Eq.~(\ref{eq21})\} can be expected, when
one considers the longitudinal fluctuation 
$\left\{(\langle X^2 \rangle-\langle X \rangle^2)/
\langle X^2\rangle_0\right\}$, for the transverse fluctuation
$\langle R_\perp^2 \rangle /\langle R_\perp^2 \rangle_0$ all theories predict a 
slower decay (proportional to $[(f\ell_p/k_BT)^{-1}]$ for large $f$
\{Eqs.~(\ref{eq22}), (\ref{eq24}), and (\ref{eq68}), respectively\},
and this slower decay in fact is not seen. The reason for this discrepancy,
however, probably is the fact that in our model only kinks by
$\pm 90^o$ are possible, and no small deflections are possible as 
in the K-P and FJC models. 
We defer a more detailed analysis of these fluctuations
to a forthcoming publication. Here we rather focus on the behavior of the 
local angular fluctuation $\langle \phi^2 \rangle$ (Fig.~\ref{fig14}).
For small forces all angles $\phi$ between a bond and the $+x$-direction
are equally probable (at the lattice we have in $d=2$ two possibilities
for $\phi=\pi/2$ or $\phi=-\pi/2$, and two possibilities for $\phi=0$
or $\pi$, respectively). Hence for $f \rightarrow 0$ we must find 
$\langle \phi^2 \rangle = (\pi^2 +\pi^2/2)/4=3\pi^2/8\approx 3.7$, and
this is compatible with the observation. For $f\ell_p/k_BT >1$ we 
observe a smooth crossover towards
\begin{equation}
  \langle \phi^2 \rangle \propto k_BT/(f\ell_p) \, , \enspace 
f\ell_p\gg k_BT
\label{eq73}
\end{equation}
and the crossover from $\langle \phi^2 \rangle \approx 3.7$ to this
decay seems to be practically independent of $\ell_p$ (as shown by the 
superposition of data for different choices of $q_b$ and hence $\ell_p$ in
Fig.~\ref{fig14}) and $L$ (compare Fig.~\ref{fig14}(a) for $L=100$ with
Fig.~\ref{fig14}(b) for $L=25600$). Although the local quantity 
$\langle \phi^2 \rangle$ thus has a remarkably simple behavior, unlike
the global quantities $\langle X \rangle /L$,
$(\langle X^2\rangle-\langle X \rangle^2)/\langle X^2\rangle_0$ and
$\langle R_\perp ^2\rangle/\langle R_\perp^2\rangle_0$, we are not aware of any
theoretical prediction relating to it.

\begin{figure*}
\begin{center}
(a)\includegraphics[scale=0.29,angle=270]{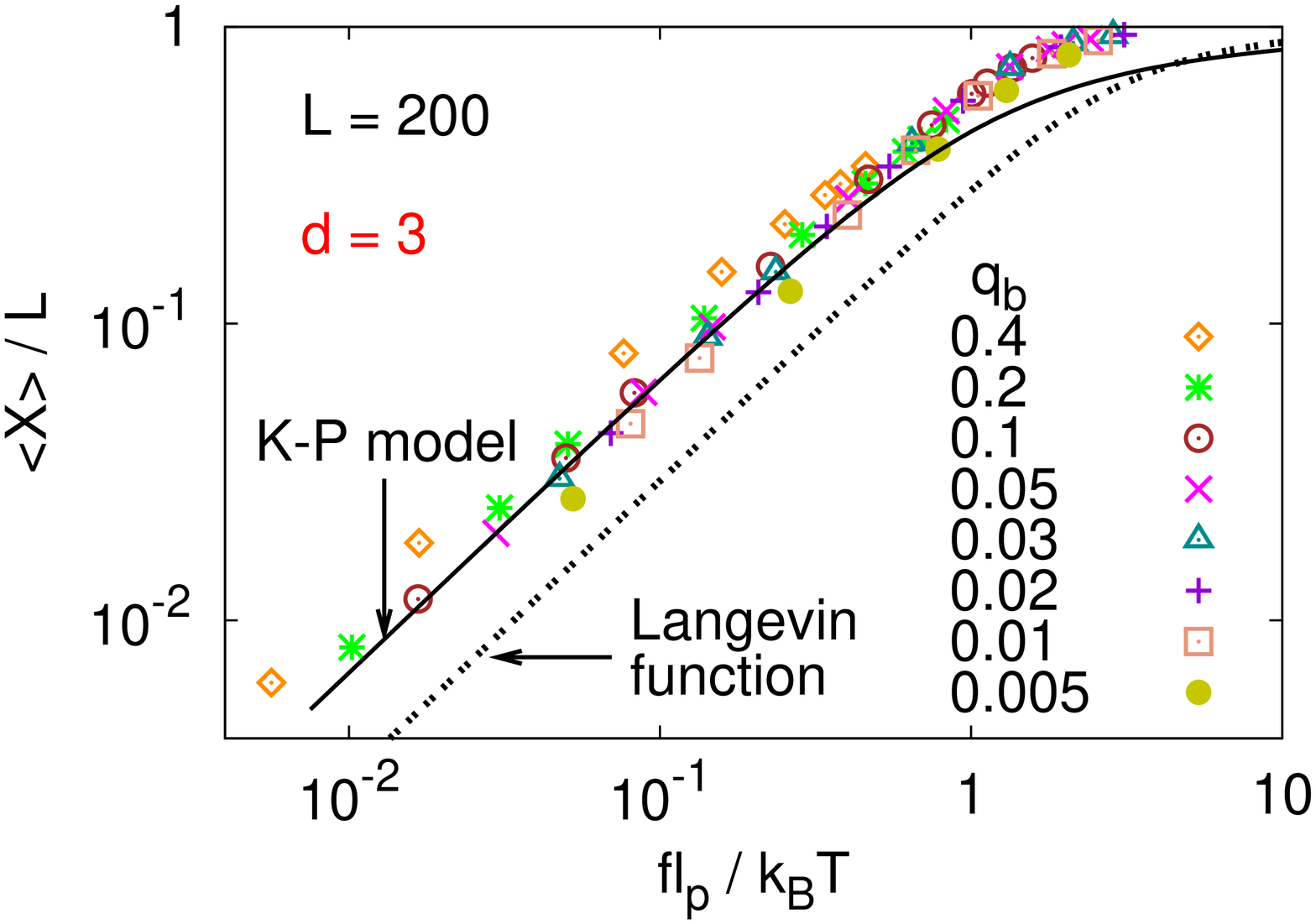}\hspace{0.4cm}
(b)\includegraphics[scale=0.29,angle=270]{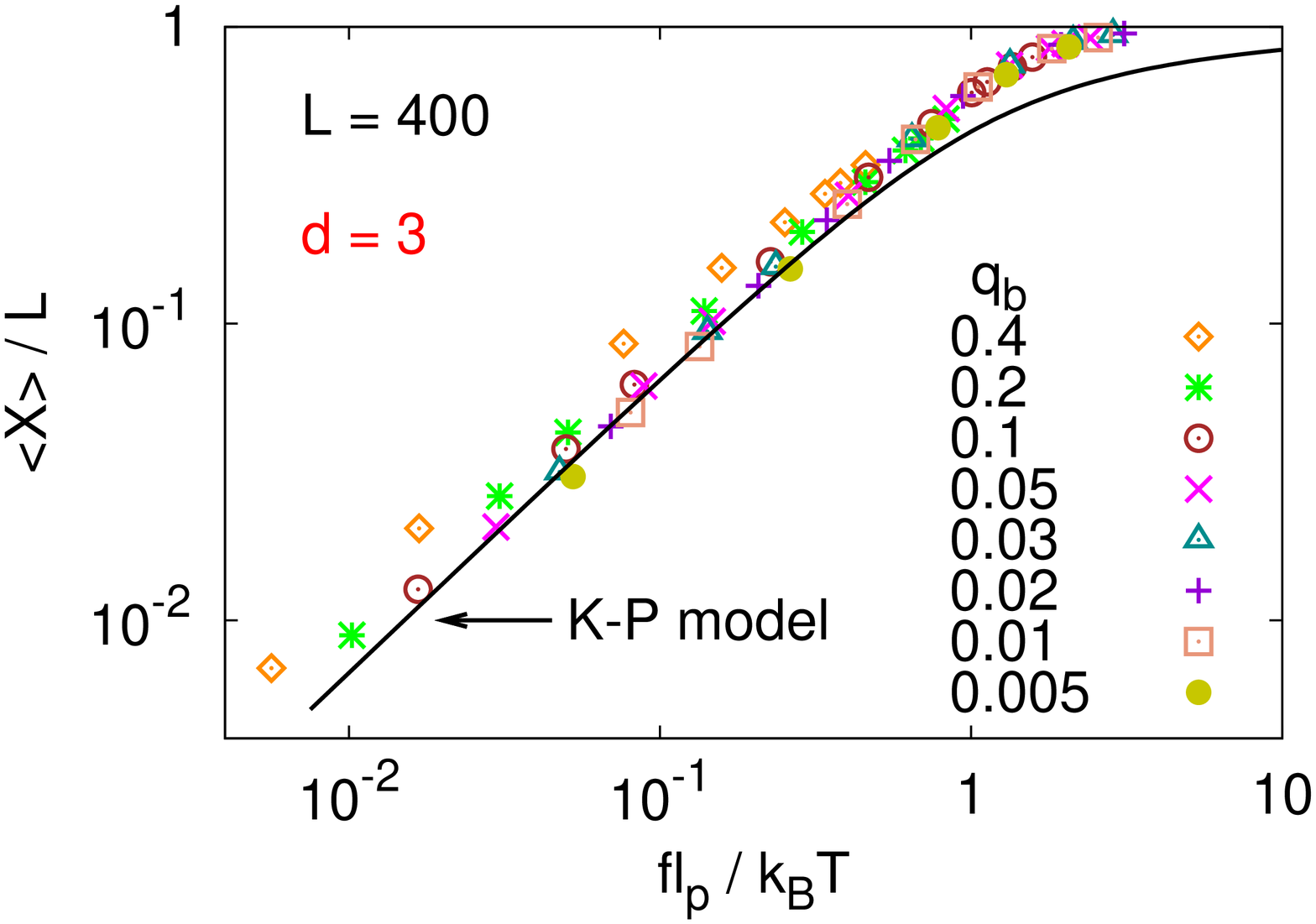}\\
(c)\includegraphics[scale=0.29,angle=270]{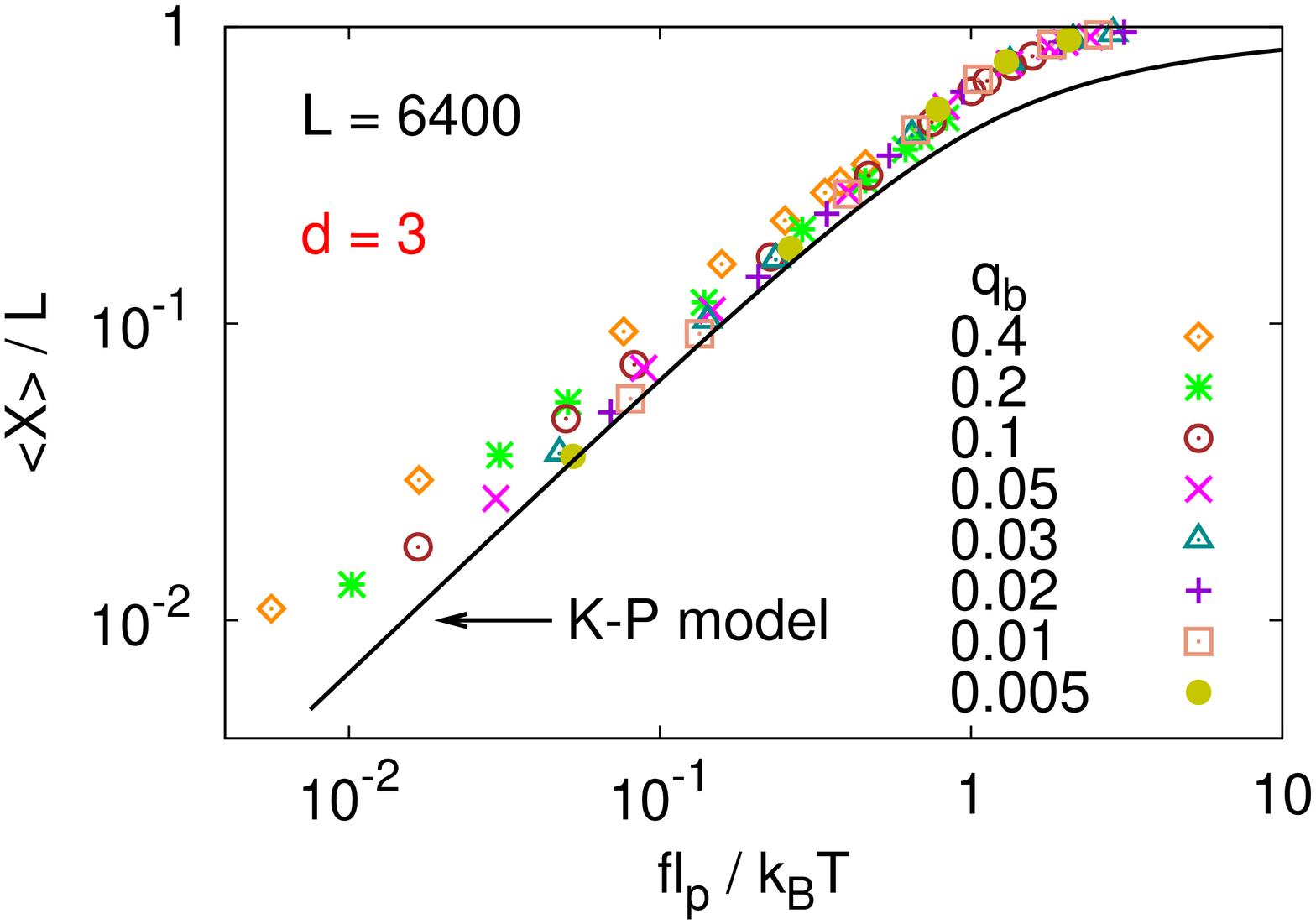}\hspace{0.4cm}
(d)\includegraphics[scale=0.29,angle=270]{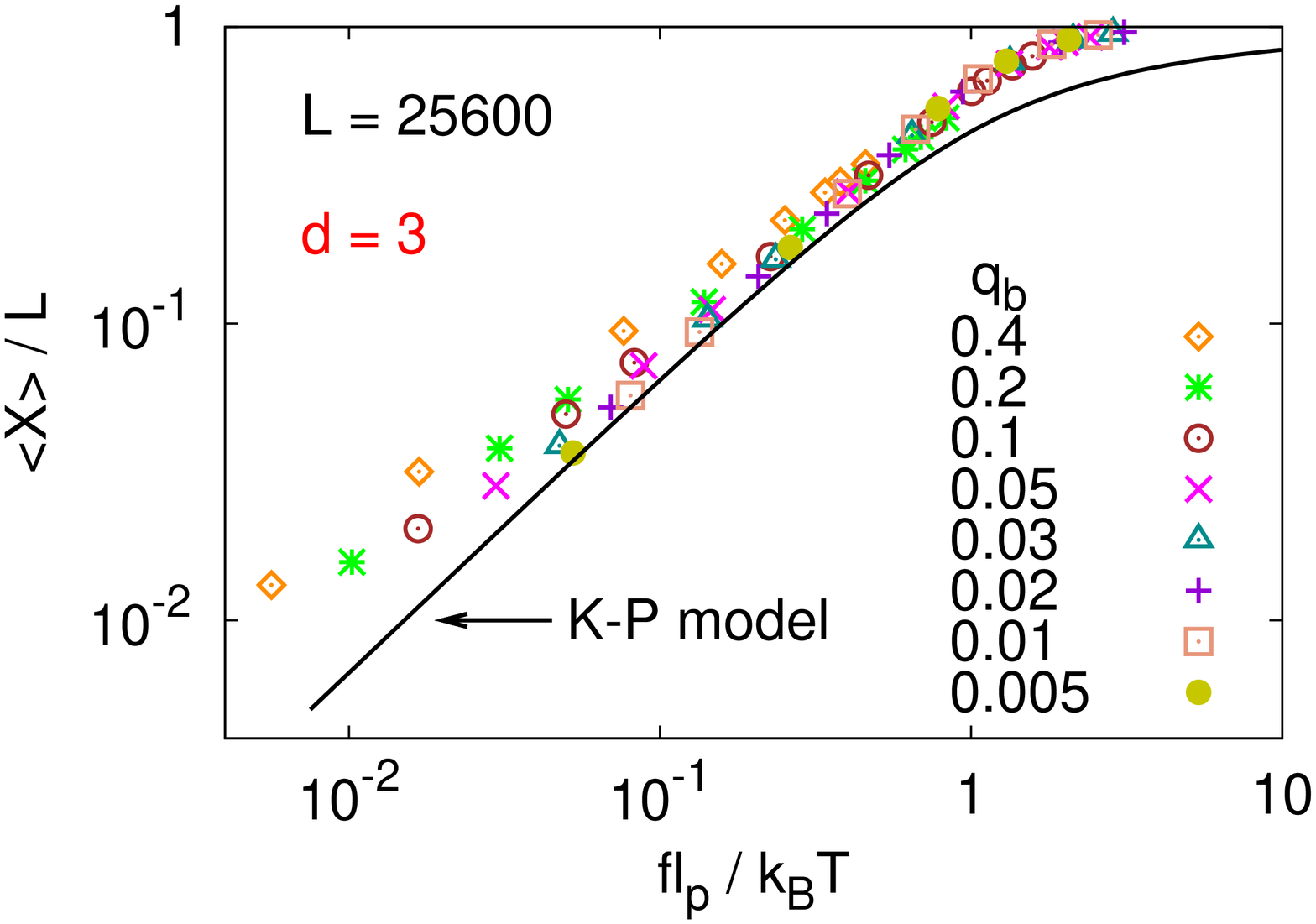}\\
\caption{Log-log plot of $\langle X \rangle /L$ versus $f\ell_p/k_BT$ for
several choices of $q_b$ as indicated, for contour length $L=200$ (a),
$L=400$ (b), $L=6400$ (c), and $L=25600$ (d). Full curve always refers to
the K-P model prediction, Eq.~(\ref{eq54}). Broken curve in (a) is the
Langevin function, Eq.~(\ref{eq4}), using $\ell_p/\ell_b = 0.71$ for
$q_b=0.4$ (Table~\ref{table2}). Data are for semiflexible chains in $d=3$.}
\label{fig15}
\end{center}
\end{figure*}

\begin{figure*}
\begin{center}
(a)\includegraphics[scale=0.29,angle=270]{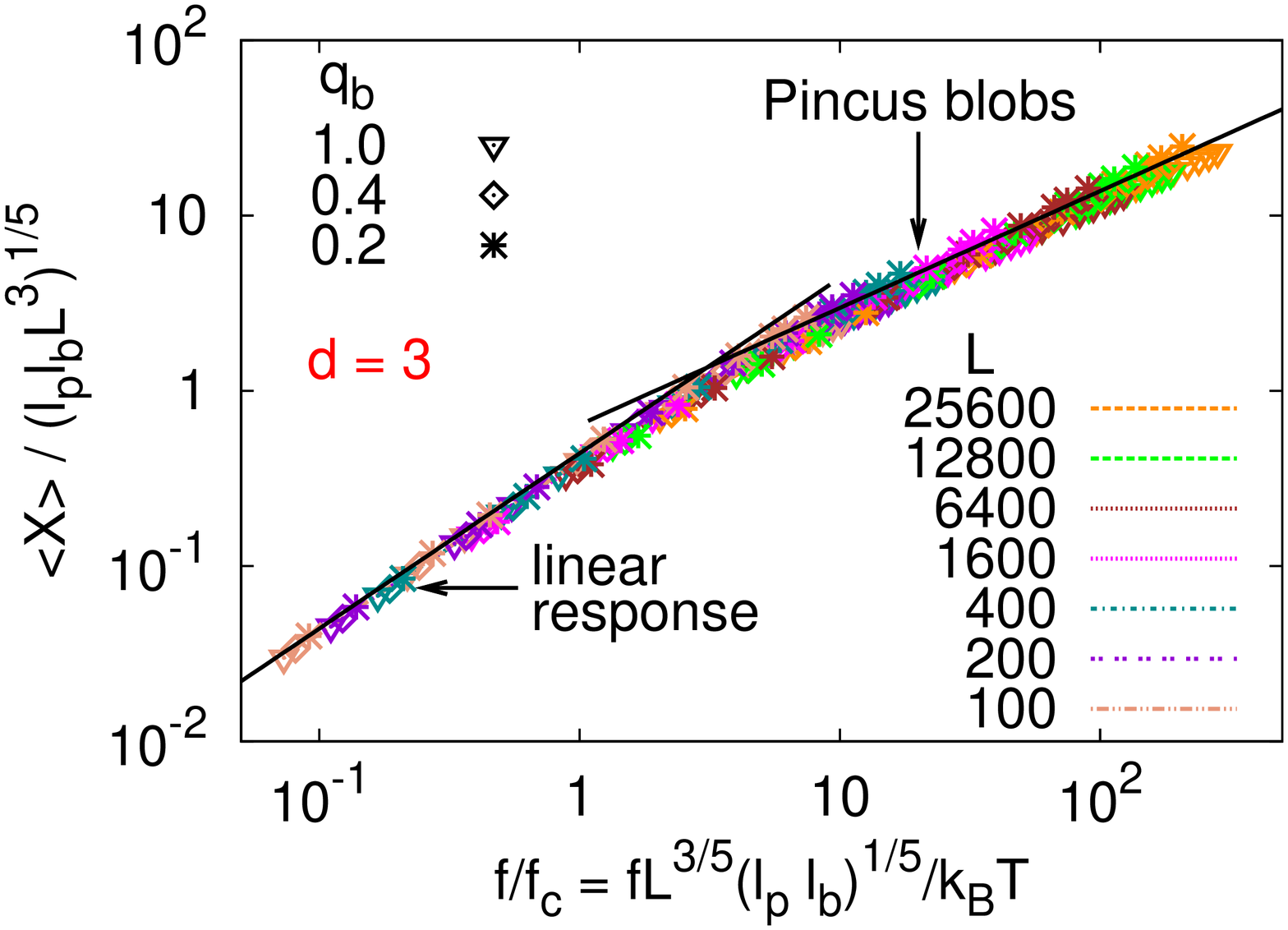}\hspace{0.4cm}
(b)\includegraphics[scale=0.29,angle=270]{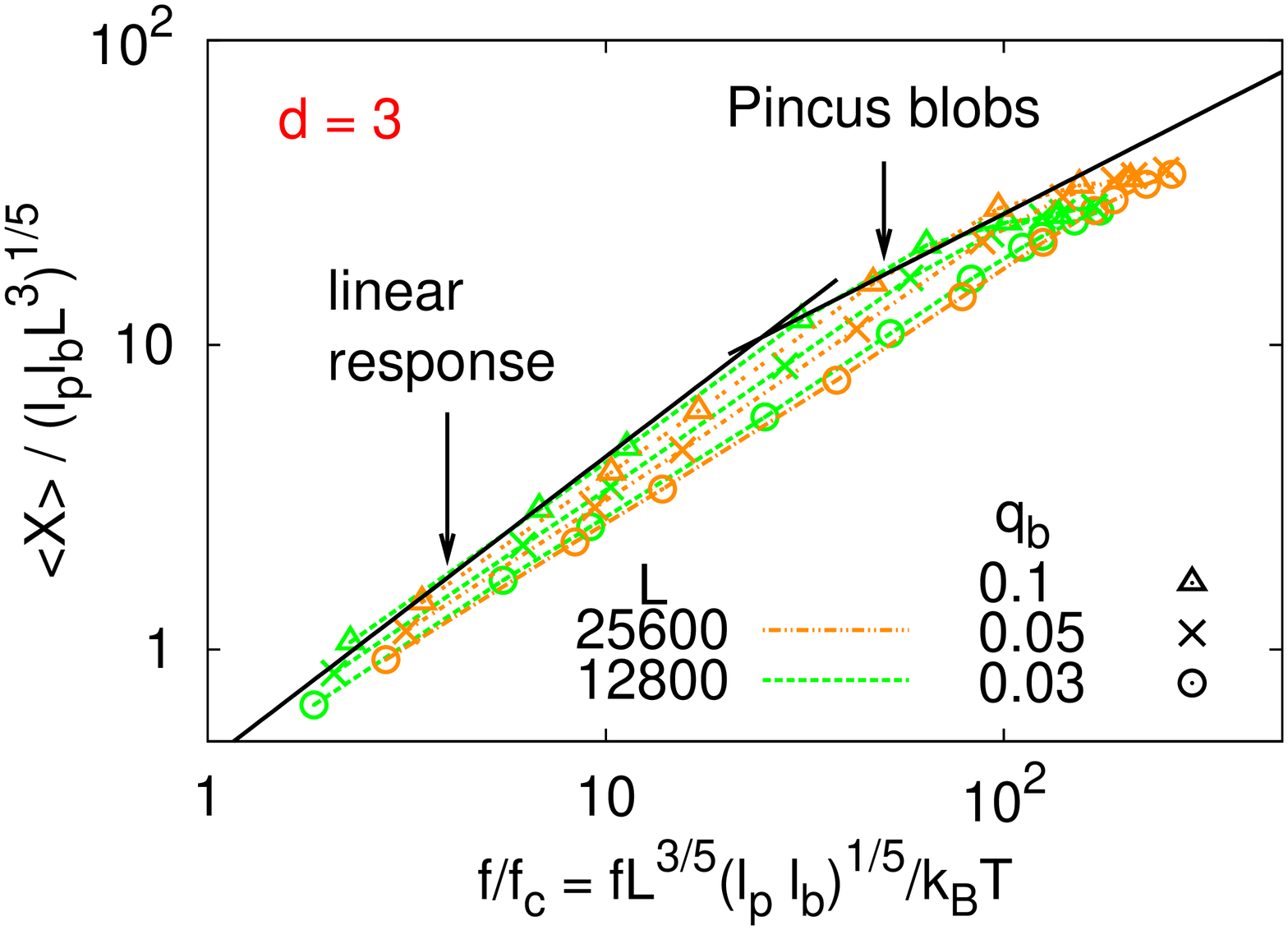}\\
(c)\includegraphics[scale=0.29,angle=270]{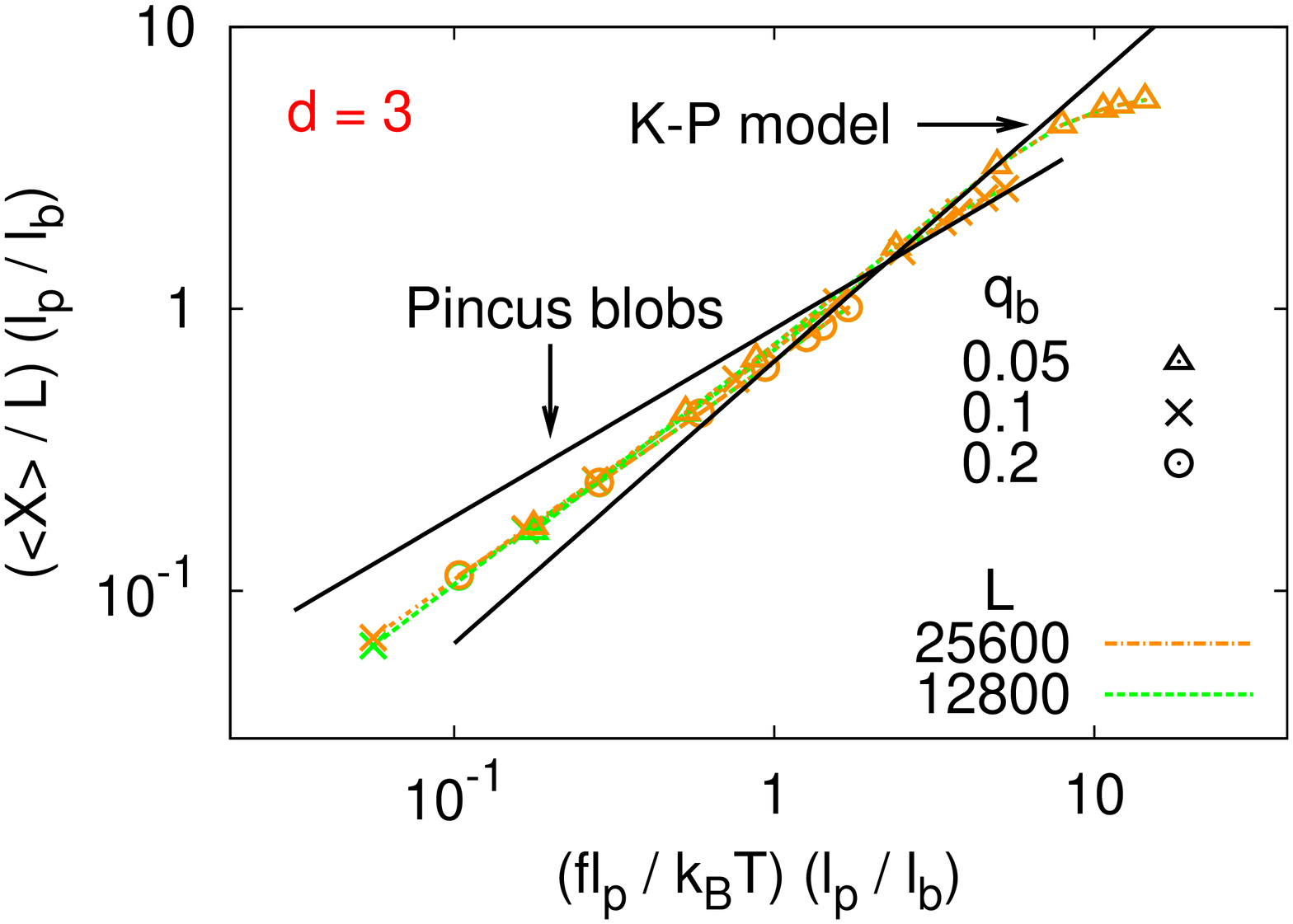}\hspace{0.4cm}
(d)\includegraphics[scale=0.29,angle=270]{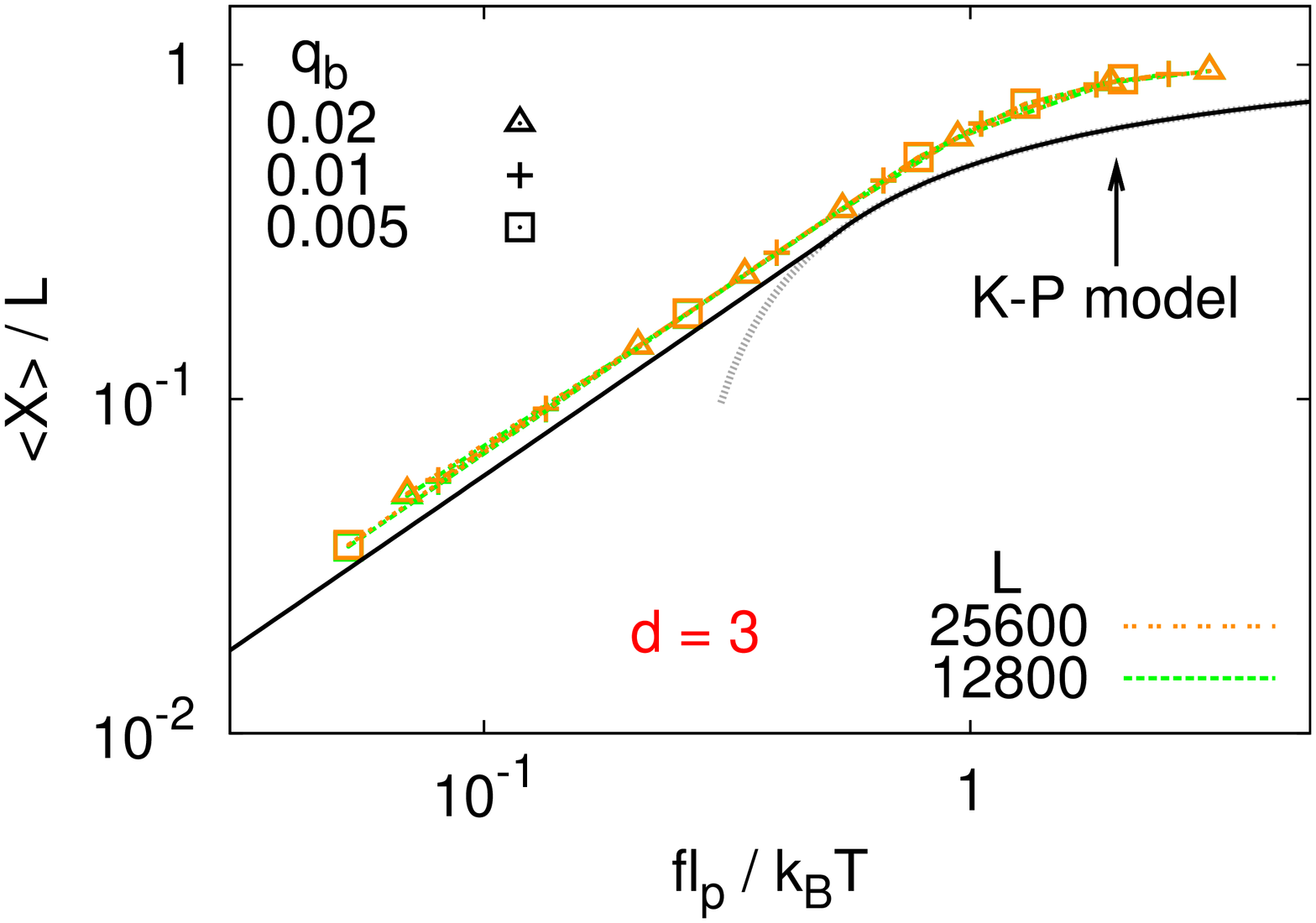}\\
\caption{(a) Log-log plot of $\langle X \rangle/(\ell_p \ell_b L^3)^{1/5}$
vs $f/f_c=fL^{3/5}(\ell_p \ell_b)^{1/5}/k_BT$, where $f_c$ is the crossover
force defined in Eq.~(\ref{eq61}), remembering $D=\ell_b$ in our model, for
the choices $q_b=1.0$, $0.4$, and $0.2$ and many choices of $L$ (a) and
for $q_b=0.1$, $0.05$, $0.03$ but only $L=25600$ and $L=12800$ (b).
Case (c) shows a plot of $(\langle X \rangle/L)(\ell_p/\ell_b)$
versus $(f\ell_p/k_BT)(\ell_p/\ell_b)$
for the choices $q_b=0.05$, $0.1$, and $0.2$, again for $L=25600$
and $L=12800$ only, to test for indications of a crossover from Pincus
blobs to the Kratky-Porod model, showing only the vicinity of the region
when this crossover should occur. Case (d) is a blow up the region
$0.05 \leq \langle X \rangle /L<0.5$, for $q_b=0.02$, $0.01$, $0.05$,
$L=25600$ and $L=12800$, to show the full K-P region for rather long
and rather stiff chains. Data are for semiflexible chains in $d=3$.}
\label{fig16}
\end{center}
\end{figure*}

\begin{figure*}
\begin{center}
(a)\includegraphics[scale=0.29,angle=270]{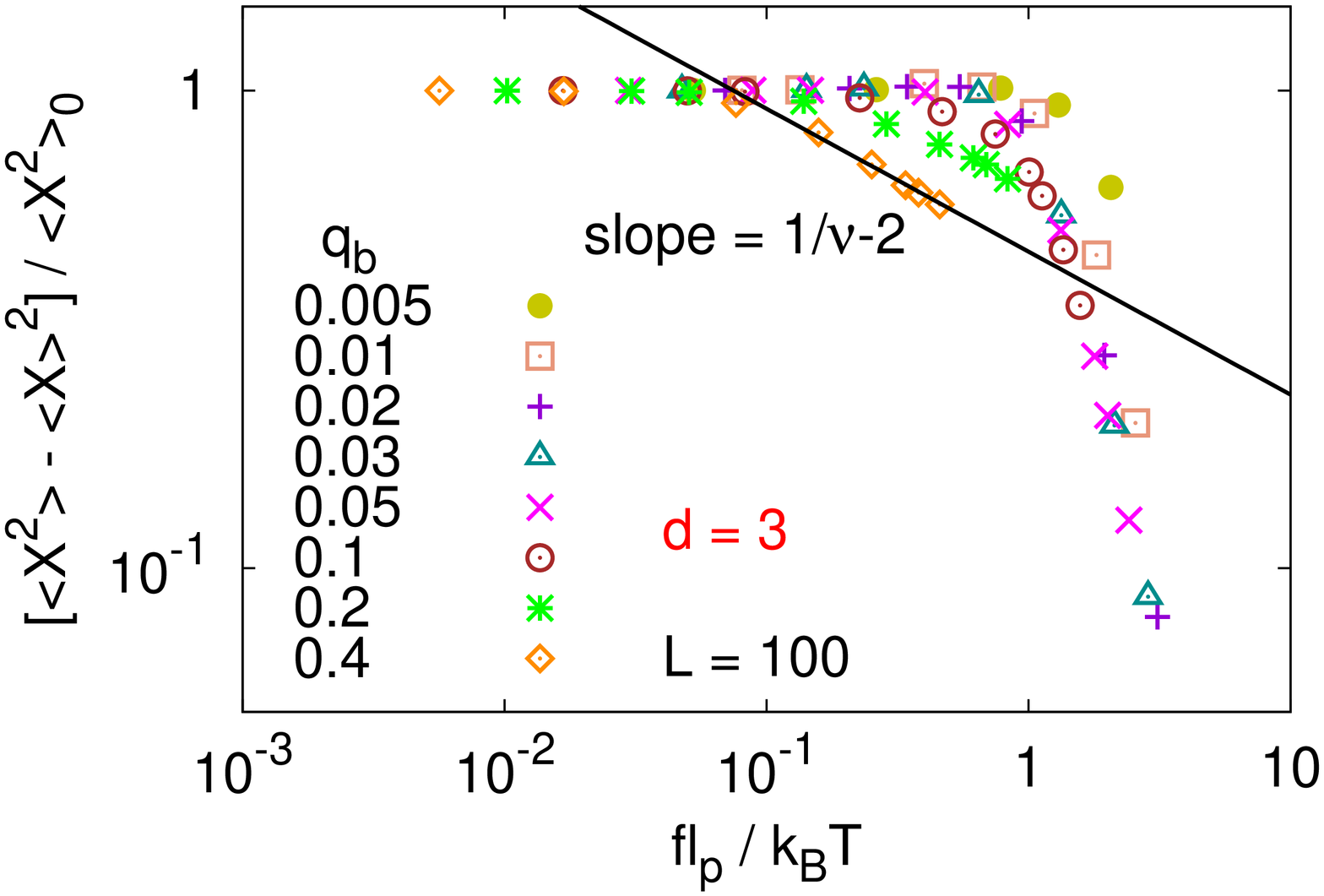}\hspace{0.4cm}
(b)\includegraphics[scale=0.29,angle=270]{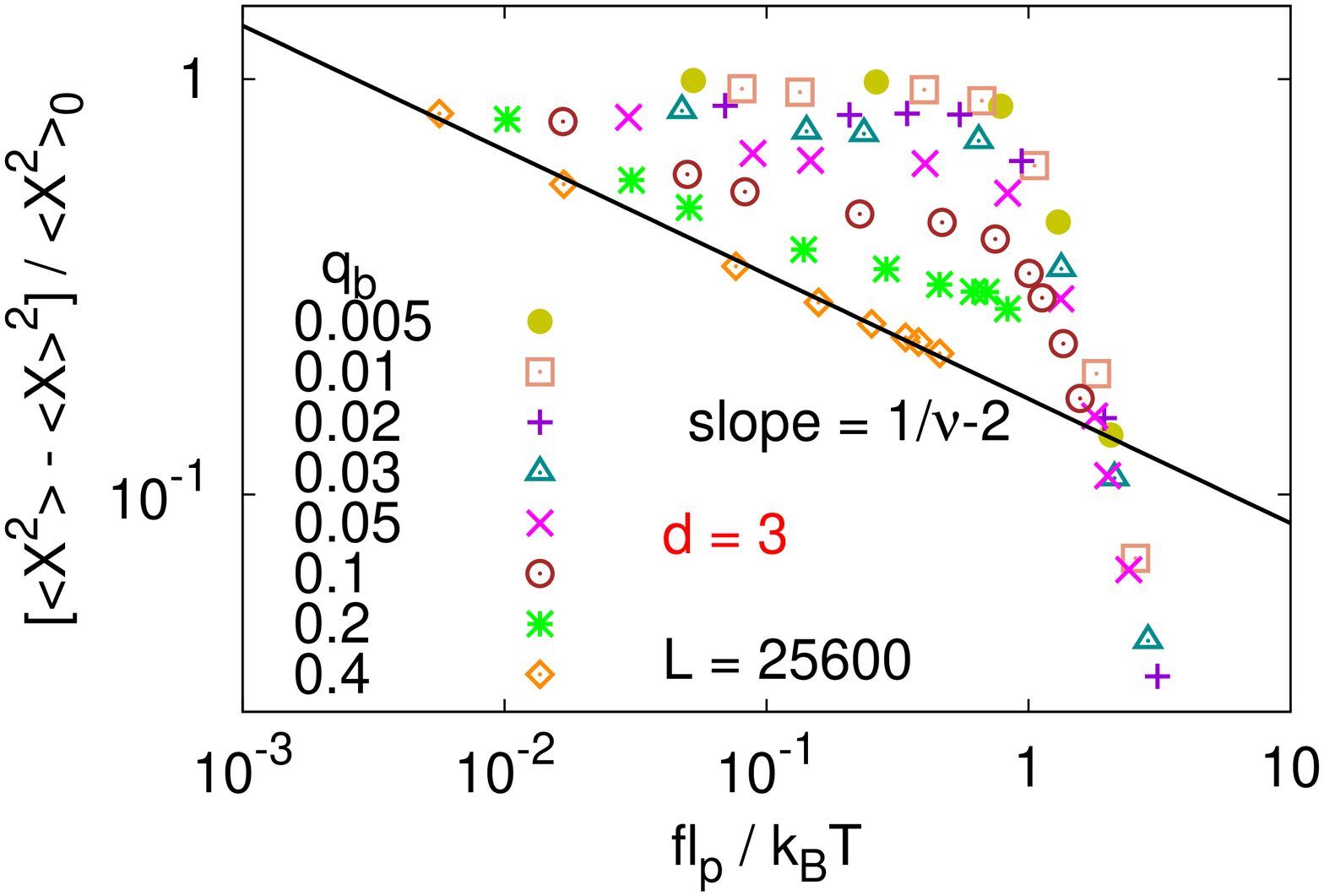}\\
(c)\includegraphics[scale=0.29,angle=270]{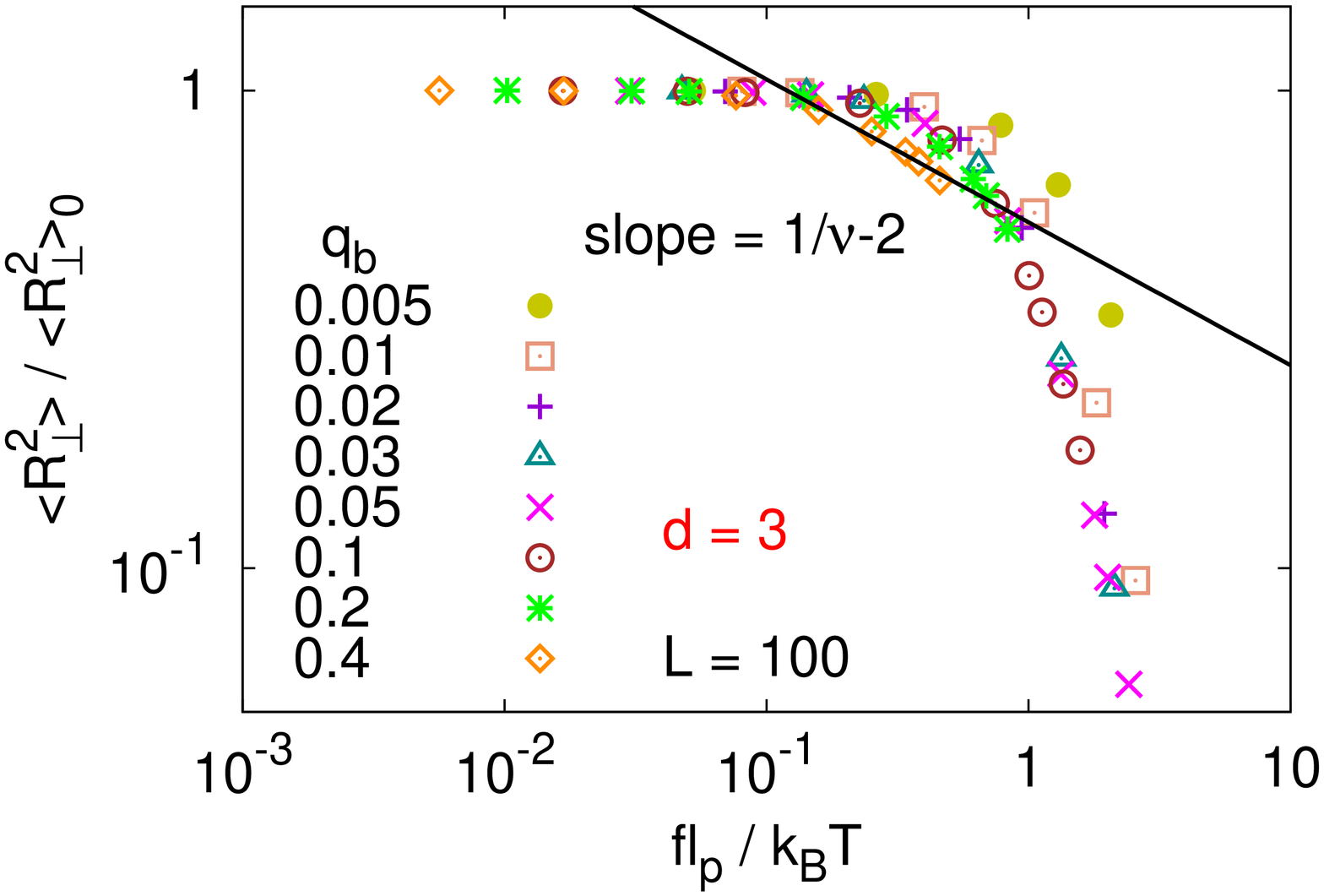}\hspace{0.4cm}
(d)\includegraphics[scale=0.29,angle=270]{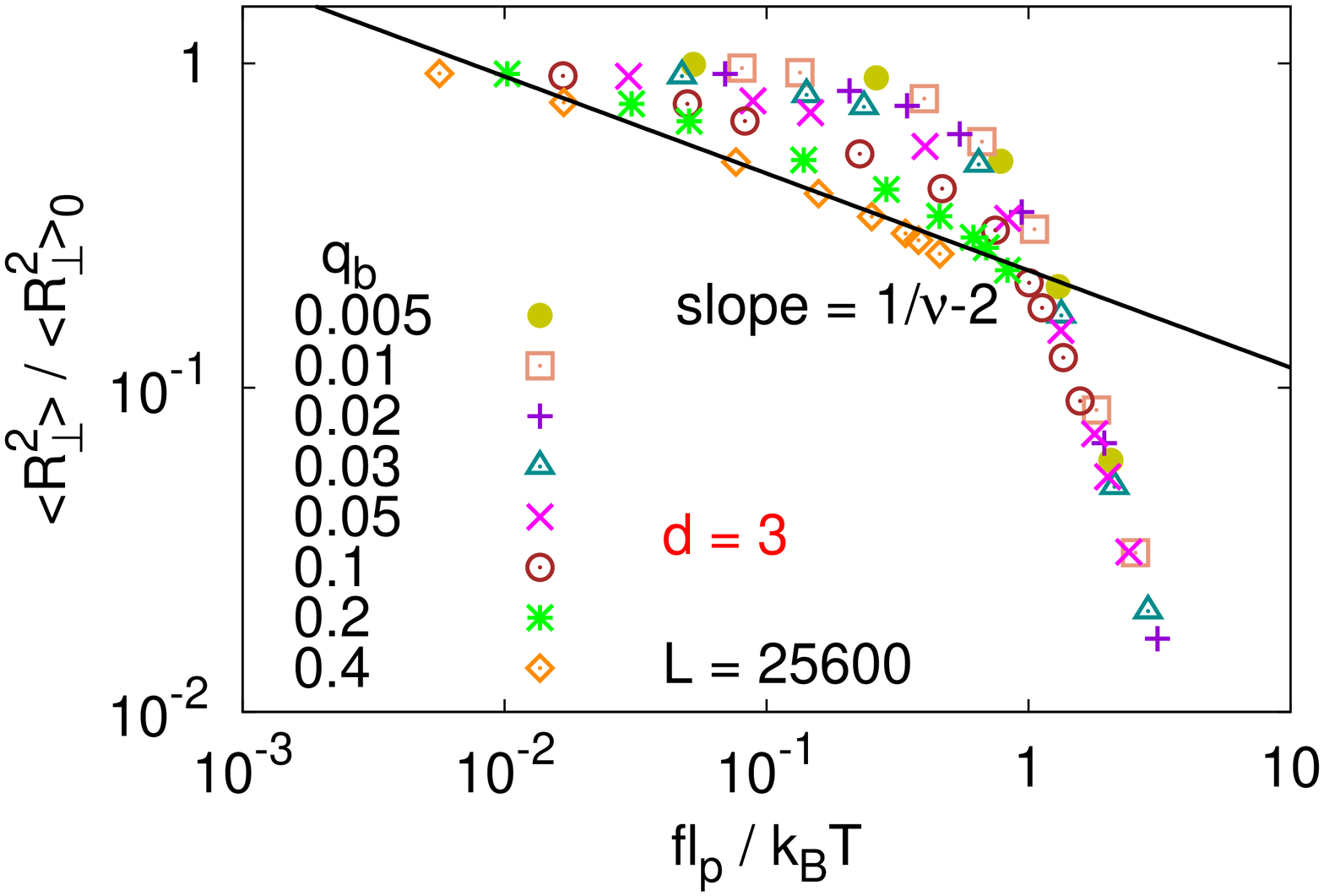}\\
\caption{Log-log plot of $[\langle X^2 \rangle - \langle X \rangle^2]
/\langle X^2 \rangle_0$ vs. $f\ell_p/k_BT$ for $L=100$ (a) and $L=25600$ (b),
and for various choices of $q_b$, as indicated. Part (c)
shows analogous data for $\langle R_\perp^2 \rangle /\langle R^2\rangle_0$
vs. $f\ell_p/k_BT$ for $L=100$ and part (d) for $L=25600$.
A straight line with slope $1/\nu-2$ ($\nu=0.588$) is shown for comparison.
Data are for semiflexible chains in $d=3$.}
\label{fig17}
\end{center}
\end{figure*}

\begin{figure*}
\begin{center}
(a)\includegraphics[scale=0.29,angle=270]{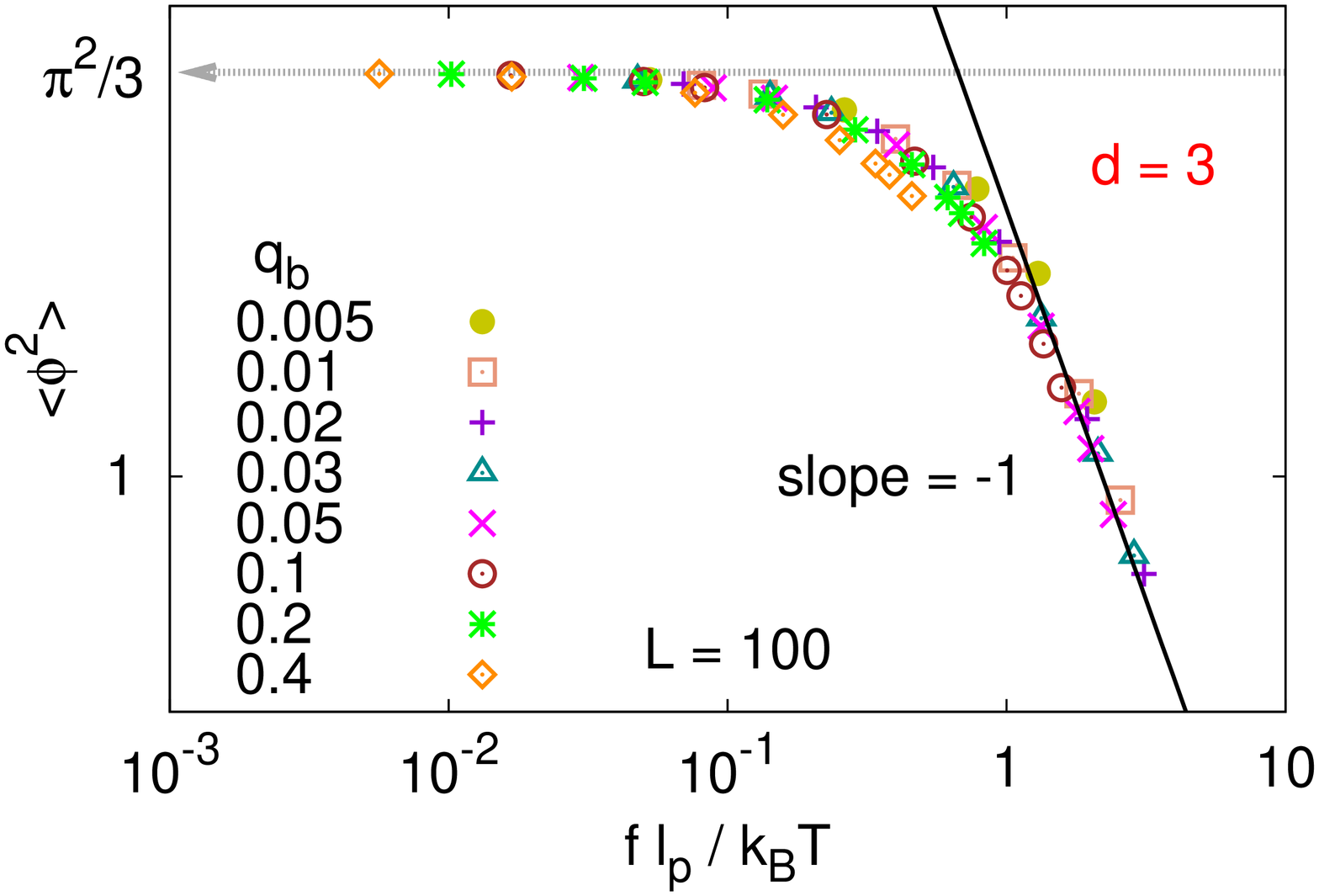}\hspace{0.4cm}
(b)\includegraphics[scale=0.29,angle=270]{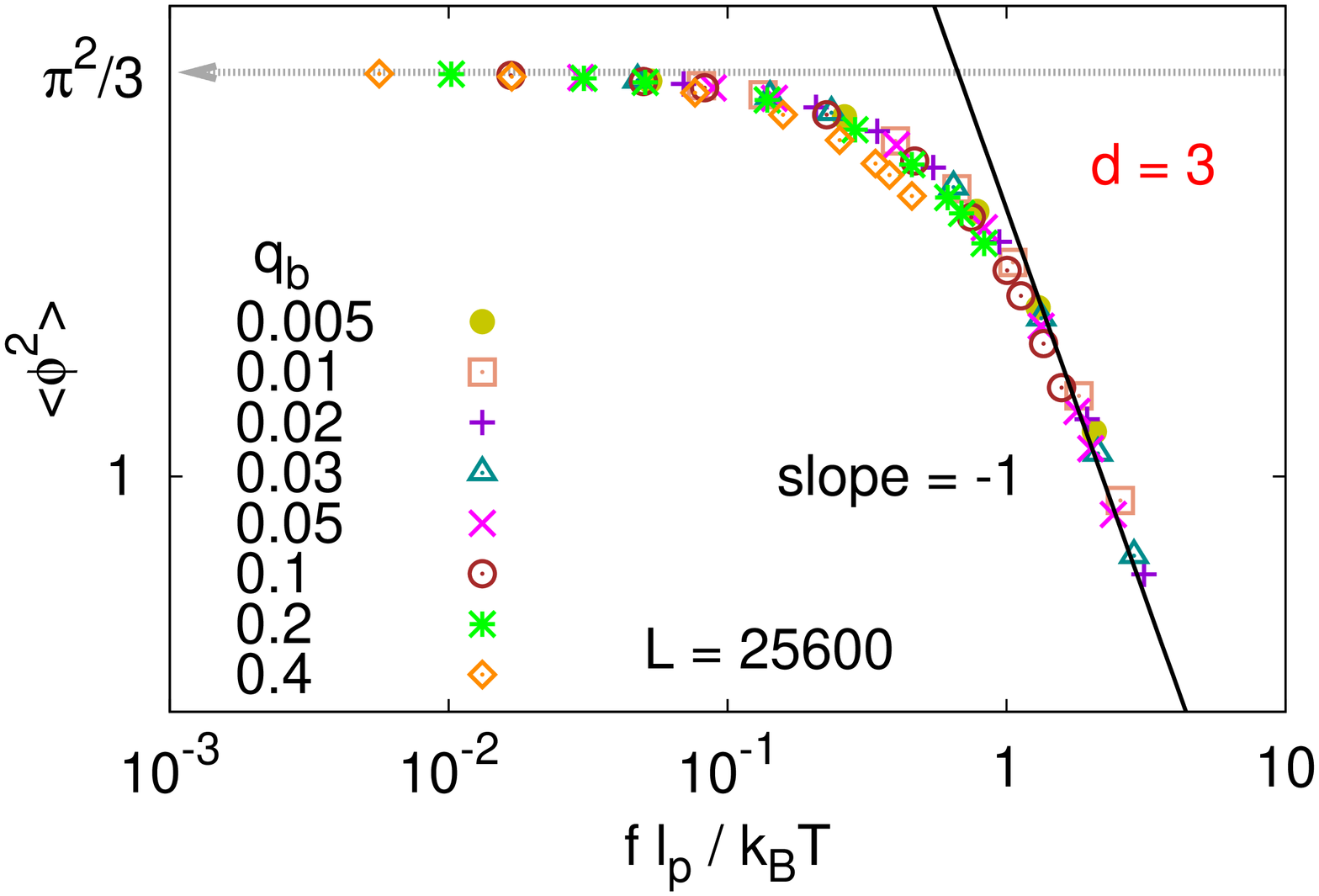}\\
\caption{Log-log plot of $\langle \phi^2 \rangle $ vs. $f\ell_p/k_BT$ for
$L=100$ (a) and $L=25600$ (b), and for many choices of $q_b$, as indicated.
$\langle \phi^2 \rangle=\pi^2/3$ as $f \rightarrow \infty$.
Data are for semiflexible chains in $d=3$.}
\label{fig18}
\end{center}
\end{figure*}

\section{Stretching semiflexible polymers in $d=3$ Dimensions}
We start by showing extension versus force curves for various choices of
the contour length $L$ in Fig.~\ref{fig15}, to provide a three-dimensional
counterpart to the data in Fig.~\ref{fig10} for two dimensions. It is 
immediately obvious that the simple Kratky-Porod prediction 
\{Eq.~(\ref{eq54})\} does a much better job than its two-dimensional
counterpart \{Eq.~(\ref{eq55})\}. 
Again, we emphasize that there are no adjustable parameters whatsoever
in our comparison, $L=N_b \ell_b$ is trivially known, and $\ell_p$ 
comes from Fig.~\ref{fig2}(d).
In fact, for $L=200$ and $L=400$ most of
the data for $0.01<\langle X \rangle <0.3$ follow the K-P prediction,
for a wide range of choices for $q_b$ and hence $\ell_p$ 
\{Table~\ref{table2}\}, only data for rather flexible chains (such as 
$q_b=0.4$, for which $\ell_p=1.2$) deviate strongly from the K-P model,
as expected. The simple Langevin function \{Eq.~(\ref{eq4})\} 
does not describe the behavior of these lattice chains with discrete
bond angles. Note that these chains are 
too short to show a well-developed Pincus blob regime yet.
For $\langle X \rangle /L >0.3$ systematic deviations from the K-P
prediction occur, which we attribute to effects due to the discreteness of
bonds and bond angles in our model. Only for very long chains (such as
$N=6400$ and $25600$) do we find more pronounced deviations from the
K-P prediction also for small relative extensions, 
$\langle X \rangle /L \leq 0.1$.

  For long chains, however, we do expect to see excluded volume
effects (manifested in Pincus blobs), as discussed in Sec.~II. Thus
Fig.~\ref{fig16} presents our data in suitably scaled form, considering
the crossover from the linear response regime to the Pincus blob regime,
both for flexible chains (a) and semiflexible ones (b), as well as the 
crossover from Pincus blobs to Kratky Porod behavior (c), and we show a
close-up of the Kratky-Porod regime for very long and at the same time rather
stiff chains (d). Of course, even with chain lengths up to $25600$ it is 
not yet possible to clearly resolve all the different power laws shown 
in Fig.~\ref{fig1} (b):
in order to be able to distinguish the various crossovers clearly from
each other, we would need very stiff chains ($\ell_p$ should then be in
the range $10^2 <\ell_p/\ell_b <10^4$), and then one would need to have 
chain lengths of many millions in order to have a well-developed 
Pincus-blob regime. Thus, we can verify the Pincus blob regime only for
rather flexible chains (Fig.~\ref{fig16}(a)), for which then 
a well-developed Kratky-Porod regime is absent. For the stiff chains, we 
can see some tendency of the data to deviate from the K-P regime in the
direction towards the Pincus blob regime (Fig.~\ref{fig16}(b)(c)), but
the latter is not fully reached because the crossover to the linear 
response takes over (Fig.~\ref{fig16}(b)). And when we study very 
stiff chains, we find deviations from the K-P model for rather small
$\langle X \rangle /L$ already, due to the discrete character of our chains.

   Fig.~\ref{fig17} shows again data for the normalized fluctuations 
of the chain linear-dimensions, and Fig.~\ref{fig18} presents a counterpart 
to Fig.~\ref{fig14}, showing a log-log plot of the local fluctuation
$\langle \phi^2 \rangle$ versus $f\ell_p/k_BT$.
While the latter (for large $L$) show again a simple crossover from the 
constant $\pi^2/3$ describing $\langle \phi^2 \rangle$ for small forces
to a power law $k_BT/f\ell_p$ for $f\ell_p/k_BT>1$, as in $d=2$ 
dimensions, the behavior of the fluctuations in the chain linear dimensions
clearly is rather complicated. Of course, there is a need to extend 
the scaling analysis, that was presented for $\langle X \rangle$ as a function
of $f\ell_p/k_BT$ in Fig.~\ref{fig1}(b) to the fluctuations 
$\langle X^2 \rangle -\langle X\rangle^2$ and $\langle R_\perp^2 \rangle$
in greater detail than we have done so far. We expect that analyzing these
fluctuations should yield additional and valuable information on the 
structure of stretched semiflexible chains, and allow to pin down the 
parameters needed to relate experimental data to theoretical models
more precisely. We plan to tackle this task in a forthcoming study.

\section{Conclusion}
In this paper, we have studied self-avoiding walks on square and simple
cubic lattices, where an energy penalty $\epsilon_b$ associated with chain
bending to model semiflexibility of the polymer chains, by extensive 
Monte Carlo simulations, using the PERM algorithm. We have obtained both
force versus extension curves and chain linear dimensions in the absence
of forces for a wide range of chain lengths $N_b$ (typically $N_b$ up
to $25600$) and chain stiffness (characterized by 
$q_b = \exp(-\epsilon_b/k_BT)$). In Sec.~II, we have attempted to present
a coherent phenomenological theoretical description, combining results
from scaling concepts with other results derived from the
Kratky-Porod model, to explain the various crossovers that can occur in the
force versus extension curves for various circumstances (Fig.~\ref{fig1}).
We have emphasized that the case of $d=2$ dimensions is rather different 
from the case $d=3$: only in the latter case one can identify a linear 
regime in the force versus extension curve that is compatible with
the Kratky-Porod model; the linear response regime both in $d=2$ and $d=3$
dimensions is strongly affected by the presence of excluded volume effects,
and for long enough chains is followed by a nonlinear (``Pincus blob")
regime for stronger forces both in $d=2$ and $d=3$ dimensions. However, for
very stiff and not too long chains in $d=3$ the chains in the absence of a 
force show Gaussian behavior, and in this case the stretched chains do not
exhibit the nonlinear Pincus blob regime, and the Kratky-Porod model holds
throughout (apart from very strong forces, where the discrete character
of polymer chains matter).

  The Monte Carlo data that we have generated do provide evidence for these 
concepts, particularly in the relatively simple case of $d=2$ dimensions.
While in $d=2$ all expected regimes of the force vs. extension curves
are confirmed and the expected scaling behavior is verified, problems remain
concerning the precise understanding of longitudinal and transverse 
fluctuations of chain linear dimensions of the chains. More work on 
these aspects (from theory, simulation, and experiment) clearly is 
desirable. We recall that imaging techniques can provide rather detailed 
information on chain configurations of semiflexible polymers adsorbed on 
substrates; we expect that our work should be useful to interpret 
such experiments.

  In the case of $d=3$ dimensions, our numerical evidence is much more 
limited: chain lengths $N_b=25600$ clearly do not suffice to fully 
resolve three 
distinct power laws (separated by smooth crossovers) in the force
versus extension curves. However, simulations for chains that are one
or two orders of magnitude larger clearly are not feasible at present.
We do obtain evidence, however, that for rather stiff thin short chains
excluded volume effects indeed are negligible, as expected,
and hence the Kratky-Porod model holds. However, when chain stiffness is 
due to thickness (persistence length $\ell_p$ being proportional to
local chain diameter $D$), the regime of Gaussian statistics disappears and
rather excluded volume effects dominate throughout, resulting in a rather 
broad regime where the force versus extension curve is 
nonlinear already for small $\langle X \rangle /L$.
Also for rather flexible chains, clear evidence for the Pincus blob 
regime is obtained (Fig.~\ref{fig16}(a)).

  In our modelling, we have approximated the interactions between monomers 
of the chain as a strictly local excluded volume interaction.
Of course, in many cases of interest the interactions are of longer range,
e.g. because of electrostatic interactions between charged groups.
Particularly for polyelectrolytes the resulting problem of an
``electrostatic persistence length" has received longstanding attention
in the 
literature~\cite{47,x,x1,x2}. Molecules such as DNA and RNA do possess
a substantial linear charge density, and the properties of such 
polyelectrolytes, in fact, will depend on the electrostatic screening due to
ions in the solution, and thus the effective persistence length will 
depend on ionic strength. In this context, the concept of an ``effective
thickness" of polyelectrolytes, that are described in terms of a ``thick
chain model"~\cite{x2}, has been used to model experimental extension
versus force curves. It will be an interesting task for the future, beyond
the scope of the present paper, to clarify the extent to which such
approaches are equivalent to the scaling concepts applied here.
In any case, it is very reassuring that very recently, after our study 
was completed, single-molecule elasticity measurements of the onset of 
excluded volume effects of stretched poly(ethylene glycol) were 
published~\cite{x3}. In this work, the Pincus blob scaling behavior
($L \propto f^{2/3}$ in $d=3$ dimensions) could be seen under several
circumstances, followed by a crossover to the linear behavior $(L\propto f)$
and subsequent saturation, compatible with the behavior predicted by the
K-P model. The interpretation given in Ref.~\cite{x3} for these experiments
is fully consistent with the description given in the present paper.
Some earlier evidence for the Pincus behavior was also found for
{DNA~\cite{x5,xx}}.

{
In our simulations, we have not considered the effects of varying the 
diameter $D$ of our chains (we have chosen $D=\ell_b=1$, the lattice spacing,
throughout). For biopolymers $D$ is a parameter of great interest as 
well~\cite{xx}, of course. In our studies of bottlebrush polymers~\cite{35,36},
however, we studied conditions for which $D\propto \ell_p$, and then the
K-P model was not useful even in $d=3$ dimensions. }

   Thus we hope that the present work will contribute to the better
understanding of both existing and future experiments. A very interesting
aspect, completely beyond the scope of the present work, are dynamic 
properties of stretched semiflexible polymers in solution, see 
e.g. Ref.~\cite{x4}. Our study should yield useful inputs for such
problems, too.

\begin{acknowledgments}
This work have been supposed by the Deutsche Forschungsgemeinschaft (DFG)
under grant No SFB 625/A3. We are particularly grateful to Wolfgang Paul
for his fruitful collaboration on some early aspects of this work.
Stimulating discussions with Ralf Everaers, Carlo Pierleoni, and
Hyuk Yu are acknowledged. We thank
the NIC J\"ulich for a generous grant of computer time on the JUROPA
supercomputer.
\end{acknowledgments}


\begin{thebibliography}{99}

\bibitem{1} P. J. Flory, \textit{Statistical Mechanics of Chain Molecules} 
(Wiley, New York, 1969).
\bibitem{1a} M. V. Volkenshtein and O. B. Ptitsyn, Sov. Phys. JETP \textbf{25},
649 (1955).
\bibitem{2} M. Fixman and J. Kovac, J. Chem. Phys. \textbf{58}, 1564 (1973).
\bibitem{3} R. G. Treloar, \textit{Physics of Rubber Elasticity}, 3rd ed. (Clarendon Press, Oxford, 1975).
\bibitem{4} P. Pincus, Macromolecules \textbf{9}, 386 (1976).
\bibitem{5} P. G. de Gennes, \textit{Scaling Concepts in Polymer Physics} (Cornell Univ. Press, Ithaca, N. Y., 1979).
\bibitem{6} I. Webman, J. L. Lebowitz, and M. H. Kalos, Phys. Rev. A \textbf{23}, 316 (1981).
\bibitem{7} J. Kovac and C. C. Crabb, Macromolecules \textbf{15}, 537 (1982).
\bibitem{8} A. Yu. Grosberg and A. R. Khokhlov, \textit{Statistical Physics of Macromolecules} (AIP Press, N. Y., 1994).
\bibitem{9} M. Wittkop, J.-U. Sommer, S. Kreitmeier and D. G\"oritz, Phys. Rev. E \textbf{49}, 5472 (1994).
\bibitem{10} P. Cifra and T. Blaha, J. Chem. Soc. Farady Trans. \textbf{91}, 2465 (1995).
\bibitem{11} J. F. Marko and E. D. Siggia, Macromolecules \textbf{28}, 8759 (1995).
\bibitem{12} K. Kroy and E. Frey, Phys. Rev. Lett. \textbf{77}, 306 (1996).
\bibitem{13} C. Pierleoni, G. Arialdi and J.-P. Ryckaert, Phys. Rev. Lett. \textbf{79}, 2990 (1997).
\bibitem{14} D. Bensimon, D. Dohmi, and M. M{\'e}zard, Europhys. Lett. \textbf{42}, 97 (1998).
\bibitem{15} J. J. Titantah, C. Pierleoni, and J.-P. Ryckaert, Phys. Rev. E \textbf{60}, 7010 (1999).
\bibitem{15A} A. Lamura, T. W. Burkhardt, and G. Gompper, Phys. Rev. E
\textbf{64}, 061801 (2001).
\bibitem{16} R. R. Netz, Macromolecules \textbf{34}, 7522 (2001).
\bibitem{17} L. Livadaru, R. R. Netz, and H. J. Kreuzer, Macromolecules \textbf{36}, 3732 (2003).
\bibitem{18} R. G. Winkler, J. Chem. Phys. \textbf{118}, 2919 (2003).
\bibitem{19} A. Rosa, T. X. Hoang, D. Marenduzzo, and A. Maritan, Macromolecules \textbf{36}, 10095 (2003).
\bibitem{20A} J. Kierfeld, O. Niamploy, V. Sa-yakanit, and R. Lipowsky, Eur. Phys. J. E \textbf{14}, 17 (2004).
\bibitem{20B} T. Hugel, M. Rief, M. Seitz, H. E. Gaub, and R. R. Netz,
Phys. Rev. Lett. \textbf{94}, 048301 (2005).
\bibitem{20} A. Prasad, Y. Hori, and J. Kondev, Phys. Rev. E \textbf{72}, 041918 (2005).
\bibitem{21} G. Morrison, C. Hyeon, N. M. Toan, B.-Y. Ha, and D. Thirumalai, Macromolecules \textbf{40}, 7343 (2007).
\bibitem{22} J. Krawczyk, I. Jensen, A. L. Owczarek, and S. Kumar, Phys. Rev. E \textbf{79}, 031912 (2009).
\bibitem{22A} A. V. Dobrynin, J.-M. Y. Carrillo, and M. Rubinstein,
Macromolecules \textbf{43}, 9181 (2010).
\bibitem{23} N. M. Toan and D. Thirumalai, Macromolecules \textbf{43}, 4394 (2010).
\bibitem{24} S. Kumar and M. S. Li, Phys. Rep. \textbf{486}, 1 (2010).
\bibitem{25} S. B. Smith, L. Finzi, and C. Bustamante, Science \textbf{258}, 
1112 (1992); M.-N. Dessinges, B. Maier, Y. Zhang, M. Peliti, D. Bensimon, 
and V. Croquette, Phys. Rev. Lett. \textbf{89}, 248102 (2002);
Y. Seol, G. M. Skinner, and K. Visscher, Phys. Rev. Lett. \textbf{93},
118102 (2004).
\bibitem{26} J. Liphardt, B. Onoa, S. B. Smith, I. Tinoco Jr. and C. Bustamante, Science \textbf{292}, 733 (2001).
\bibitem{27} M. Grandbois, M. Beyer, M. Rief, H. Clausen-Schaumann and H. E. Gaub, Science \textbf{283}, 1727 (1999).
\bibitem{28} M. Rief, M. Gautel, F. Oesterhelt, J. M. Fernandez and H. E. Gaub, Science \textbf{276}, 1109 (1997).
\bibitem{29} N. Gunari, M. Schmidt, and A. Janshoff, Macromolecules \textbf{39}, 2219 (2006).
\bibitem{30} A. Halperin and E. B. Zhulina, Europhys. Lett. \textbf{15}, 417 (1991).
\bibitem{31} D. Marenduzzo, A. Maritan, A. Rosa, and F. Seno, Eur. Phys. J. E \textbf{15}, 83 (2004).
\bibitem{32} D. Marenduzzo, A. Maritan, A. Rosa, and F. Seno, Phys. Rev. Lett. \textbf{90}, 088301 (2003).
\bibitem{33} S. Kumar and G. Mishra, Phys. Rev. E \textbf{78}, 011907 (2008).
\bibitem{34} S. Stepanow, Eur. Phys. J. B \textbf{39}, 499 (2004).
\bibitem{35A} A. A. Gorbunov and A. M. Skvortsov, J. Chem. Phys. \textbf{98}, 5961 (1993).
\bibitem{36A} S. Bhattacharya, V. G. Rostiashvilli, A. Milchev, and T. A. Vilgis, Macromolecules \textbf{42}, 2236 (2009).
\bibitem{37A} A. M. Skvortsov, L. I. Klushin, G. J. Fleer, and F. A. Leermakers, J. Chem. Phys. \textbf{132}, 064110 (2010).
\bibitem{35} H.-P. Hsu, W. Paul, and K. Binder, Macromolecules \textbf{43}, 3094 (2010).
\bibitem{36} H.-P. Hsu, W. Paul, and K. Binder, EPL \textbf{92}, 28003 (2010);
Macromol. Theory \& Simul. \textbf{20}, 510 (2011).

\bibitem{37} H.-P. Hsu, W. Paul, and K. Binder, EPL \textbf{95}, 68004 (2011).
\bibitem{38} L. Sch\"afer, A. Ostendorf, and J. Hager, J. Phys. A\textbf{32}, 7875 (1999).
\bibitem{39} O. Kratky and G. Porod, J. Colloid Sci. \textbf{4}, 35 (1949).
\bibitem{40} N. Saito, K. Takahashi, and Y. Yunoki, J. Phys. Soc. Japan \textbf{22}, 219 (1967).
\bibitem{41} L. Sch\"afer and K. Elsner, Eur. Phys. J. E \textbf{13}, 225 (2004).
\bibitem{42} R. Everaers, A. Milchev and V. Yamakov, Eur. Phys. J. E \textbf{8}, 3 (2002).
\bibitem{43} M. Rubinstein and R. H. Colby, \textit{Polymer
Physics} (Oxford Univ. Press, Oxford, 2003).
\bibitem{44} J. C. LeGuillou and J. Zinn-Justin, Phys. Rev.
B \textbf{21}, 3976 (1980).
\bibitem{45} J. Des Cloizeaux and G. Jannink, \textit{Polymers in
Solution: Their Modelling and Structure} (Clarendon Press, Oxford,
1990).
\bibitem{Lam} {P.-M. Lam, Biopolymers \textbf{64}, 57 (2002).}
\bibitem{46} S. F. Edwards and P. Singh, J. Chem. Soc., Faraday
Trans. \textbf{2}, 75, 1001 (1979).
\bibitem{47} R. G. Winkler, P. Reineker and L. Harnau, J. Chem.
Phys. \textbf{101}, 8119 (1994).
\bibitem{48} K. Kremer and K. Binder, Computer Phys.
Rep.~\textbf{7}, 259 (1988).
\bibitem{49} D W. Schaefer, J. F. Joanny, and P. Pincus,
Macromolecules \textbf{13}, 1280 (1980).
\bibitem{50} R. R. Netz and D. Andelman, Phys. Rep. \textbf{380},
1 (2003).
\bibitem{51} T. Norisuye and H. Fujita, Polymer J. \textbf{14},
143 (1982).
\bibitem{52} T. Odijk, Macromolecules \textbf{16}, 1340 (1983).
\bibitem{53} T. Odijk, Macromolecules \textbf{17}, 502 (1984).
\bibitem{54} J. Wilhelm and E. Frey, Phys. Rev. Lett. \textbf{77},
2581 (1996).
\bibitem{55} D. Thirumalai and B.-Y. Ha, cond-mat/9705200
\bibitem{56} J. K. Bhattacharjee, D. Thirumalai, and J. D.
Bryngelson, cond-mat/9709345.
\bibitem{57} J. Samuel and S. Sinha, Phys. Rev. E \textbf{66},
050801 (2002).
\bibitem{58} A. Dhar and D. Chaudhuri, Phys. Rev. Lett.
\textbf{89}, 065502 (2002).
\bibitem{59} C. R. Cantor and P. R. Schimmel, \textit{Biophysical
Chemistry, Part III, The Behavior of Biological Macromolecules}
(W. H. Freeman, San Francisco, 1980).
\bibitem{60} R. Lavery, A. Lebrun, J.-F. Allemand, D. Bensimon and
V. Croquette, J. Phys.: Condens. Matter \textbf{14}, R383 (2002).
\bibitem{61} P. Virnau, Y. Kantor, and M. Kardar, J. Am. Chem.
Soc. \textbf{127}, 15102 (2005).
\bibitem{62} A. Borgia, P. M. Williams, and J. Clarke, Annu. Rev.
Biochem. \textbf{77}, 101 (2008).
\bibitem{63} P. Virnau, A. Mallam, and S. Jackson, 
J. Phys.: Condens. Matter \textbf{23}, 033101 (2011).
\bibitem{64} P. Grassberger, Phys. Rev. E \textbf{56}, 3682 (1997).
\bibitem{65} U. Bastolla and P. Grassberger, J. Stat. Phys.
\textbf{89}, 1061 (1997).
\bibitem{66} H.-P. Hsu and P. Grassberger, J. Stat. Phys. 
\textbf{144}, 597 (2011).
\bibitem{Yoshinaga} N. Yoshinaga, K. Yoshikawa, and S. Kidoaki,
J. Chem. Phys. \textbf{116}, 9926 (2002).
\bibitem{x} U. Micka and K. Kremer, Phys. Rev. E \textbf{54}, 2653 (1996).
\bibitem{x1} M. Ullner, B. J\"onsson, C. Peterson, O. Sommelius, and
B. S\"oderberg, J. Chem. Phys. \textbf{107}, 1279 (1997);
M. Ullner, and C. E. Woodward, Macromolecules \textbf{35}, 1437 (2002).
\bibitem{x2} N. M. Toan and C. Micheletti, J. Phys.: Condens. Matter
\textbf{18}, S269 (2006).
\bibitem{x3} A. Dittmore, D. B. McIntosh, S. Halliday, and O. A. Saleh,
Phys. Rev. Lett. \textbf{107}, 148301 (2011).
\bibitem{x5} O. A. Saleh, D. B. McIntosh, P. Pincus, and N. Ribeck,
Phys. Rev. Lett. \textbf{102}, 068301 (2009);
D. B. McIntosh, N. Ribeck, and O. A. Saleh, Phys. Rev. E \textbf{80},
041803 (2009).
\bibitem{xx} {N. M. Toan, D. Marenduzzo, and C. Micheletti,
Biophy. J. \textbf{89}, 80 (2005).}
\bibitem{x4} J. W. Hatfield and S. R. Quake, Phy. Rev. Lett. \textbf{82},
3548 (1999).
\end{thebibliography}
\end{document}